\documentclass{article}
\usepackage{amssymb,amsmath,graphicx}
\usepackage{amsfonts,natbib}
\usepackage[caption=false]{subfig}
\usepackage{epsfig,latexsym,graphicx}
\newtheorem{theorem}{Theorem}[section]

\newtheorem{corollary}[theorem]{Corollary}

\newtheorem{example}{Example}[section]
\input{tcilatex}

\begin{document}

\title{A new class of robust two-sample Wald-type tests
}

\author{Abhik Ghosh$^{1\ast}$, Nirian Martin$^{2}$, Ayanendranath Basu$^{1}$ and Leandro Pardo$^{2}$
\\ $^1$ Indian Statistical Institute, Kolkata, India
\\ $^2$ Complutense University, Madrid, Spain
\\ $^{*}$Corresponding author; Email: abhianik@gmail.com 
}
\date{}
\maketitle

\begin{abstract}
Parametric hypothesis testing associated with two independent samples arises frequently in several applications 
in biology, medical sciences, epidemiology, reliability and many more. 
In this paper, we propose robust Wald-type tests for testing such two sample problems 
using the minimum density power divergence estimators of the underlying parameters.
In particular, we consider the simple two-sample hypothesis concerning the full parametric homogeneity of the samples
as well as the general two-sample (composite) hypotheses involving nuisance parameters also. 
The asymptotic and theoretical robustness properties of the proposed Wald-type tests have been developed
for both the simple and general composite hypotheses. Some particular cases of testing against one-sided alternatives 
are discussed with specific attention to testing the effectiveness of a treatment in clinical trials. 
Performances of the proposed tests have also been illustrated numerically through appropriate real data examples.

\end{abstract}

\bigskip

\section{Introduction}\label{SEC:intro}

Testing of parametric hypothesis is an important paradigm of statistical inference.
In many real life applications like medical sciences, biology, epidemiology, sociology, reliability etc.,
we need to compare data from two independent samples through appropriate two-sample tests of hypotheses.
Examples include but not limited to comparing mean of any biomarkers or success of any treatment  
between control and treatment groups, comparing lifetime of two populations in reliability etc.

Mathematically, let $\left( \mathcal{X}\text{, }\beta _{\mathcal{X}},\text{ }P_{\boldsymbol{{\boldsymbol{\theta}} }}
\right)_{\boldsymbol{{\boldsymbol{\theta}} }\mathbf{\in}\boldsymbol{\Theta }}$
be the statistical space associated with the random variable $X{,}$ where $%
\beta _{\mathcal{X}}$ is the $\sigma$-field of Borel subsets $A\subset 
\mathcal{X}$ and $\left\{ P_{\boldsymbol{{\boldsymbol{\theta}} }}\right\} _{\boldsymbol{%
		{\boldsymbol{\theta}} }\in \boldsymbol{\Theta }}$ is a family of probability distributions
defined on the measurable space $\left( \mathcal{X}\text{, }\beta _{\mathcal{%
		X}}\right) $ where $\boldsymbol{\Theta }$ is an open subset of $\mathbb{R}%
^{p}$, with $p\geq 1.$ Probability measures $P_{\boldsymbol{{\boldsymbol{\theta}} }}$ are
assumed to be described by densities $f_{\boldsymbol{{\boldsymbol{\theta}} }}\left(
x\right) =dP_{\boldsymbol{{\boldsymbol{\theta}} }}/d\mu \left( x\right) ,$ where $\mu $ is
a $\sigma $-finite measure on $\left( \mathcal{X}\text{, }\beta _{\mathcal{X}%
}\right) .$ We shall denote by 
$\mathcal{F}=\left\{ f_{\boldsymbol{{\boldsymbol{\theta}} }}:~\boldsymbol{{\boldsymbol{\theta}} \in \Theta\subset}\mathbb{R}^{p}\right\}$
a set of  parametric model densities.
On the basis of two independent random samples $X_{1},...,X_{n}$ and 
$Y_{1},....,Y_{m}$ of sizes $n$ and $m$, respectively, from two densities 
$f_{\mathbf{{\boldsymbol{\theta}} }_{1}}\left( x\right) $ and $f_{\mathbf{{\boldsymbol{\theta}} }_{2}}\left(x\right)$ 
belonging to $\mathcal{F}$, we can solve the problem of complete homogeneity 
by testing 
\begin{equation}
	H_{0}:\boldsymbol{{\boldsymbol{\theta}} }_{1}=\boldsymbol{{\boldsymbol{\theta}} }_{2}\text{ versus }H_{1}:%
	\boldsymbol{{\boldsymbol{\theta}} }_{1}\neq \boldsymbol{{\boldsymbol{\theta}} }_{2}.  \label{0.2}
\end{equation}
The classical test statistics for solving the problem of testing given in (%
\ref{0.2}) are the likelihood ratio test, Wald  test and Rao test and the
unknown parameters are estimated on the basis of the maximum likelihood
estimator (MLE). Some new tests statistics have been presented in the
literature based on divergence measures; see for instance \cite{Basu/etc:2011}
and \cite{Pardo:2006}. It is well-known that the MLE is a BAN estimator, i.e., is
efficient asymptotically but at the same time it has serious problems of
robustness. In order to avoid that problem, it has been introduced in the
statistical literature some procedures of testing based on estimators with
good behavior in relation to the robustness. In \cite{Basu/etc:2013} the
problem considered in (\ref{0.2}) was studied on the basis of the density
power divergence. They introduced a family of test statistics based on the
density power divergence between $f_{\mathbf{{\boldsymbol{\theta}} }_{1}}$ and $f_{\mathbf{%
		{\boldsymbol{\theta}} }_{2}}$ when the parameters are estimated considering the minimum
density power divergence estimator (MDPDE) of \cite{Basu/etc:1998}. 
For more details about MDPDE see Section \ref{SEC:MDPDE}. 

After solving the problem considered in (\ref{0.2}), we will be able to test
in normal populations, i.e., $f_{\mathbf{{\boldsymbol{\theta}} }_{1}}\equiv N(\mu
_{1},\sigma _{1})$ and $f_{\mathbf{{\boldsymbol{\theta}} }_{2}}\equiv N(\mu _{2},\sigma
_{2}),$ the following problem of complete homogeneity  
\[
H_{0}:(\mu _{1},\sigma _{1})=\left( \mu _{2},\sigma _{2}\right) \text{
	versus }H_{1}:(\mu _{1},\sigma _{1})\neq \left( \mu _{2},\sigma _{2}\right) .
\]%
But there are other interesting problems of homogeneity, for instance, to
test  
\begin{equation}
	H_{0}:\mu _{1}=\mu _{2}\text{ versus }H_{1}:\mu _{1}\neq \mu _{2}\nonumber
\end{equation}%
when the variance is the same but unknown, i.e., general composite hypothesis with two samples. 
This problem with normal population has been considered in \cite{Basu/etc:2015b} 
on the basis of introducing a family of test statistics based on the
density power divergence measure and estimating the unknown parameters using
the MDPDE. The results obtained were excellent in relation to the robustness
and efficiency since some tests were presented in which the lost of
efficiency in relation to the size and the power was not important but the
increase of robustness was very significant. One can think that the problems
considered previously can be solved in that way in a very satisfactory way
and from a theoretical point of view it is true but from a practical point
of view sometimes it is not very easy to get the density power divergence
measure between  $f_{\mathbf{{\boldsymbol{\theta}} }_{1}}$ and $f_{\mathbf{{\boldsymbol{\theta}} }_{2}}.$
In this paper we are going to present a family of test statistics which are
easy to calculate based on the MDPDE for any general two-sample problems with any parametric distribution. 
These test statistics are called Wald-type test statistics and their usefulness have been illustrated 
in the literature of one sample testing problems by \cite{Basu/etc:2016} and \cite{Ghosh/etc:2016}. 
In the present paper, not only we shall present the asymptotic distribution of the
Wald-type test statistics for two-sample problems but also a theoretical study of 
the robustness properties of them along with suitable examples and numerical illustrations. 

The rest of the paper is organized as follows:
In Section \ref{SEC:MDPDE} we present some results in relation to
the MDPDE that will be necessary for the rest of the paper. 
Section \ref{SEC:Simple_problem} is devoted to present the family of Wald-type tests for solving the problem of
complete homogeneity, i.e., the problem considered in (\ref{0.2}). 
After defining the Wald-type tests based on MDPDE, we study its asymptotic
distribution as well as the theoretical robustness properties with examples. 
In Section \ref{SEC:General_problem}, we present a family of Wald-type tests 
for general composite hypotheses in two sample context. 
We shall derive the asymptotic distribution of the Wald-type
tests introduced as well as its robustness properties. 
Illustrations will be provided for the special case of testing partial homogeneity 
in presence of nuisance parameters like, for example, 
testing equality of two normal means with unknown (nuisance) variances.
In Section \ref{SEC:One-sided_problem}, we briefly describe the extensions for testing the two-sample 
hypotheses against one-sided alternatives for some particular cases.
Section \ref{SEC:real_data} will present several real life applications of our proposal
with interesting data examples from applied sciences like medical, biology, reliability etc. 
Appropriate simulation studies with some comments on the choice of tuning parameters
will be presented in Section \ref{SEC:numerical}. 
The paper ends with a short concluding remark in Section \ref{SEC:conclusion}.
For brevity in presentation, the proofs of all the results have been moved to Appendix \ref{APP:Proof}.

\subsection{Minimum density power divergence estimator: Asymptotic properties and robustness}
\label{SEC:MDPDE}

Given any two densities $f_{\mathbf{{\boldsymbol{\theta}} }_{1}}$ and $f_{\mathbf{{\boldsymbol{\theta}} }%
	_{2}}$ from $\mathcal{F}$, the density power divergence with a nonnegative
tuning parameter $\beta $, is defined as \citep{Basu/etc:1998}
\begin{equation}
	d_{\beta }(f_{\mathbf{{\boldsymbol{\theta}} }_{1}},f_{\mathbf{{\boldsymbol{\theta}} }_{2}})=\left\{ 
	\begin{array}{ll}
		\int \left\{ f_{\mathbf{{\boldsymbol{\theta}} }_{2}}^{1+\beta }(x)-\left( 1+\frac{1}{\beta }%
		\right) f_{\mathbf{{\boldsymbol{\theta}} }_{2}}^{\beta }(x)f_{\mathbf{{\boldsymbol{\theta}} }_{1}}(%
		\boldsymbol{x})+\frac{1}{\beta }f_{\mathbf{{\boldsymbol{\theta}} }_{1}}^{1+\beta }(x)\right\} dx, 
		& \text{for}\mathrm{~}\beta >0, \\[2ex]
		\int f_{\mathbf{{\boldsymbol{\theta}} }_{1}}(x)\mathrm{ln}\left( \frac{f_{\mathbf{{\boldsymbol{\theta}} }%
				_{1}}(x)}{f_{\mathbf{{\boldsymbol{\theta}} }_{2}}(x)}\right) dx, & \text{for}\mathrm{~}%
		\beta =0.%
	\end{array}%
	\right.   \label{0.6}
\end{equation}%
The divergence corresponding to $\beta =0$ may be derived from the general
case by taking the continuous limit as $\beta \rightarrow 0^+$, and 
the resulting $d_{0}(f_{\mathbf{{\boldsymbol{\theta}} }_{1}},f_{\mathbf{{\boldsymbol{\theta}} }_{2}})$ turns out to
be the Kullback-Leibler divergence.

Let $G$ represent the distribution function corresponding to the density $g$ that generates the data. 
We model it by the model density $f_{\boldsymbol{\theta}} \in \mathcal{F}$ and 
we are interested in the estimation of $\boldsymbol{{\boldsymbol{\theta}}}$ based on an observed random sample from $g$. 
The corresponding minimum density power divergence functional at $G$ with tuning parameter $\beta$,
denoted by $\boldsymbol{U}_{\beta }(G)$, is defined as 
$d_{\beta }(g,f_{\boldsymbol{U}_{\beta }(G)})
=\displaystyle\min_{\boldsymbol{{\boldsymbol{{\boldsymbol{\theta}}}} }\in\Theta }d_{\beta }(g,f_{\boldsymbol{{\boldsymbol{\theta}} }})$.  
Therefore the MDPDE of $\boldsymbol{{\boldsymbol{\theta}} }$ with tuning parameter $\beta$ is given by
$\widehat{\boldsymbol{{\boldsymbol{\theta}} }}_{\beta }=\boldsymbol{U}_{\beta }(G_{n}),$
where $G_{n}$ is the empirical distribution function associated with the observed
random sample $X_{1},\ldots ,X_{n}$ from the population with density $g$. 
As the last term of equation (\ref{0.6}) does not depend on $\boldsymbol{{\boldsymbol{\theta}}}$, 
$\widehat{\boldsymbol{{\boldsymbol{\theta}} }}_{\beta }$ is given by%
\begin{eqnarray}
\widehat{\boldsymbol{{\boldsymbol{\theta}} }}_{\beta }&=&\arg \min_{\boldsymbol{{\boldsymbol{\theta}} }\in\Theta }
\left\{ \int f_{\boldsymbol{{\boldsymbol{\theta}} }}^{1+\beta }({x})d{x}-\left( 1+\frac{1}{\beta }\right) 
\frac{1}{n}\sum_{i=1}^{n}f_{		\boldsymbol{{\boldsymbol{\theta}} }}^{\beta }(X_{i})\right\} ,  
~~~~\mbox{ if }\beta >0,\label{0.9}\\
\mbox{and }~~~~~~~~~~~~~~
\widehat{\boldsymbol{{\boldsymbol{\theta}} }}_{\beta }&=&\arg \min_{\boldsymbol{{\boldsymbol{\theta}} }\in\Theta }
\left\{ -\frac{1}{n}\sum_{i=1}^{n}\mathrm{ln}f_{\boldsymbol{{\boldsymbol{\theta}} }}(X_{i})\right\},
~~~~~~~~~~~~~~~~~~~~~~~~~~~~~~~~~\mbox{ if }\beta =0.  \label{0.10}
\end{eqnarray}%
Notice that $\widehat{\boldsymbol{{\boldsymbol{\theta}} }}_{\beta }$ for $\beta =0$ 
coincides with the maximum likelihood estimator (MLE). Denoting

\begin{equation*}
	V_{\boldsymbol{{\boldsymbol{\theta}} }}\left( x\right) =\int f_{\boldsymbol{{\boldsymbol{\theta}} }%
	}^{1+\beta }(\boldsymbol{x})d\boldsymbol{x-}\left( 1+\frac{1}{\beta }\right)
	f_{\boldsymbol{{\boldsymbol{\theta}} }}^{\beta }(x),
\end{equation*}%
expression (\ref{0.9}) can be written as 
$	\widehat{\boldsymbol{{\boldsymbol{\theta}} }}_{\beta }=\arg \displaystyle\min_{\boldsymbol{{\boldsymbol{\theta}} }\in
		\Theta }\frac{1}{n}\displaystyle\sum_{i=1}^{n}V_{\boldsymbol{{\boldsymbol{\theta}} }}(X_{i}).$
It shows that the MDPDE is an M-estimator. 

The functional $\boldsymbol{T}_{\beta }(G)$ is Fisher consistent; it takes
the value $\boldsymbol{{\boldsymbol{\theta}} }$$_{0}$, the true value of the parameter,
when the true density is a member of the model, i.e. $g=f_{\boldsymbol{%
		{\boldsymbol{\theta}} }_{0}}$. Let us assume $g=f_{\boldsymbol{{\boldsymbol{\theta}} }_{0}}$, and define
the quantities%
\begin{equation}
	\boldsymbol{J}_{\beta }\left( \boldsymbol{{\boldsymbol{\theta}} }\right) =\int \boldsymbol{%
		u}_{\boldsymbol{{\boldsymbol{\theta}} }}(\boldsymbol{x})\boldsymbol{u}_{\boldsymbol{{\boldsymbol{\theta}} }%
	}^{T}(\boldsymbol{x})f_{\boldsymbol{{\boldsymbol{\theta}} }}^{1+\beta }(\boldsymbol{x})d%
	\boldsymbol{x},\text{\quad }\boldsymbol{K}_{\beta }\left( \boldsymbol{{\boldsymbol{\theta}} 
	}\right) =\int \boldsymbol{u}_{\boldsymbol{{\boldsymbol{\theta}} }}(\boldsymbol{x})%
	\boldsymbol{u}_{\boldsymbol{{\boldsymbol{\theta}} }}^{T}(\boldsymbol{x})f_{\boldsymbol{%
			{\boldsymbol{\theta}} }}^{1+2\beta }(\boldsymbol{x})d\boldsymbol{x}-\boldsymbol{\xi }%
	_{\beta }\left( \boldsymbol{{\boldsymbol{\theta}} }\right) \boldsymbol{\xi }_{\beta
	}^{T}\left( \boldsymbol{{\boldsymbol{\theta}} }\right) ,  \label{0.11}
\end{equation}%
where 	
$\boldsymbol{\xi }_{\beta }\left( \boldsymbol{{\boldsymbol{\theta}} }\right) 
=\int \boldsymbol{u}_{\boldsymbol{{\boldsymbol{\theta}} }}(\boldsymbol{x})
f_{\boldsymbol{{\boldsymbol{\theta}} }}^{1+\beta}(\boldsymbol{x})d\boldsymbol{x}$
and $\boldsymbol{u}_{\boldsymbol{{\boldsymbol{\theta}} }}(\boldsymbol{x})
=\frac{\partial }{\partial \boldsymbol{{\boldsymbol{\theta}} }}\mathrm{ln}f_{\boldsymbol{{\boldsymbol{\theta}} }}(\boldsymbol{x})$.
Then, following \cite{Basu/etc:1998,Basu/etc:2011}, it can be shown that, 
under Assumptions (D1)--(D5) of \citet[][p.~304]{Basu/etc:2011} to be referred as ``Basu et al.~conditions" in the rest of the paper,  
\begin{equation}
	n^{1/2}({\boldsymbol{\widehat{\boldsymbol{{\boldsymbol{\theta}} }}}}_{\beta }-\boldsymbol{%
		{\boldsymbol{\theta}} }_{0})\underset{n\rightarrow \infty }{\overset{\mathcal{L}}{%
			\longrightarrow }} {N}(\boldsymbol{0}_{p},\boldsymbol{{\boldsymbol{\Sigma}} }_{\beta
	}(\boldsymbol{{\boldsymbol{\theta}} }_{0})),  \label{0.12}
\end{equation}%
where 
$\boldsymbol{{\boldsymbol{\Sigma}} }_{\beta }(\boldsymbol{{\boldsymbol{\theta}} })
=\boldsymbol{J}_{\beta }^{-1}(\boldsymbol{{\boldsymbol{\theta}} })
\boldsymbol{K}_{\beta }(\boldsymbol{{\boldsymbol{\theta}} })
\boldsymbol{J}_{\beta }^{-1}(\boldsymbol{{\boldsymbol{\theta}} }).$
%
It is a simple exercise to see that for $\beta =0,$ $\boldsymbol{J}_{\beta
	=0}\left( \boldsymbol{{\boldsymbol{\theta}} }\right) =\boldsymbol{K}_{\beta =0}\left( 
\boldsymbol{{\boldsymbol{\theta}} }\right) =\boldsymbol{I}_{F}\left( \boldsymbol{{\boldsymbol{\theta}} }%
\right) ,$ being $\boldsymbol{I}_{F}\left( \boldsymbol{{\boldsymbol{\theta}} }\right) $ the
Fisher information matrix associated to the model under consideration.
Therefore we obtain the classical well known result, 
\begin{equation}
	n^{1/2}({\boldsymbol{\widehat{\boldsymbol{{\boldsymbol{\theta}} }}}}-\boldsymbol{{\boldsymbol{\theta}} }%
	_{0})\underset{n\rightarrow \infty }{\overset{\mathcal{L}}{\longrightarrow }}%
	{N}(\boldsymbol{0}_{p},\boldsymbol{I}_{F}^{-1}(\boldsymbol{{\boldsymbol{\theta}} }%
	_{0})).  \nonumber
\end{equation}

Next, the influence function can be used to study the robustness of the MDPDE like any other estimator or test statistic. 
If the influence function is bounded, the corresponding
estimator or the statistic is said to have infinitesimal robustness.
Therefore, the influence function particularly can be used to quantify
infinitesimal robustness of an estimator or a statistic by measuring the
approximate impact on an additional observation to the underlying data. More
simply, the influence function $\mathcal{IF}\left( x,\boldsymbol{U}_{\beta
},F_{\boldsymbol{{\boldsymbol{\theta}} }_{0}}\right) $ is the first derivative of an
estimator or statistic viewed as a functional $\boldsymbol{U}_{\beta }$ and
it describes the normalized influence on the estimate or statistic of an
infinitesimal observation $x$.
We pay special attention to the robustness of the family of Wald-type test statistics
introduced in this paper. To do that it is necessary to study the robustness of the MDPDEs. 
In \cite{Basu/etc:1998} it was established that
the influence function of the minimum density power divergence functional is 
\begin{equation}
	\mathcal{IF}\left( x,\boldsymbol{U}_{\beta },F_{\boldsymbol{{\boldsymbol{\theta}} }%
		_{0}}\right) =\lim_{\varepsilon \rightarrow 0}\frac{\boldsymbol{U}_{\beta
		}\left( F_{\varepsilon }\right) -\boldsymbol{U}_{\beta }\left( F_{%
		\boldsymbol{{\boldsymbol{\theta}} }_{0}}\right) }{\varepsilon }=\boldsymbol{J}_{\beta
}^{-1}(\boldsymbol{{\boldsymbol{\theta}} }_{0})\left( \boldsymbol{u}_{\boldsymbol{{\boldsymbol{\theta}} }%
}\left( x\right) f_{\boldsymbol{{\boldsymbol{\theta}} }_{0}}^{\beta }(x)-\boldsymbol{\xi }%
\left( \boldsymbol{{\boldsymbol{\theta}} }_{0}\right) \right) ,  \label{0.15}
\end{equation}
where $F_{\varepsilon }=(1-\varepsilon )F_{\boldsymbol{{\boldsymbol{\theta}} }_{0}}+\varepsilon \wedge _{x}$ 
is the $\varepsilon$-contaminated distribution of $F_{\boldsymbol{{\boldsymbol{\theta}} }_{0}}$, 
the distribution function corresponding to $f_{\boldsymbol{\theta}}$, 
with respect to the point mass distribution $\wedge _{x}$ at $x$. 
If we assume that $\boldsymbol{J}_{\beta }($$\boldsymbol{{\boldsymbol{\theta}} }$$_{0})$ and 
$\boldsymbol{\xi}\left( \boldsymbol{{\boldsymbol{\theta}} }_{0}\right)$ are finite, 
the influence function is a bounded function of $x$ whenever 
$\boldsymbol{u}_{\boldsymbol{{\boldsymbol{\theta}} }}\left( x\right) f_{\boldsymbol{{\boldsymbol{\theta}} }_{0}}^{\beta }(x)$ is bounded.
And this is the case for most common parametric models at $\beta>0$ implying the robustness of MDPDEs with $\beta>0$.

\section{A Simple Two-Sample Problem}
\label{SEC:Simple_problem}

Let $X_{1},...,X_{n}$ and $Y_{1},...,Y_{m}$ be two samples of sizes $n$ and $m$ respectively 
from two populations having distribution belonging to $\mathcal{F}$
with parameters $\boldsymbol{{\boldsymbol{\theta}} }_{1}$ and $\boldsymbol{{\boldsymbol{\theta}} }_{2}$.
The most common problem under this setup is to test the complete homogeneity
of the two populations. But we have two different situations depending if
some of the parameters are known. To clarify this point we can be interested
in testing the homogeneity of two normal populations with means and
variances unknown or with known variances. 
The problem of testing the homogeneity of two normal populations with unknown and common variance will be studied in the next Section. 
In general notation, we shall assume that,%
\begin{equation*}
	\boldsymbol{{\boldsymbol{\theta}} }_{1}=\left(\theta_{1,1},...,\theta_{1,r},\theta_{1,r+1},...,\theta_{1,p}\right)^{T}
	=\left( ^{\ast }\boldsymbol{{\boldsymbol{\theta}} }_{1}^{T},^{0}\boldsymbol{{\boldsymbol{\theta}} }_{1}^{T}\right) ^{T}
~\mbox{ and }~\boldsymbol{{\boldsymbol{\theta}} }_{2}=\left(\theta_{21,1},...,\theta_{2,r},\theta_{2,r+1},...,\theta_{2,p}\right)^{T}
	=\left( ^{\ast }\boldsymbol{{\boldsymbol{\theta}} }_{2}^{T},^{0}\boldsymbol{{\boldsymbol{\theta}} }_{2}^{T}\right) ^{T}~~~~~
\end{equation*}%
with $^{0}\boldsymbol{{\boldsymbol{\theta}} }_{1}$ and $^{0}\boldsymbol{{\boldsymbol{\theta}} }_{2}$ known $(p-r)$-vectors. 
Based on $X_{1},...,X_{n}$ we can get the MLE, 
$^{\ast }\widehat{\boldsymbol{{\boldsymbol{\theta}} }}_{1},$ of $^{\ast }\boldsymbol{{\boldsymbol{\theta}} }_{1}$ and based
on $Y_{1},...,Y_{m}$ the MLE, $^{\ast }\widehat{\boldsymbol{{\boldsymbol{\theta}} }}_{2},$
of $^{\ast }\boldsymbol{{\boldsymbol{\theta}} }_{2}$. Assuming $^{\ast }\boldsymbol{{\boldsymbol{\theta}} }%
_{1}=^{\ast }\boldsymbol{{\boldsymbol{\theta}} }_{2}$ we can obtain an estimator, $_{\left(
	o\right) }^{\ast }\widehat{\boldsymbol{{\boldsymbol{\theta}} }}_{1}$ of the common value $%
^{\ast }\boldsymbol{{\boldsymbol{\theta}} }_{1}$ by using the two random samples
$X_{1},...,X_{n}$ and $Y_{1},...,Y_{m}$ together. It is well-known that, under 
$^{\ast }\boldsymbol{{\boldsymbol{\theta}} }_{1}=^{\ast }\boldsymbol{{\boldsymbol{\theta}} }_{2}$,
\begin{equation}
	\sqrt{\frac{mn}{m+n}}\left( ^{\ast }\widehat{\boldsymbol{{\boldsymbol{\theta}} }}%
	_{1}-^{\ast }\widehat{\boldsymbol{{\boldsymbol{\theta}} }}_{2}\right) \underset{%
		n\rightarrow \infty }{\overset{\mathcal{L}}{\longrightarrow }}\mathcal{N}(%
	\boldsymbol{0}_{r},\omega \boldsymbol{I}_{F}^{-1}(^{\ast }\boldsymbol{%
		{\boldsymbol{\theta}} }_{1},^{0}\boldsymbol{{\boldsymbol{\theta}} }_{1})+\left( 1-\omega \right) 
	\boldsymbol{I}_{F}^{-1}(^{\ast }\boldsymbol{{\boldsymbol{\theta}} }_{1},^{0}\boldsymbol{%
		{\boldsymbol{\theta}} }_{2}))  \label{0.16}
\end{equation}%
with 
\begin{equation*}
	\omega =\lim_{m,n\rightarrow \infty }\frac{m}{m+n}.
\end{equation*}%
Based on (\ref{0.16}) we can consider the Wald test for testing%
\begin{equation*}
	H_{0}:^{\ast }\boldsymbol{{\boldsymbol{\theta}} }_{1}=^{\ast }\boldsymbol{{\boldsymbol{\theta}} }_{2}\text{
		versus }H_{1}:^{\ast }\boldsymbol{{\boldsymbol{\theta}} }_{1}\neq ^{\ast }\boldsymbol{%
		{\boldsymbol{\theta}} }_{2},
	\label{EQ:7two_sample}
\end{equation*}%
\begin{eqnarray}
\mbox{given by }~~~~~~W_{m,n} &=&\frac{mn}{m+n}\left( ^{\ast }\widehat{\boldsymbol{{\boldsymbol{\theta}} }}%
	_{1}-^{\ast }\widehat{\boldsymbol{{\boldsymbol{\theta}} }}_{2}\right) ^{T}\left( \frac{m%
		\boldsymbol{I}_{F}^{-1}(^{\ast }\widehat{\boldsymbol{{\boldsymbol{\theta}} }}_{1},^{0}\boldsymbol{%
			{\boldsymbol{\theta}} }_{1})}{m+n}+\frac{n\boldsymbol{I}_{F}^{-1}(^{\ast }\widehat{\boldsymbol{%
			{\boldsymbol{\theta}} }}_{2},^{0}\boldsymbol{{\boldsymbol{\theta}} }_{2})}{m+n}\right) ^{-1}\left( ^{\ast }%
	\widehat{\boldsymbol{{\boldsymbol{\theta}} }}_{1}-^{\ast }\widehat{\boldsymbol{{\boldsymbol{\theta}} }}%
	_{2}\right)   ~~~~~~~~~~~~~~~~~~~~~~\nonumber
	\\
	&=&mn\left( ^{\ast }\widehat{\boldsymbol{{\boldsymbol{\theta}} }}_{1}-^{\ast }\widehat{%
		\boldsymbol{{\boldsymbol{\theta}} }}_{2}\right) ^{T}\left( m\boldsymbol{I}_{F}^{-1}(^{\ast }%
	\widehat{\boldsymbol{{\boldsymbol{\theta}}}}_{1},^{0}\boldsymbol{{\boldsymbol{\theta}} }_{1})+n\boldsymbol{I}%
	_{F}^{-1}(^{\ast }\widehat{\boldsymbol{{\boldsymbol{\theta}}}}_{2},^{0}\boldsymbol{{\boldsymbol{\theta}} }%
	_{2})\right) ^{-1}\left( ^{\ast }\widehat{\boldsymbol{{\boldsymbol{\theta}} }}_{1}-^{\ast }%
	\widehat{\boldsymbol{{\boldsymbol{\theta}} }}_{2}\right) .  \notag
\end{eqnarray}%
We can observe that in the case that $r=p$ we have 
$\boldsymbol{I}_{F}^{-1}(^{\ast }\boldsymbol{{\boldsymbol{\theta}} }_{1},^{0}\boldsymbol{%
		{\boldsymbol{\theta}} }_{1})=\boldsymbol{I}_{F}^{-1}(^{\ast }\boldsymbol{{\boldsymbol{\theta}} }_{1},^{0}%
	\boldsymbol{{\boldsymbol{\theta}} }_{2})=\boldsymbol{I}_{F}^{-1}(\boldsymbol{{\boldsymbol{\theta}} }_{0}),$
with $\boldsymbol{{\boldsymbol{\theta}} }_{1}=\boldsymbol{{\boldsymbol{\theta}} }_{2}=\boldsymbol{{\boldsymbol{\theta}} }%
_{0}$ and the Wald test is given by 
\begin{equation}
W_{m,n}=\frac{mn}{m+n}\left( \widehat{\boldsymbol{{\boldsymbol{\theta}} }}_{1}-\widehat{%
\boldsymbol{{\boldsymbol{\theta}} }}_{2}\right) ^{T}\boldsymbol{I}_{F}(^{(0)}\widehat{\boldsymbol{{\boldsymbol{\theta}}}})
\left( \widehat{\boldsymbol{{\boldsymbol{\theta}} }}_{1}-\widehat{\boldsymbol{{\boldsymbol{\theta}} }}_{2}\right),
\label{EQ:Wald-TS}
\end{equation}
where $^{(0)}\widehat{\boldsymbol{{\boldsymbol{\theta}}}}$ denotes the 
MLE of $\boldsymbol{{\boldsymbol{\theta}}}_0$ based on the pooled sample.

Based on (\ref{0.16}) in the case of two normal populations,  with known
variances $\sigma _{1}^{2}$ and $\sigma _{2}^{2}$, we can test  
$	H_{0}:\mu _{1}=\mu _{2}.$
In this case 
\begin{equation*}
	W_{m,n}=mn\frac{\left( \widehat{\mu }_{1}-\widehat{\mu }_{2}\right) ^{2}}{%
		m\sigma _{1}^{2}+n\sigma _{2}^{2}}=\frac{\left( \widehat{\mu }_{1}-\widehat{%
			\mu }_{2}\right) ^{2}}{\frac{\sigma _{1}^{2}}{n}+\frac{\sigma _{2}^{2}}{m}}.
\end{equation*}%
Although it has several nice optimum properties, it is highly non-robust in
presence of outliers even in any one sample. Here, we will generalize this
Wald test to make it robust by replacing the MLE by the corresponding MDPDEs.

In the following we shall present the results for $r=p$, i.e., to test for the hypothesis in (\ref{0.2}). 
The case $r<p$ can be studied in a similar way.

Let us assume $^{(1)}\widehat{{\boldsymbol{\theta}}}_{\beta}$  and $^{(2)}\widehat{{\boldsymbol{\theta}}}_{\beta}$  
denote the MDPDEs of ${\boldsymbol{\theta}}_1$ and ${\boldsymbol{\theta}}_2$ respectively, 
obtained by minimizing the DPD with tuning parameter $\beta$ for each of the two samples separately. 
Further, under the null hypothesis $H_0: {\boldsymbol{\theta}}_1 = {\boldsymbol{\theta}}_2 = {\boldsymbol{\theta}}_0$ in (\ref{0.2}),  
we can consider the two samples pooled together as one i.i.d.~sample of size $m+n$ from a population 
having density function $f_{{\boldsymbol{\theta}}_0}$; 
let $^{(0)}\widehat{{\boldsymbol{\theta}}}_{\beta}$ denote the corresponding MDPDE 
of ${\boldsymbol{\theta}}_0$ with tuning parameter $\beta$ based on the pooled sample. 
Note that, all the three estimators $^{(1)}\widehat{{\boldsymbol{\theta}}}_{\beta}$ , $^{(2)}\widehat{{\boldsymbol{\theta}}}_{\beta}$  
and $^{(0)}\widehat{{\boldsymbol{\theta}}}_{\beta}$ should coincide with ${\boldsymbol{\theta}}_0$ asymptotically 
under $H_0$ with probability tending to one. 
Assuming identifiability of the model family, the difference between the two estimators $^{(1)}\widehat{{\boldsymbol{\theta}}}_{\beta}$ 
and $^{(2)}\widehat{{\boldsymbol{\theta}}}_{\beta}$ gives us an idea of 
the distinction between the two samples and hence indicate any departure from the null hypothesis.
So, we define a generalized Wald-type test statistics by
\begin{eqnarray}
\label{EQ:7_2SDT}
T_{m,n}^{(\beta)}
= \frac{nm}{n+m} ~ \left( ^{(1)}\widehat{{\boldsymbol{\theta}}}_{\beta} - ^{(2)}\widehat{{\boldsymbol{\theta}}}_{\beta}\right)^T
{\boldsymbol{\Sigma}}_{\beta}\left( ^{(0)}\widehat{{\boldsymbol{\theta}}}_{\beta}\right)^{-1}
\left( ^{(1)}\widehat{{\boldsymbol{\theta}}}_{\beta} - ^{(2)}\widehat{{\boldsymbol{\theta}}}_{\beta}\right).
\end{eqnarray}
Note that, at $\beta=0$, all the MDPDEs used coincide with corresponding MLEs and hence the
generalized Wald-type test statistic $T_{m,n}^{(\beta)}$ coincides with the classical Wald test statistic $W_{m,n}$ in (\ref{EQ:Wald-TS}).

\subsection{Asymptotic Properties}

In order to perform any statistical test, we first need to derive the asymptotic distribution of 
the test statistics under $H_0$. 
Using the asymptotic properties of the MDPDEs presented in Section \ref{SEC:MDPDE}, 
we can easily obtain the asymptotic null distribution of the proposed test statistics  $T_{m,n}^{(\beta)}$ 
which is presented in the following theorem. 
Throughout the rest of the paper, we will assume Conditions (A)--(D) of \citet[][p.~429]{Lehmann:1983}
about the assumed model family which we will refer as ``Lehmann conditions".
Also, we consider the following assumption.

\noindent
\textbf{Assumption (A):}
\begin{enumerate}
\item $\frac{m}{m+n} \rightarrow \omega \in (0, 1)$ as $m,n \rightarrow \infty$
\item The asymptotic variance-covariance matrix ${\boldsymbol{\Sigma}}_{\beta}({\boldsymbol{\theta}})$ 
of the MDPDE with tuning parameter $\beta$
is continuous in ${\boldsymbol{\theta}}$.
\end{enumerate}

\begin{theorem}
Suppose the model density satisfies the Lehmann and Basu et al.~conditions, and Assumption (A) holds.
Then the asymptotic distribution of $T_{m,n}^{(\beta)}$ under the null hypothesis in (\ref{EQ:7two_sample}) 
is $\chi_p^2$, the chi-square distribution with $p$ degrees of freedom.
\label{THM:2sample_simple_asympNull}
\end{theorem}

The asymptotic null distribution of the test in \cite{Basu/etc:2013} is a linear combination of chi-square
distribution and hence it is somewhat difficult to obtain the critical values of their test in practice.
On the contrary, our proposed tests have a simple chi-square limit under the null hypothesis and hence 
are much easier to perform. Our proposal provides, in this sense, an advantageous procedure for testing.

However, when the null hypothesis is not correct, i.e., ${\boldsymbol{\theta}}_1 \neq {\boldsymbol{\theta}}_2$, then the pooled estimator 
$^{(0)}\widehat{{\boldsymbol{\theta}}}_{\beta}$ no longer converges to ${\boldsymbol{\theta}}_1$ or ${\boldsymbol{\theta}}_2$; 
rather it will then converges in probability to 
a new value ${\boldsymbol{\theta}}_3$, say, which is a function of ${\boldsymbol{\theta}}_1$, ${\boldsymbol{\theta}}_2$ and $\omega$. For example, 
if the estimators are additive in sample data, e.g.~sample mean, then we will have 
${\boldsymbol{\theta}}_3 = (1-\omega) {\boldsymbol{\theta}}_1 + \omega{\boldsymbol{\theta}}_2$. Define 
$
l_{{\boldsymbol{\theta}} _{3},\beta }^{\ast }({\boldsymbol{\theta}} _{1},{\boldsymbol{\theta}} _{2})
=({\boldsymbol{\theta}} _{1}-{\boldsymbol{\theta}}_{2})^{T}{\boldsymbol{\Sigma}} _{\beta }({\boldsymbol{\theta}} _{3})^{-1}({\boldsymbol{\theta}} _{1}-{\boldsymbol{\theta}} _{2}).
$ 
Then we have the following result.

\begin{theorem}
  \label{THM:7two_sample_power1}
Suppose the model density satisfies the Lehmann and Basu et al.~conditions, and Assumption (A) holds. 
Then, as $m, n \rightarrow\infty$, we have for any ${\boldsymbol{\theta}}_1\neq{\boldsymbol{\theta}}_2$ 
\begin{eqnarray}
\sqrt{\frac{mn}{m+n}}
\left[l_{^{(0)}\widehat{{\boldsymbol{\theta}}}_{\beta},\beta}^{\ast }(^{(1)}\widehat{{\boldsymbol{\theta}}}_{\beta},^{(2)}\widehat{{\boldsymbol{\theta}}}_{\beta}) 
- l_{{\boldsymbol{\theta}} _{3},\beta }^{\ast }({\boldsymbol{\theta}} _{1},{\boldsymbol{\theta}} _{2})\right]
\underset{m, n\rightarrow \infty }{\overset{\mathcal{L}}{\longrightarrow }}
N\left(0, 4\sigma_{{\boldsymbol{\theta}}_3,\beta}^2({\boldsymbol{\theta}}_1,{\boldsymbol{\theta}}_2)\right),
\end{eqnarray}  
where
$\sigma_{{\boldsymbol{\theta}}_3,\beta}^2({\boldsymbol{\theta}}_1,{\boldsymbol{\theta}}_2)=({\boldsymbol{\theta}}_1 - {\boldsymbol{\theta}}_2)^T{\boldsymbol{\Sigma}}_{\beta}({\boldsymbol{\theta}}_3)^{-1}
\left[\omega{\boldsymbol{\Sigma}}_{\beta}({\boldsymbol{\theta}}_1) + (1-\omega){\boldsymbol{\Sigma}}_{\beta}({\boldsymbol{\theta}}_2)\right]
{\boldsymbol{\Sigma}}_{\beta}({\boldsymbol{\theta}}_3)^{-1}({\boldsymbol{\theta}}_1 - {\boldsymbol{\theta}}_2).$
\end{theorem}

This theorem leads to an approximation to the power function  
$\pi_{m,n,\alpha}^{(\beta)} ({\boldsymbol{\theta}}_1, {\boldsymbol{\theta}}_2) = P\left(T_{m,n}^{(\beta)} > \chi_{p,\alpha}^2\right)$
of the proposed Wald-type tests for testing (\ref{EQ:7two_sample}) at the significance level $\alpha$,
where $\chi_{p,\alpha}^2$ denotes the $(1-\alpha)$-th quantile of the $\chi_p^2$ distribution.

\begin{corollary}
Under the assumption of Theorem \ref{THM:7two_sample_power1}, we have
  \begin{eqnarray}
  \pi_{m,n,\alpha}^{(\beta)} ({\boldsymbol{\theta}}_1, {\boldsymbol{\theta}}_2) 
= 1 - \Phi_n \left( \frac{\sqrt{\frac{n+m}{nm}}}  {2\sigma_{{\boldsymbol{\theta}}_3,\beta}({\boldsymbol{\theta}}_1, {\boldsymbol{\theta}}_2)} 
\left[ \chi_{p,\alpha}^2-  \frac{nm}{n+m}   l_{{\boldsymbol{\theta}} _{3},\beta }^{\ast }({\boldsymbol{\theta}} _{1},{\boldsymbol{\theta}} _{2}) \right] \right), 
  ~~~~~~~~ {\boldsymbol{\theta}}_1 \ne {\boldsymbol{\theta}}_2,\nonumber
  \end{eqnarray}
for a sequence of distributions $\Phi_n(\cdot)$ tending uniformly to the standard normal distribution $\Phi(\cdot)$.
\end{corollary}

The corollary also helps us to determine the sample size requirement for our proposed test to achieve any pre-specified power level. 
Further, we have $\pi_{m,n,\alpha}^{(\beta)} ({\boldsymbol{\theta}}_1, {\boldsymbol{\theta}}_2)\rightarrow 1$ 
for any  ${\boldsymbol{\theta}}_1 \ne {\boldsymbol{\theta}}_2$ as $ m, n \rightarrow \infty$. 
Hence the proposed test with rejection rule
$ \left\{T_{m,n}^{(\beta)} >  \chi_{p,\alpha}^2\right\}$  is consistent.

\begin{corollary}\label{Corr:simple_const}
Under the assumption of Theorem \ref{THM:7two_sample_power1}, the proposed Wald-type test is consistent in the Fraser's sense.
\end{corollary}

Next, we look at the performance of the proposed test under the contiguous alternatives. 
Now, in case of two sample problem, we can have different types of contiguous alternatives.
For example, we can assume ${\boldsymbol{\theta}}_2$ to be fixed and ${\boldsymbol{\theta}}_1$ converging to ${\boldsymbol{\theta}}_2$ so that
$H_{1,n}' : {\boldsymbol{\theta}}_1={\boldsymbol{\theta}}_{1,n} = {\boldsymbol{\theta}}_2 + n^{-\frac{1}{2}}{\boldsymbol{\Delta}}_1$ 
for some $p$-vector ${\boldsymbol{\Delta}}_1$ of non-zero reals such that
${\boldsymbol{\theta}}_2 + n^{-\frac{1}{2}}{\boldsymbol{\Delta}}_1\in \Theta$. 
Conversely, we can have ${\boldsymbol{\theta}}_1$ to be fixed and 
$H_{1,m}'' : {\boldsymbol{\theta}}_2={\boldsymbol{\theta}}_{2,m} = {\boldsymbol{\theta}}_1 + m^{-\frac{1}{2}}{\boldsymbol{\Delta}}_2$ 
for some ${\boldsymbol{\Delta}}_2\in \mathbb{R}^p-\{\boldsymbol{0}\}$ with 
${\boldsymbol{\theta}}_1 + m^{-\frac{1}{2}}{\boldsymbol{\Delta}}_2\in \Theta$.
Here, we consider a general form of the contiguous alternative given by 
\begin{equation}
H_{1,n,m}: {\boldsymbol{\theta}}_1={\boldsymbol{\theta}}_{1,n} = {\boldsymbol{\theta}}_0 + n^{-\frac{1}{2}}{\boldsymbol{\Delta}}_1,~~
{\boldsymbol{\theta}}_2={\boldsymbol{\theta}}_{2,m} = {\boldsymbol{\theta}}_0 + m^{-\frac{1}{2}}{\boldsymbol{\Delta}}_2, ~~~~~~
({\boldsymbol{\Delta}}_1, {\boldsymbol{\Delta}}_2)\in \mathbb{R}^p\times\mathbb{R}^p-\{(\boldsymbol{0}_p,\boldsymbol{0}_p)\},
\label{EQ:Contiguous_alternative}
\end{equation} 
for some fixed ${\boldsymbol{\theta}}_0\in \Theta$. Note that, putting ${\boldsymbol{\Delta}}_2=\boldsymbol{0}$ 
in (\ref{EQ:Contiguous_alternative}) we get $H_{1,n}'$ back from $H_{1,n,m}$, 
whereas ${\boldsymbol{\Delta}}_1=\boldsymbol{0}$ yields $H_{1,m}''$. 
The following theorem gives the asymptotic distribution of the proposed test statistics
$T_{m,n}^{(\beta)}$ under this general contiguous alternatives $H_{1,m,n}$.

\begin{theorem}
Suppose the model density satisfies the Lehmann and Basu et al.~conditions and the assumption (A) holds.
Then the asymptotic distribution of $T_{m,n}^{(\beta)}$ 
under the contiguous alternative $H_{1,n,m}$ given by (\ref{EQ:Contiguous_alternative}) 
is $\chi^2_p(\delta_{\beta})$, the non-central chi-square distribution with $p$ degrees of freedom 
and non-centrality parameter 
$\delta_{\beta} =\boldsymbol{W}({\boldsymbol{\Delta}}_1,{\boldsymbol{\Delta}}_2)^T
{\boldsymbol{\Sigma}}_{\beta}({\boldsymbol{\theta}}_0)^{-1}\boldsymbol{W}({\boldsymbol{\Delta}}_1,{\boldsymbol{\Delta}}_2)$ 
with
$\boldsymbol{W}({\boldsymbol{\Delta}}_1,{\boldsymbol{\Delta}}_2)
=\left[\sqrt{\omega}{\boldsymbol{\Delta}}_1 - \sqrt{1-\omega}{\boldsymbol{\Delta}}_2\right]$.
\label{THM:2sample_simple_asympCont1}
\end{theorem}

We can easily obtain the asymptotic power $\pi_{\beta}({\boldsymbol{\Delta}}_1,{\boldsymbol{\Delta}}_2)$ under the contiguous alternatives $H_{1,n,m}$ from the above theorem. In particular, denoting the distribution function of a random variable $Z$ by $F_Z$, we have
\begin{eqnarray}
\pi_{\beta}({\boldsymbol{\Delta}}_1,{\boldsymbol{\Delta}}_2) = 1 - F_{\chi^2_p(\delta_{\beta})}(\chi_{p,\alpha}^2).
\label{EQ:simpel_contgPower}
\end{eqnarray}

\begin{example}[Testing equality of two Normal means with known equal variances]\label{EXM:1}
\normalfont{
We first present the simplest possible case of testing two normal means with known equal variance $\sigma^2$.
Here the model family is $\mathcal{F}=\{N(\theta,\sigma^2):\theta\in\mathbb{R}\}$ with $\sigma$ being known.
In this case, the asymptotic variance ${{\Sigma}}_{\beta}({{\theta}})$ of the MDPDE with tuning parameter $\beta$
is given by ${{\Sigma}}_{\beta}({{\theta}}) = \left(1+\frac{\beta^2}{1+2\beta}\right)^{3/2}\sigma^2$. 
Hence, our generalized Wald-type test statistics has much simpler form in this case given by
$$
T_{m,n}^{(\beta)} = \frac{mn}{m+n}\left(1+\frac{\beta^2}{1+2\beta}\right)^{-3/2}
\left(\frac{^{(1)}\widehat{{{\theta}}}_{\beta} - ^{(2)}\widehat{{{\theta}}}_{\beta}}{\sigma}\right)^2,
$$ 
and it has $\chi_1^2$ asymptotic distribution under $H_0$.
Note that, at $\beta=0$, this test statistic coincides with the classical Wald-test statistic
$W_{m,n} = \frac{mn}{m+n} \left(\frac{^{(1)}\widehat{{{\theta}}}_0 - ^{(2)}\widehat{{{\theta}}}_0}{\sigma}\right)^2
=\frac{mn}{m+n} \left(\frac{\bar{X} - \bar{Y}}{\sigma}\right)^2$,
where $\bar{X}$ and $\bar{Y}$ are the sample means of $X_1, \ldots, X_m$ and $Y_1, \ldots, Y_n$ respectively.

Clearly, these tests are consistent for any $\beta\geq 0$ by Corollary \ref{Corr:simple_const}. 
Further, the asymptotic power of the proposed test under contiguous  alternatives $H_{1,m,n} $
can be easily obtained as
$$
\pi_{\beta}({{\Delta}}_1,{{\Delta}}_2) = 1 - F_{\chi_1^2(\delta_{\beta})}(\chi_{1,\alpha}^2),
$$
with $\delta_{\beta}=\left(1+\frac{\beta^2}{1+2\beta}\right)^{-3/2}\sigma^{-2}W({{\Delta}}_1,{{\Delta}}_2)^2.$
Table \ref{TAB:ContPower} presents the values of $\pi_{\beta}({{\Delta}}_1, {{\Delta}}_2)$ over $\beta\in[0,1]$ 
for different values of $W({{\Delta}}_1,{{\Delta}}_2)$. 
Note that, whenever $W({{\Delta}}_1,{{\Delta}}_2)=0$, the alternative coincides with null and hence we get back the level of the test
and as $W({{\Delta}}_1,{{\Delta}}_2)$ increases the power also increases as expected.  
Clearly, this asymptotic power decreases as $\beta$ increases but this loss 
is not significant at small positive values of $\beta$. This fact is quite intuitive as the 
classical Wald-test at $\beta=0$ is asymptotically most powerful under pure model.
But, as we will see in the next two subsections, we can gain much higher robustness with respect to the outliers
at the cost of this small loss in asymptotic power.
}
\end{example}

\begin{table}[h] 
\caption{Asymptotic contiguous power of the proposed Wald-type test at 95\% level 
for testing equality of two normal means as in Example \ref{EXM:1} with known common $\sigma^2=1$}
\centering
\begin{tabular}{|l|ccccccc|} \hline 
& \multicolumn{7}{c|}{$\beta$} \\ 	
$W({{\Delta}}_1,{{\Delta}}_2)$ &	0	&	0.1	&	0.3	&	0.5	&	0.7	&	0.9	&	1	\\\hline\hline
0	&	0.050	&	0.050	&	0.050	&	0.050	&	0.050	&	0.050	&	0.050	\\
1	&	0.170	&	0.169	&	0.160	&	0.150	&	0.140	&	0.131	&	0.127	\\
2	&	0.516	&	0.511	&	0.484	&	0.449	&	0.413	&	0.380	&	0.364	\\
3	&	0.851	&	0.847	&	0.821	&	0.784	&	0.742	&	0.698	&	0.677	\\
5	&	0.999	&	0.999	&	0.998	&	0.996	&	0.992	&	0.985	&	0.981	\\\hline
\end{tabular}
\label{TAB:ContPower}
\end{table}

\subsection{Influence Function of the Wald-type Test Statistics}
\label{SEC:Simple_IF_test}

The robustness of any two sample test is relatively complicated compared to the one sample case because, 
here, one may have contamination in either of the two sample or even in both the samples.
Let us first derive the Hampel's influence function (IF) of the two sample Wald-type test statistics
to study the robustness of the proposed test. 
Consider the set-up of previous subsection and denote $G_1=F_{{\boldsymbol{\theta}}_1}$ and $G_2=F_{{\boldsymbol{\theta}}_2}$. 
Then, ignoring the multiplier $\frac{nm}{n+m}$, 
we can define the statistical functional corresponding to the proposed Wald-type test statistics $T_{m,n}^{(\beta)}$ as 
$$
T_{\beta}(G_1, G_2) = \left(\boldsymbol{U}_{\beta}(G_1)-\boldsymbol{U}_{\beta}(G_2)\right)^T
{\boldsymbol{\Sigma}}_{\beta}^{-1}({\boldsymbol{\theta}}_0)
\left(\boldsymbol{U}_{\beta}(G_1)-\boldsymbol{U}_{\beta}(G_2)\right),
$$ 
where $\boldsymbol{U}_{\beta}$ is the MDPDE functional defined in Section \ref{SEC:MDPDE}.

Now consider the contaminated distributions $G_{1,\varepsilon} = (1-\varepsilon) G_1 + \varepsilon \wedge_x$ 
and $G_{2,\varepsilon} = (1-\varepsilon) G_2 + \varepsilon \wedge_y$ where $\varepsilon$ is 
the contaminated proportion and $x$, $y$ are the point of contamination in the two samples respectively.
Then the Hampel's first-order influence function of our test functional, when the contamination 
is only in the first sample, is given by 
\begin{eqnarray}
IF^{(1)}(x; T_{\beta}, G_1, G_2) 
&=& \left.\frac{\partial}{\partial\varepsilon}T_{\beta}(G_{1,\varepsilon}, G_2)
\right|_{\varepsilon=0} 
= 2(\boldsymbol{U}_{\beta}(G_1)-\boldsymbol{U}_{\beta}(G_2))^T 
{\boldsymbol{\Sigma}}_{\beta}^{-1}({\boldsymbol{\theta}}_0) \mathcal{IF}(x; \boldsymbol{U}_{\beta}, G_1).\nonumber
\end{eqnarray}
Similarly, if there is contamination only in the second sample, then the corresponding 
IF is given by 
\begin{eqnarray}
IF^{(2)}(y; T_{\beta}, G_1, G_2) 
&=& \left.\frac{\partial}{\partial\varepsilon}T_{\beta}(G_1,G_{2,\varepsilon})
\right|_{\varepsilon=0} 
= - 2(\boldsymbol{U}_{\beta}(G_1)-\boldsymbol{U}_{\beta}(G_2))^T 
{\boldsymbol{\Sigma}}_{\beta}^{-1}({\boldsymbol{\theta}}_0) \mathcal{IF}(y; \boldsymbol{U}_{\beta}, G_1).\nonumber
\end{eqnarray}
Finally, if we assume that the contamination is in both the samples,
Hampel's IF turns out to be 
\begin{eqnarray}
IF(x, y; T_{\beta}, G_1, G_2) 
&=& \left.\frac{\partial}{\partial\varepsilon}T_{\beta}(G_{1,\varepsilon}, G_{2,\varepsilon})
\right|_{\varepsilon=0} 
=2(\boldsymbol{U}_{\beta}(G_1)-\boldsymbol{U}_{\beta}(G_2))^T 
{\boldsymbol{\Sigma}}_{\beta}^{-1}({\boldsymbol{\theta}}_0)\boldsymbol{D}_{\beta}(x,y), \nonumber
\end{eqnarray}
where $\boldsymbol{D}_{\beta}(x,y) = \left[\mathcal{IF}(x; \boldsymbol{U}_{\beta}, G_1)-\mathcal{IF}(y; \boldsymbol{U}_{\beta}, G_2)\right].$
Now, in particular, if we assume the null hypothesis to be true with $G_1=G_2=F_{{\boldsymbol{\theta}}_1}$, 
then $\boldsymbol{U}_{\beta}(G_1)=\boldsymbol{U}_{\beta}(G_2)={\boldsymbol{\theta}}_1$.
Therefore, all the above three types of influence function will be zero at the null hypothesis 
in (\ref{EQ:7two_sample}), which implies that the Wald-type tests are robust for all $\beta\geq 0$.
This is clearly not informative about the robustness of the tests as we all know the non-robust nature of 
$T_{m,n}^{(0)}$ (which is the classical Wald test statistic $W_{m,n}$).

Therefore, we need to consider the second order influence function for this case of two sample problem.
When there is contamination only in the first sample, the corresponding second order IF is given by 
\begin{eqnarray}
IF^{(1)}_2(x; T_{\beta}, G_1, G_2) &=& 
\frac{\partial^2}{\partial^2\varepsilon}T_{\beta}(G_{1,\varepsilon},G_2)
\big|_{\varepsilon=0} \nonumber \\
&=&  2(\boldsymbol{U}_{\beta}(G_1)-\boldsymbol{U}_{\beta}(G_2))^T 
{\boldsymbol{\Sigma}}_{\beta}^{-1}({\boldsymbol{\theta}}_0) \mathcal{IF}_2(x; \boldsymbol{U}_{\beta}, G_1) \nonumber\\
&& + 2\mathcal{IF}(x; \boldsymbol{U}_{\beta}, G_1)^T {\boldsymbol{\Sigma}}_{\beta}^{-1}({\boldsymbol{\theta}}_0) 
\mathcal{IF}(x; \boldsymbol{U}_{\beta}, G_1). 
\nonumber
\end{eqnarray}
For the particular case of null distribution ${\boldsymbol{\theta}}_1={\boldsymbol{\theta}}_2$, it simplifies to 
\begin{equation}
IF^{(1)}_2(x; T_{\beta},  F_{{\boldsymbol{\theta}}_1},F_{{\boldsymbol{\theta}}_1}) 
= 2\mathcal{IF}(x; \boldsymbol{U}_{\beta}, F_{{\boldsymbol{\theta}}_1})^T 
{\boldsymbol{\Sigma}}_{\beta}^{-1}({\boldsymbol{\theta}}_0) \mathcal{IF}(x; \boldsymbol{U}_{\beta}, F_{{\boldsymbol{\theta}}_1}).\nonumber
\end{equation}
Similarly, if the contamination is in the second sample only, 
then the second order IF  simplifies to
\begin{equation}
IF^{(2)}_2(y; T_{\beta},  F_{{\boldsymbol{\theta}}_1},F_{{\boldsymbol{\theta}}_1}) 
= 2\mathcal{IF}(y; \boldsymbol{U}_{\beta}, F_{{\boldsymbol{\theta}}_1})^T 
{\boldsymbol{\Sigma}}_{\beta}^{-1}({\boldsymbol{\theta}}_0)\mathcal{IF}(y; \boldsymbol{U}_{\beta}, F_{{\boldsymbol{\theta}}_1}). \nonumber
\end{equation}
Note that these two IFs are bounded with respect to the contamination points $x$ or $y$ 
if and only if the IF of the corresponding MDPDE used is bounded; 
but it is the case for all $\beta>0$ under most common parametric models. Hence for any $\beta>0$, the proposed 
test gives robust inference with respect to contamination in any one of the samples. 
However, at $\beta=0$ the MDPDE becomes the non-robust MLE having unbounded influence function 
and so using that estimator makes the classical Wald test statistic to be highly non-robust also.

Finally for the case of contamination in both samples, the corresponding second order 
IF is given by 
\begin{eqnarray}
IF_2(x, y; T_{\beta}, G_1, G_2) 
&=& \left.\frac{\partial^2}{\partial^2\varepsilon}T_{\beta}(G_{1,\varepsilon}, G_{2,\varepsilon})
\right|_{\varepsilon=0} \nonumber\\
&=& 2(\boldsymbol{U}_{\beta}(G_1)-\boldsymbol{U}_{\beta}(G_2))^T {\boldsymbol{\Sigma}}_{\beta}^{-1}({\boldsymbol{\theta}}_0) 
\left[\mathcal{IF}_2(x; \boldsymbol{U}_{\beta}, G_1)-\mathcal{IF}_2(y; \boldsymbol{U}_{\beta}, G_2)\right]
\nonumber \\&& ~~ 
+ 2\boldsymbol{D}_{\beta}(x,y)^T{\boldsymbol{\Sigma}}_{\beta}^{-1}({\boldsymbol{\theta}}_0)\boldsymbol{D}_{\beta}(x,y). \nonumber
\end{eqnarray}
In particular, at the null hypothesis $\boldsymbol{\theta}_1=\boldsymbol{\theta}_2$, we have
\begin{eqnarray}
&& IF_2(x, y; T_{\beta}, F_{{\boldsymbol{\theta}}_1}, F_{{\boldsymbol{\theta}}_1}) 
=2\boldsymbol{D}_{\beta}(x,y)^T{\boldsymbol{\Sigma}}_{\beta}^{-1}({\boldsymbol{\theta}}_0)\boldsymbol{D}_{\beta}(x,y).~~~~~~~
\nonumber
\end{eqnarray}
Note that if $x=y$ then $D_{\beta}(x,y)=0$ and hence this second order influence function is zero implying the 
robustness of the proposed test with any values of the parameter; this is expected intuitively as 
the same contamination in both the samples nullifies each other for testing 
the equivalence of the two samples as in (\ref{EQ:7two_sample}). 
However, if $x\neq y$, then the influence function of our test is bounded if and only if 
the difference $D_{\beta}(x,y)$ between the influence functions of the MDPDEs used is bounded.
This happens whenever the IF of the MDPDE is bounded, i.e., at $\beta>0$.

\begin{figure}[h]
\centering
\subfloat[Influence function]{
\includegraphics[width=0.35\textwidth]{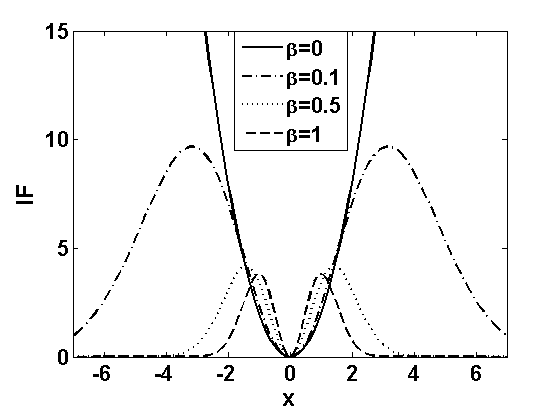}
\label{FIG:IF2_testStat}} ~
\subfloat[Gross error sensitivity]{
\includegraphics[width=0.35\textwidth]{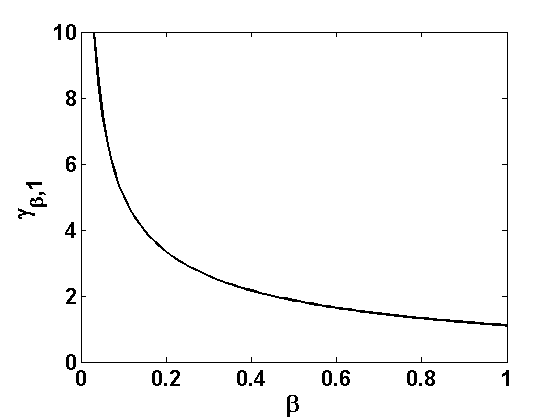}
\label{FIG:GES_test}}
\caption{Second order influence function of the proposed Wald-type test statistics and corresponding gross error sensitivity 
	$\gamma_{\beta,1}$ under contamination only in first sample 
	for testing equality of two normal means as in Example \ref{EXM:3.2} with known common $\sigma^2=1$}
	\label{FIG:IF2_test}%
\end{figure}

\begin{figure}[h]
\centering
\subfloat[$\beta=0$]{
\includegraphics[width=0.33\textwidth]{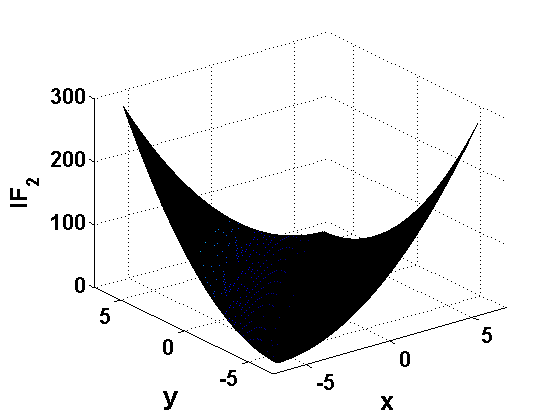}
\label{FIG:IF2_testStat00}} ~
\subfloat[$\beta=0.1$]{
	\includegraphics[width=0.33\textwidth]{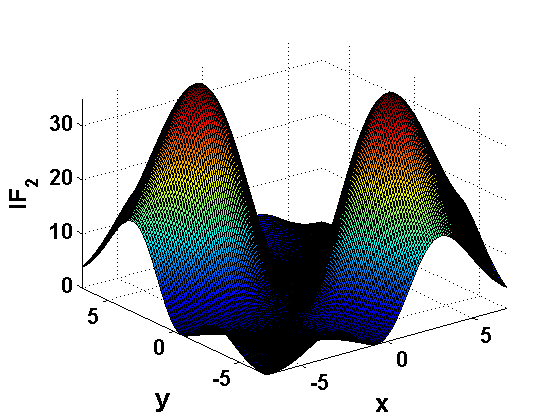}
	\label{FIG:IF2_testStat01}} ~
\subfloat[$\beta=0.5$]{
	\includegraphics[width=0.33\textwidth]{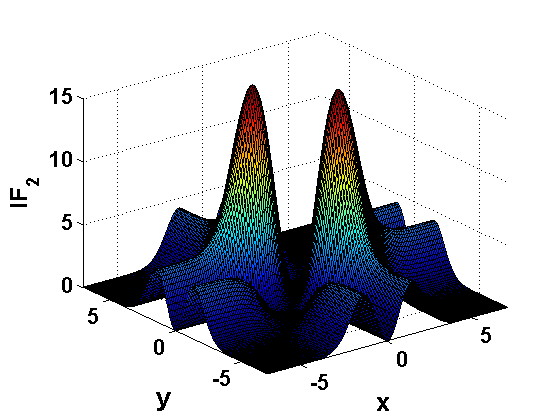}
\label{FIG:IF2_testStat05}}
\caption{Second order influence function of the proposed Wald-type test statistics under contamination in both the samples 
		for testing equality of two normal means as in Example \ref{EXM:3.2} with known common $\sigma^2=1$}
	\label{FIG:IF2_test2}%
\end{figure}

\begin{example}[Continuation of Example \ref{EXM:1}]\label{EXM:3.2}
\normalfont{
Let us again consider the previous example on testing two normal means as in Example \ref{EXM:1}.
We have seen that the proposed Wald-type tests are consistent for all $\beta \geq 0$
but their power against contiguous alternatives decreases slightly as $\beta$ increases. 
Now let us verify the claimed robustness of these tests.

Clearly, the first order IFs of the test statistics will always be zero. 
For contamination only in the first sample, the second order IF of the test statistic $T_{\beta}$
at the null hypothesis in (\ref{EQ:7two_sample}) has a simpler form given by 
$$
IF_2^{(1)}(x; T_{\beta}, F_{{{\theta}}_1}, F_{{{\theta}}_1})=\frac{2}{\sigma^2}
\left(1+2\beta\right)^{3/2}(x-{{\theta}}_1)^2 e^{-\frac{\beta(x-{{\theta}}_1)^2}{\sigma^2}}.
$$ 
Figure \ref{FIG:IF2_testStat} presents the plot of this second order IF for different values of $\beta\in [0,1]$.
It is evident from the figure that the second order IF is unbounded at $\beta=0$ implying the non-robustness
of the classical Wald test statistic; but it is bounded for all $\beta >0$ implying the robustness of our proposals. 
Further, Figure \ref{FIG:GES_test} presents the plot of the maximum possible influence of infinitesimal contamination 
on the test statistics, known as the ``gross error sensitivity", computed as
$$
\gamma_{\beta,1} = \sup_x\left|\left|IF_2^{(1)}(x; T_{\beta}, F_{{{\theta}}_1}, F_{{{\theta}}_1})\right|\right|
=\frac{2}{\sigma^3\sqrt{\beta}}\left(1+\frac{\beta}{1+\beta}\right)e^{-\frac{\sqrt{\beta}}{\sigma}}.
$$
It clearly shows that the robustness of our proposed test statistics increases as $\beta$ increases
(since $\gamma_{\beta,1}$ decreases). Thus, just like the trade-off between efficiency and robustness of MDPDE,
the parameter $\beta$ again controls the trade-off between asymptotic contiguous power and robustness
for the proposed MDPDE based test statistics.

Similar inferences can also be drawn for contamination only in the second sample.

Next consider the case when there is contamination in both the samples. In this case, the second order IF is given by
$$
IF_2(x,y; T_{\beta}, F_{{{\theta}}_1}, F_{{{\theta}}_1})=\frac{2}{\sigma^2}
\left(1+2\beta\right)^{3/2}\left[(x-{{\theta}}_1) e^{-\frac{\beta(x-{{\theta}}_1)^2}{2\sigma^2}}
- (y-{{\theta}}_1) e^{-\frac{\beta(y-{{\theta}}_1)^2}{2\sigma^2}}\right]^2.
$$
The plot of $IF_2(x,y; T_{\beta}, F_{{{\theta}}_1}, F_{{{\theta}}_1})$ 
have been presented in Figure \ref{FIG:IF2_test2},
which clearly show the robust nature of our proposals at $\beta>0$ and 
the non-robust nature of the classical Wald test (at $\beta=0$) unless $x=y$.
By looking at the maximum possible influence in this case, we can again see that, 
even under contamination in both the samples, the robustness of our proposed Wald-type test statistics increases
as $\beta$ increases.

}
\end{example}

\subsection{Power and Level Influence Functions}
\label{SEC:Simple_IF_power}

The robustness of a test statistic, although necessary, 
may not be sufficient in all the cases  since the performance of any test 
is finally measured through its level and power.
In this section, we consider the effect of contamination on the asymptotic power and level of the proposed Wald-type tests.
Due to consistency, the asymptotic power against any fixed alternative will be one. 
So, we again consider the contiguous alternatives $H_{1,m,n}$ given by (\ref{EQ:Contiguous_alternative})
along with contamination over these alternatives. 
Following Hampel et al.~(1986), the effect of contaminations should tend to zero,
as the alternatives tend to the null (i.e., ${\boldsymbol{\theta}}_{1,n}\rightarrow{\boldsymbol{\theta}}_0$ and 
${\boldsymbol{\theta}}_{2,m}\rightarrow{\boldsymbol{\theta}}_0$ as $m,n\rightarrow\infty$) at the same rate 
to avoid confusion between the neighborhoods of the two hypotheses
(also see \cite{Huber/Carol:1970}, \cite{Heritier/Ronchetti:1994}, \cite{Toma/Broniatowski:2011}, 
\cite{Ghosh/etc:2015,Ghosh/etc:2016} for some one sample applications). 
Further, in case of the present two sample problem, the contamination can be 
in any one sample or in both the samples. 
When the contamination is only in the first sample, 
we consider the corresponding contamination distribution for the first population as
$$
F_{1,n,\varepsilon,x}^L = \left(1-\frac{\varepsilon}{\sqrt{n}}\right) F_{{\boldsymbol{\theta}}_0} 
+ \frac{\varepsilon}{\sqrt{n}} \wedge_x ~~~~
F_{1,n,\varepsilon,x}^P = \left(1-\frac{\varepsilon}{\sqrt{n}}\right) F_{{\boldsymbol{\theta}}_{1,n}} 
+ \frac{\varepsilon}{\sqrt{n}} \wedge_x,
$$
for the level and power calculations respectively along with the usual uncontaminated distributions
for the second population.
Then the corresponding level influence function (LIF) and the power influence function (PIF)
at the null ${\boldsymbol{\theta}}_1={\boldsymbol{\theta}}_2={\boldsymbol{\theta}}_0$ are given by
$$
LIF^{(1)}(x; T_{\beta}, F_{{\boldsymbol{\theta}}_0}) = \lim_{m,n \rightarrow \infty} 
~ \frac{\partial}{\partial \varepsilon} 
P_{(F_{1,n,\varepsilon,x}^L,F_{{\boldsymbol{\theta}}_0})}(T_{m,n}^{(\beta)} > \chi_{p,\alpha}^2) 
\big|_{\varepsilon=0},
~
$$
$$
PIF^{(1)}(x; T_{\beta}, F_{{\boldsymbol{\theta}}_0}) = \lim_{m, n \rightarrow \infty} 
~ \frac{\partial}{\partial \varepsilon} 
P_{(F_{1,n,\varepsilon,x}^P,F_{{\boldsymbol{\theta}}_{2,m}})}(T_{m,n}^{(\beta)} > \chi_{p,\alpha}^2)\big|_{\varepsilon=0}.
$$
Similarly, when contamination is assumed to be only in the second sample, then we take the uncontaminated 
distributions for the first population and the contaminated distribution for the second population as
$$
F_{2,m,\varepsilon,y}^L = \left(1-\frac{\varepsilon}{\sqrt{m}}\right) F_{{\boldsymbol{\theta}}_0} 
+ \frac{\varepsilon}{\sqrt{m}} \wedge_y ~~~~
F_{2,m,\varepsilon,y}^P = \left(1-\frac{\varepsilon}{\sqrt{m}}\right) F_{{\boldsymbol{\theta}}_{2,m}} 
+ \frac{\varepsilon}{\sqrt{m}} \wedge_y,
$$
for the level and power calculations respectively.
Corresponding LIF and PIF at the null ${\boldsymbol{\theta}}_1={\boldsymbol{\theta}}_2={\boldsymbol{\theta}}_0$ are given by
$$
LIF^{(2)}(y; T_{\beta}, F_{{\boldsymbol{\theta}}_0}) = \lim_{m, n \rightarrow \infty} 
~ \frac{\partial}{\partial \varepsilon} 
P_{(F_{{\boldsymbol{\theta}}_0},F_{2,m,\varepsilon,y}^L)}(T_{m,n}^{(\beta)} > \chi_{p,\alpha}^2)\big|_{\varepsilon=0},
~
$$
$$
PIF^{(2)}(y; T_{\beta}, F_{{\boldsymbol{\theta}}_0}) = \lim_{m, n \rightarrow \infty} 
~ \frac{\partial}{\partial \varepsilon} 
P_{(F_{{\boldsymbol{\theta}}_{1,n}},F_{2,m,\varepsilon,y}^P)}(T_{m,n}^{(\beta)} > \chi_{p,\alpha}^2)\big|_{\varepsilon=0}.
$$
Finally, while considering contamination in both the samples with above contaminated distributions,
we define the corresponding LIF and PIF as
$$
LIF(x,y; T_{\beta}, F_{{\boldsymbol{\theta}}_0}) = \lim_{m, n \rightarrow \infty} 
~ \frac{\partial}{\partial \varepsilon} 
P_{(F_{1,n,\varepsilon,x}^L,F_{2,m,\varepsilon,y}^L)}(T_{m,n}^{(\beta)} > \chi_{p,\alpha}^2)\big|_{\varepsilon=0},
~
$$
$$
PIF(x,y; T_{\beta}, F_{{\boldsymbol{\theta}}_0}) = \lim_{m, n \rightarrow \infty} 
~ \frac{\partial}{\partial \varepsilon} 
P_{(F_{1,n,\varepsilon,x}^P,F_{2,m,\varepsilon,y}^P)}(T_{m,n}^{(\beta)} > \chi_{p,\alpha}^2)\big|_{\varepsilon=0}.
$$

First let us derive the asymptotic distribution of the proposed Wald-type test statistics $T_{m,n}^{(\beta)}$ 
under the contaminated distributions. Let us define 
$\widetilde{{\boldsymbol{\Delta}}}_i = {\boldsymbol{\Delta}}_i 
+ \varepsilon \mathcal{IF}(x_i; \boldsymbol{U}_{\beta}, F_{{\boldsymbol{\theta}}_0})$ 
for $i=1, 2$ with $x_1=x$ and $x_2=y$. 
Then we have the following theorem.

\begin{theorem}
Suppose the model density satisfies the Lehmann and Basu et al.~conditions and Assumption (A) holds.
Then the asymptotic distribution of $T_{m,n}^{(\beta)}$ under any contaminated 
contiguous alternative distributions $(D_1, D_2)$ is 
$\chi_p^2\left(\lambda\right)$ where $\lambda$ is the parameter of non-centrality given by 
$\lambda = \widetilde{\boldsymbol{W}}_{\varepsilon}^T{\boldsymbol{\Sigma}}_{\beta}({\boldsymbol{\theta}}_0)^{-1}
\widetilde{\boldsymbol{W}}_{\varepsilon}$,
where
\begin{eqnarray}
 \begin{array}{r l l}
 \widetilde{\boldsymbol{W}}_{\varepsilon}= & \boldsymbol{W}\left(\widetilde{{\boldsymbol{\Delta}}}_1,{{\boldsymbol{\Delta}}}_2\right),
 & \mbox{if } ~(D_1, D_2)=(F_{1,n,\varepsilon,x}^P,F_{{\boldsymbol{\theta}}_{2,m}}),\\
 = & \boldsymbol{W}\left({{\boldsymbol{\Delta}}}_1,\widetilde{{\boldsymbol{\Delta}}}_2\right), 
 & \mbox{if } ~(D_1, D_2)=(F_{{\boldsymbol{\theta}}_{1,n}},F_{2,m,\varepsilon,y}^P),\\
 = & \boldsymbol{W}\left(\widetilde{{\boldsymbol{\Delta}}}_1,\widetilde{{\boldsymbol{\Delta}}}_2\right), 
 & \mbox{if } ~(D_1, D_2)=(F_{1,n,\varepsilon,x}^P,F_{2,m,\varepsilon,y}^P).
 \end{array}
 \label{EQ:ncp_simple_contcontg}
\end{eqnarray}
\label{THM:2sample_simple_asympCont}
\end{theorem}

From the above theorem, we get the asymptotic power of the proposed Wald-type tests 
under the contaminated contiguous alternatives as
$$
\pi_{\beta}({\boldsymbol{\Delta}}_1, {\boldsymbol{\Delta}}_2; \varepsilon)=P_{(D_1, D_2)}\left(T_{m,n}^{(\beta)} > \chi^2_{p,\alpha}\right)
=1 - F_{\chi_p^2\left(
	\widetilde{\boldsymbol{W}}_{\varepsilon}^T{\boldsymbol{\Sigma}}_{\beta}({\boldsymbol{\theta}}_0)^{-1}
	\widetilde{\boldsymbol{W}}_{\varepsilon}\right)}(\chi^2_{p,\alpha}).
$$
Using infinite series expansion of a non-central chi-square distribution function \citep{Kotz/etc:1967b},
we get
\begin{eqnarray}
\pi_{\beta}({\boldsymbol{\Delta}}_1, {\boldsymbol{\Delta}}_2; \varepsilon)
&=&\sum_{v=0}^{\infty} C_v\left(\widetilde{\boldsymbol{W}}_{\varepsilon}, {\boldsymbol{\Sigma}}_{\beta}({\boldsymbol{\theta}}_0)^{-1}\right)
P\left(\chi^2_{p+2v} > \chi^2_{p,\alpha}\right),\nonumber\\
\mbox{where }~~~~~~~~~~~~~~~~~~~~~~~~~~~~~~~~~~~~~~
&& C_v(\boldsymbol{t}, \boldsymbol{A}) = \frac{(\boldsymbol{t}^T\boldsymbol{A}\boldsymbol{t})^v}{v! 2^v} 
e^{-\frac{1}{2}\boldsymbol{t}^T\boldsymbol{A}\boldsymbol{t}}.
~~~~~~~~~~~~~~~~~~~~~~~~~~~~~~~~~~~~~~~~~~~~~~~~~~\nonumber
\end{eqnarray}
In particular, substituting $\varepsilon=0$ in the above theorem, 
we get back Theorem \ref{THM:2sample_simple_asympCont1} on the asymptotic contiguous power
of our tests and hence expression (\ref{EQ:simpel_contgPower}) can be written as 
$$
\pi_{\beta}({\boldsymbol{\Delta}}_1,{\boldsymbol{\Delta}}_2) = \pi_{\beta}({\boldsymbol{\Delta}}_1, {\boldsymbol{\Delta}}_2; \boldsymbol{0}_p)
=\sum_{v=0}^{\infty} C_v\left(\boldsymbol{W}({\boldsymbol{\Delta}}_1,\boldsymbol{\Delta}_2), {\boldsymbol{\Sigma}}_{\beta}({\boldsymbol{\theta}}_0)^{-1}\right)
P\left(\chi^2_{p+2v} > \chi^2_{p,\alpha}\right).
$$
Further, substituting ${\boldsymbol{\Delta}}_1={\boldsymbol{\Delta}}_2=\boldsymbol{0}_p$, we get the asymptotic level of our 
Wald-type tests under the contamination as $\alpha_{\varepsilon}=\pi_{\beta}(\boldsymbol{0},\boldsymbol{0}; \varepsilon)$.

Now we can define the power influence functions of our proposed tests which is nothing but 
$\left.\frac{\partial}{\partial\varepsilon}\pi_{\beta}({\boldsymbol{\Delta}}_1,{\boldsymbol{\Delta}}_2; \varepsilon)\right|_{\varepsilon=0}$ 
under standard regularity conditions.
Using the infinite series expression of a non-central chi-square distribution function,
we can derive an explicit form of the PIFs
as presented in the following theorem.

\begin{theorem}
Suppose the model density satisfies the Lehmann and Basu et al.~conditions, and Assumption (A) holds.
Then the power influence functions of our proposed Wald-type tests are given by 
\begin{eqnarray}
PIF^{(1)}(x; T_{\beta}, F_{{\boldsymbol{\theta}}_0})
&=& \sqrt{\omega}K_p^*\left(\delta_{\beta}\right)
\boldsymbol{W}({\boldsymbol{\Delta}}_1,\boldsymbol{\Delta}_2)^T{\boldsymbol{\Sigma}}_{\beta}({\boldsymbol{\theta}}_0)^{-1}
\mathcal{IF}(x; \boldsymbol{U}_{\beta}, F_{{\boldsymbol{\theta}}_0}),\nonumber\\
PIF^{(2)}(y; T_{\beta}, F_{{\boldsymbol{\theta}}_0})
&=& \sqrt{1-\omega}K_p^*\left(\delta_{\beta}\right)
\boldsymbol{W}({\boldsymbol{\Delta}}_1,\boldsymbol{\Delta}_2)^T{\boldsymbol{\Sigma}}_{\beta}({\boldsymbol{\theta}}_0)^{-1}
\mathcal{IF}(y; \boldsymbol{U}_{\beta}, F_{{\boldsymbol{\theta}}_0}),\nonumber\\
PIF(x,y; T_{\beta}, F_{{\boldsymbol{\theta}}_0})
&=& K_p^*\left(\delta_{\beta}\right)
\boldsymbol{W}({\boldsymbol{\Delta}}_1,\boldsymbol{\Delta}_2)^T{\boldsymbol{\Sigma}}_{\beta}({\boldsymbol{\theta}}_0)^{-1}
\boldsymbol{W}\left(\mathcal{IF}(x; \boldsymbol{U}_{\beta}, F_{{\boldsymbol{\theta}}_0}), 
\mathcal{IF}(x; \boldsymbol{U}_{\beta}, F_{{\boldsymbol{\theta}}_0})\right),\nonumber
\end{eqnarray}
where $\delta_{\beta}$ and $\boldsymbol{W}({\boldsymbol{\Delta}}_1,\boldsymbol{\Delta}_2)$ 
are as defined in Theorem \ref{THM:2sample_simple_asympCont1} and 
$$K_p^*(s)=e^{-\frac{s}{2}}\sum_{v=0}^{\infty} \frac{s^{v-1}}{v! 2^v}\left(2v-s\right)
P\left(\chi^2_{p+2v} > \chi^2_{p,\alpha}\right).
$$
\label{THM:2sample_simple_PIF}
\end{theorem}

Note that the PIFs are also a function of the influence function of the MDPDE used and hence they are bounded
whenever $\beta>0$. Thus the proposed tests will be robust for all $\beta>0$. 
However, at $\beta=0$,  these PIFs will be unbounded (unless there is contamination at the same points $x=y$
in both the samples) which proves the non-robust nature of the classical Wald test.

Note that, although there is no direct relationship between the IF of test statistics 
with the corresponding PIF in general, in this present case they are seen to be related indirectly 
via the IF of the MDPDE. So, using a robust MDPDE with $\beta>0$ in the proposed Wald-type tests will make both the test
statistics and its asymptotic power robust under infinitesimal contamination.

Finally, we can find the level influence function of the proposed Wald-type tests either starting from $\alpha_{\varepsilon}$
and following the same steps as in the case of PIFs or just by substituting 
${\boldsymbol{\Delta}}_1={\boldsymbol{\Delta}}_2=\boldsymbol{0}$
in the expression of the PIFs given in Theorem \ref{THM:2sample_simple_PIF}. 
In either case, since $\boldsymbol{W}({\boldsymbol{0}}, {\boldsymbol{0}})=\boldsymbol{0}$, it turns out that
\begin{equation}
LIF^{(1)}(x; T_{\beta}, F_{{\boldsymbol{\theta}}_0})=0, ~~~
LIF^{(2)}(y; T_{\beta}, F_{{\boldsymbol{\theta}}_0})=0, ~~~
LIF(x,y; T_{\beta}, F_{{\boldsymbol{\theta}}_0})=0,
\end{equation}
provided the corresponding IF of $\boldsymbol{U}_{\beta}$ is bounded, which is true at $\beta >0$.
Hence the asymptotic level of our Wald-type tests is always stable with respect infinitesimal contamination. 
This fact was also expected as we are using the asymptotic critical values for testing.

\begin{figure}[h]
	\centering
	\subfloat[Contamination in only first samples]{
		\includegraphics[width=0.35\textwidth]{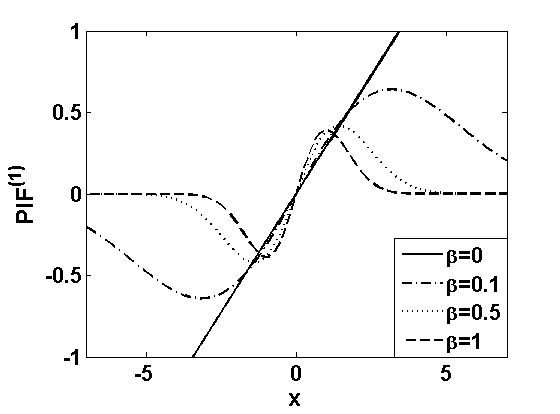}
		\label{FIG:PIF1}} ~
	\subfloat[Contamination in both samples, $\beta=0$]{
		\includegraphics[width=0.35\textwidth]{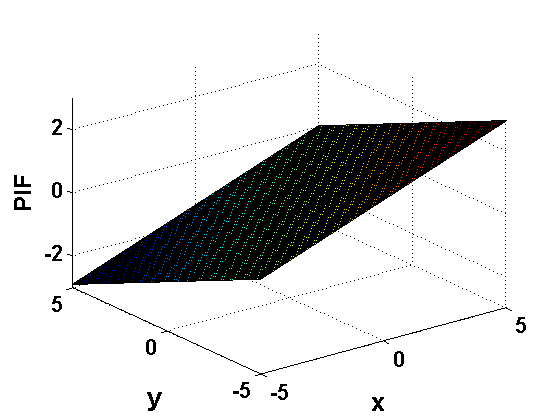}
		\label{FIG:PIF2_00}} \\
	\subfloat[Contamination in both samples, $\beta=0.1$]{
		\includegraphics[width=0.35\textwidth]{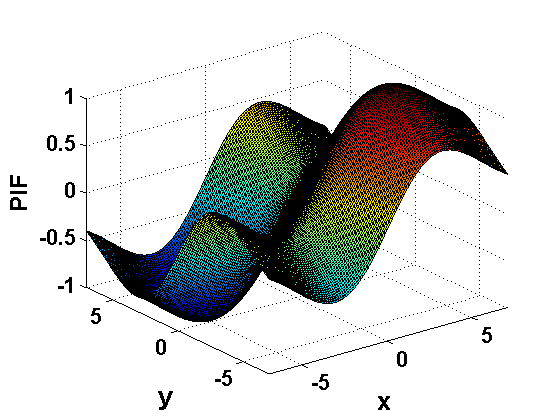}
		\label{FIG:PIF2_01}} ~
	\subfloat[Contamination in both samples, $\beta=0.5$]{
		\includegraphics[width=0.35\textwidth]{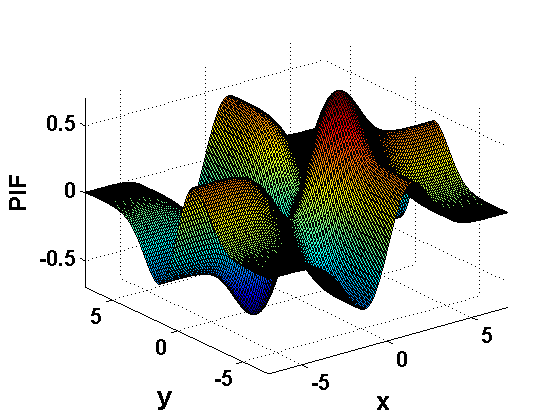}
		\label{FIG:PIF2_05}}
	\caption{Power influence functions of the proposed Wald-type test statistics at 95\% level 
		for testing equality of two normal means as in Example \ref{EXM:3.3} with known common $\sigma^2=1$,
		$W(\Delta_1,\Delta_2)=2$ and $\omega=0.5$ ($n=m$).}
	\label{FIG:PIF_test2}%
\end{figure}

\begin{example}[Continuation of Examples \ref{EXM:1} and \ref{EXM:3.2}]\label{EXM:3.3}
\normalfont{
Let us again consider the problem of testing for normal means as in Examples \ref{EXM:1} and \ref{EXM:3.2}.
As seen above, the level influence function is always zero implying the level robustness 
of our proposed Wald-type test for all $\beta>0$.
Next, to study the power robustness, we compute the functions 
$PIF^{(1)}(x; T_{\beta}, F_{{{\theta}}_0})$ and $PIF(x,y; T_{\beta}, F_{{{\theta}}_0})$
numerically for different values of $\beta$ with $\theta_0=0$ 
and plot them over the contamination points $x$ and $y$ in Figure \ref{FIG:PIF_test2}.
$PIF^{(2)}(y; T_{\beta}, F_{{{\theta}}_0})$ has the same nature as $PIF^{(1)}(x; T_{\beta}, F_{{{\theta}}_0})$.
The figures clearly show the robustness of the proposed Wald-type tests with $\beta>0$,
where the robustness increases (i.e., maximum possible PIF decreases) as $\beta$ increases.
Further, all the PIFs at $\beta=0$ are unbounded implying the non-robust nature of the classical Wald test.
}
\end{example}

%

\section{General Composite Hypotheses with Two Samples}
\label{SEC:General_problem}

In the previous section, we have considered the simplest two sample problem which tests
for equality of all the model parameters. However, in practice, 
we need to test many different complicated hypotheses which cannot be solved just
by considering the Wald-type test statistic $T_{m,n}^{(\beta)}$ defined in the previous section.
For example, in many real life problems, we are only interested in a proper subset of 
the parameters ignoring the rest as nuisance parameters;
example includes popular mean test taking variance parameter unknown and nuisance.
Further, in case of testing for multiplicative heteroscedasticity of two samples,
we have to test if the ratio of variance parameters equals a pre-specified limit 
with means being unknown and nuisance.
Neither of them belongs to the problem considered in the previous section.

In this section, we will consider a general class of hypotheses involving 
two independent samples, which would include most of the above real life testing problems. 
Suppose ${\boldsymbol{\psi}}({\boldsymbol{\theta}}_1, {\boldsymbol{\theta}}_2)$ denote a general function from $\mathbb{R}^p\times\mathbb{R}^p$ to $\mathbb{R}^r$. 
Then, considering the set-up of the previous section, we want to develop a family of robust tests 
for the general class of hypothesis given by 
\begin{equation}
H_0 : {\boldsymbol{\psi}}({\boldsymbol{\theta}}_1, {\boldsymbol{\theta}}_2)=\boldsymbol{0}_r ~~~ \mbox{against} ~~~~ 
 H_1 : {\boldsymbol{\psi}}({\boldsymbol{\theta}}_1, {\boldsymbol{\theta}}_2) \neq \boldsymbol{0}_r.
\label{EQ:7two_sample_Gen}
\end{equation}
In particular, the problem of testing normal mean with unknown variance can be seen as a particular 
case of the above general set-up with ${\boldsymbol{\psi}}((\mu_1, \sigma_1^2), (\mu_2, \sigma_2^2)) = \mu_1 - \mu_2$.
Further, to test for multiplicative heteroscedasticity, we can take
 ${\boldsymbol{\psi}}((\mu_1, \sigma_1^2), (\mu_2, \sigma_2^2)) = \frac{\sigma_1^2}{\sigma_2^2} - C_0$ 
 for some known constant $C_0$ and apply the above general set-up. It is interesting to note that,  
this general class of hypotheses in (\ref{EQ:7two_sample_Gen}) also contains the simple hypothesis 
in (\ref{EQ:7two_sample}) as its special case with ${\boldsymbol{\psi}}({\boldsymbol{\theta}}_1, {\boldsymbol{\theta}}_2) = {\boldsymbol{\theta}}_1-{\boldsymbol{\theta}}_2$.

Now, to define a robust Wald-type test statistics for this general set-up, we again consider the 
MDPDEs of ${\boldsymbol{\theta}}_1$ and ${\boldsymbol{\theta}}_2$ with tuning parameter $\beta$ 
as given by $^{(1)}\widehat{{\boldsymbol{\theta}}}_{\beta}$  and $^{(2)}\widehat{{\boldsymbol{\theta}}}_{\beta}$ 
based on the individual samples separately. 
Note that, whenever $H_0$ is true, we should have 
${\boldsymbol{\psi}}(^{(1)}\widehat{{\boldsymbol{\theta}}}_{\beta}, ^{(2)}\widehat{{\boldsymbol{\theta}}}_{\beta})\approx \boldsymbol{0}_r$ 
in large sample and so its observed value provide the indication of any departure from the null hypothesis.
Using its asymptotic variance-covariance matrix as a normalizing factor, 
we define the corresponding Wald-type test statistic as
\begin{equation}
\label{EQ:7_2SDT_gen}
 \widetilde{T_{m,n}^{(\beta)}}
= \frac{nm}{n+m} ~ {\boldsymbol{\psi}}\left( ^{(1)}\widehat{{\boldsymbol{\theta}}}_{\beta} , ^{(2)}\widehat{{\boldsymbol{\theta}}}_{\beta}\right)^T
\widetilde{{\boldsymbol{\Sigma}}_{\beta}}( ^{(1)}\widehat{{\boldsymbol{\theta}}}_{\beta}, ^{(2)}\widehat{{\boldsymbol{\theta}}}_{\beta})^{-1}
{\boldsymbol{\psi}}\left( ^{(1)}\widehat{{\boldsymbol{\theta}}}_{\beta} , ^{(2)}\widehat{{\boldsymbol{\theta}}}_{\beta}\right),
\end{equation}
where 
$\widetilde{{\boldsymbol{\Sigma}}_{\beta}}({\boldsymbol{\theta}}_1, {\boldsymbol{\theta}}_2) 
=\omega {\boldsymbol{\Psi}}_1({\boldsymbol{\theta}}_1,{\boldsymbol{\theta}}_2)^T{\boldsymbol{\Sigma}}_{\beta}({\boldsymbol{\theta}}_1){\boldsymbol{\Psi}}_1({\boldsymbol{\theta}}_1,{\boldsymbol{\theta}}_2)
+ (1-\omega) {\boldsymbol{\Psi}}_2({\boldsymbol{\theta}}_1,{\boldsymbol{\theta}}_2)^T{\boldsymbol{\Sigma}}_{\beta}({\boldsymbol{\theta}}_2){\boldsymbol{\Psi}}_2({\boldsymbol{\theta}}_1,{\boldsymbol{\theta}}_2)$ with 
$$
{\boldsymbol{\Psi}}_i({\boldsymbol{\theta}}_1, {\boldsymbol{\theta}}_2) = \frac{\partial}{\partial{\boldsymbol{\theta}}_i}{\boldsymbol{\psi}}({\boldsymbol{\theta}}_1, {\boldsymbol{\theta}}_2)^T, ~~~~ i=1, 2.
$$
Note that, at $\beta=0$, the Wald-type test statistics 
$ \widetilde{T_{m,n}^{(0)}}$ is again nothing but the classical Wald test statistics 
for the general hypothesis (\ref{EQ:7two_sample_Gen}) and hence our proposal is indeed 
a generalization of the classical Wald test.

Interestingly, although the general hypothesis contains the hypothesis (\ref{EQ:7two_sample}) as its special case, 
the Wald-type test statistics $ \widetilde{T_{m,n}^{(\beta)}}$ with ${\boldsymbol{\psi}}({\boldsymbol{\theta}}_1, {\boldsymbol{\theta}}_2) = {\boldsymbol{\theta}}_1-{\boldsymbol{\theta}}_2$ 
is not the same as the Wald-type test statistics $T_{m,n}^{(\beta)}$ considered in the previous section. 
However, whenever  $\boldsymbol{\Sigma}_{\beta}({\boldsymbol{\theta}})$ is linear in the parameters, 
these two Wald-type test statistics coincide asymptotically with probability tending to one. 
In this section, we present the properties of the statistics $\widetilde{T_{m,n}^{(\beta)}}$ 
with general ${\boldsymbol{\psi}}$-function satisfying the following assumption.
\\

\noindent\textbf{Assumption (B):}
\begin{itemize}
\item  ${\boldsymbol{\Psi}}_i({\boldsymbol{\theta}}_1, {\boldsymbol{\theta}}_2)$, $i=1, 2$, exist, have rank $r$ and are continuous with respect to its arguments.
\end{itemize}
 

\subsection{Asymptotic Properties}

We again start with the asymptotic null distribution of the proposed Wald-type test 
statistics $\widetilde{T_{m,n}^{(\beta)}}$ in order to obtain the required critical values for the test.

\begin{theorem}
Suppose the model density satisfies the Lehmann and Basu et al.~conditions and Assumptions (A) and (B) hold. 
Then, under the null hypothesis in (\ref{EQ:7two_sample_Gen}), $\widetilde{T_{m,n}^{(\beta)}}$ 
asymptotically follows a $\chi_r^2$ distribution.
\label{THM:2sample_gen_asympNull}
\end{theorem}
Therefore, the level-$\alpha$ critical region for the proposed test based on $\widetilde{T_{m,n}^{(\beta)}}$
for testing (\ref{EQ:7two_sample_Gen}) is given by $$\widetilde{T_{m,n}^{(\beta)}} > \chi^2_{r,\alpha}.$$
Next, in order to consider  an approximation to the asymptotic power for 
this general test based on $\widetilde{T_{m,n}^{(\beta)}}$, 
we are going to use the following function
$$
  \widetilde{l^*}({\boldsymbol{\theta}}_1,{\boldsymbol{\theta}}_2) 
  = {\boldsymbol{\psi}}({\boldsymbol{\theta}}_1, {\boldsymbol{\theta}}_2)^T\widetilde{{\boldsymbol{\Sigma}}_{\beta}}({\boldsymbol{\theta}}_1,{\boldsymbol{\theta}}_2)^{-1}{\boldsymbol{\psi}}({\boldsymbol{\theta}}_1, {\boldsymbol{\theta}}_2).
$$

\begin{theorem}
Suppose the model density satisfies the Lehmann and Basu et al.~conditions and Assumptions (A)-(B) hold. 
Then, whenever ${\boldsymbol{\psi}}({\boldsymbol{\theta}}_1, {\boldsymbol{\theta}}_2)\neq \boldsymbol{0}_r$, we have 
\begin{eqnarray}
\sqrt{\frac{mn}{m+n}}
\left[\widetilde{l^*}( ^{(1)}\widehat{{\boldsymbol{\theta}}}_{\beta},^{(2)}\widehat{{\boldsymbol{\theta}}}_{\beta})
- \widetilde{l^*}({\boldsymbol{\theta}}_1,{\boldsymbol{\theta}}_2)\right] 
\underset{m, n\rightarrow \infty }{\overset{\mathcal{L}}{\longrightarrow }}  
  N\left(0, 4\widetilde{l^*}({\boldsymbol{\theta}}_1,{\boldsymbol{\theta}}_2)\right), ~\mbox{ as }~m, n \rightarrow\infty.
\end{eqnarray}  
\label{THM:2sample_Gen_powerApprox}
\end{theorem}

Note that, from the above theorem, we can easily obtain an approximation to the power function of the 
proposed  level-$\alpha$ Wald-type tests based on $\widetilde{T_{m,n}^{(\beta)}}$  as
  \begin{eqnarray}
  \widetilde{\pi_{m,n,\alpha}}^{(\beta)} ({\boldsymbol{\theta}}_1, {\boldsymbol{\theta}}_2) = 
  P\left(\widetilde{T_{m,n}^{(\beta)}} > \chi_{r,\alpha}^2\right) 
  = 1 - \Phi_n \left( \frac{\sqrt{\frac{n+m}{nm}}}  {2\sqrt{\widetilde{l^*}({\boldsymbol{\theta}}_1, {\boldsymbol{\theta}}_2)}} 
  \left[ \chi_{r,\alpha}^2-  \frac{nm}{n+m}\widetilde{l^*}({\boldsymbol{\theta}}_1,{\boldsymbol{\theta}}_2) \right] \right), \nonumber
  \end{eqnarray}
for a sequence of distributions $\Phi_n(\cdot)$ tending uniformly to the standard normal distribution $\Phi(\cdot)$,
 whenever ${\boldsymbol{\psi}}({\boldsymbol{\theta}}_1, {\boldsymbol{\theta}}_2)\neq \boldsymbol{0}_r$. 
In such cases, it can be easily checked that 
$\widetilde{\pi_{m,n,\alpha}}^{(\beta)} ({\boldsymbol{\theta}}_1, {\boldsymbol{\theta}}_2)\rightarrow 1$ as $m,n \rightarrow \infty$.
This proves the consistency of our proposed tests.

\begin{corollary}\label{Corr:gen_const}
Under the assumptions of Theorem \ref{THM:2sample_Gen_powerApprox}, 
the proposed Wald-type tests based on $\widetilde{T_{m,n}^{(\beta)}}$ are consistent.
\end{corollary}

Now, let us study the performance of the proposed general two-sample Wald-type tests under the contiguous alternative hypotheses. 
As discussed in the previous section, there could be different choices 
for the contiguous alternative hypotheses for any general null hypothesis. 
Here, following the similar idea as in the alternatives in (\ref{EQ:Contiguous_alternative}),  
we consider the general form of the contiguous alternatives given by 
\begin{equation}
  H_{1,n,m}: {\boldsymbol{\theta}}_1={\boldsymbol{\theta}}_{1,n} = {\boldsymbol{\theta}}_{10} + n^{-\frac{1}{2}}{\boldsymbol{\Delta}}_1,~~
  {\boldsymbol{\theta}}_2={\boldsymbol{\theta}}_{2,m} = {\boldsymbol{\theta}}_{20} + m^{-\frac{1}{2}}{\boldsymbol{\Delta}}_2, ~~~~~~
({\boldsymbol{\Delta}}_1, {\boldsymbol{\Delta}}_2)\in \mathbb{R}^p\times\mathbb{R}^p-\{(\boldsymbol{0}_p,\boldsymbol{0}_p)\},
  \label{EQ:Contiguous_alternative_gen}
\end{equation} 
for some fixed $\left({\boldsymbol{\theta}}_{10}, {\boldsymbol{\theta}}_{20}\right)\in \Theta_0
=\{ \left({\boldsymbol{\theta}}_{1}, {\boldsymbol{\theta}}_{2}\right) \in \Theta \times \Theta 
: \boldsymbol{\psi}({\boldsymbol{\theta}}_{1}, {\boldsymbol{\theta}}_{2})=0\}$. 
The asymptotic distribution of $\widetilde{T_{m,n}^{(\beta)}}$ under these alternatives $H_{1,m,n}$
has been presented in the following theorem.

\begin{theorem}
Suppose the model density satisfies the Lehmann and Basu et al.~conditions and Assumptions (A)-(B) hold.
Then the asymptotic distribution of $\widetilde{T_{m,n}^{(\beta)}}$ under $H_{1,n,m}$ in 
(\ref{EQ:Contiguous_alternative_gen}) is $\chi^2_r(\widetilde{\delta_{\beta}})$,
where 
$$
\widetilde{\delta_{\beta}} = \boldsymbol{W}_{\boldsymbol{\psi}}({\boldsymbol{\Delta}}_1,{\boldsymbol{\Delta}}_2)^T
\widetilde{{\boldsymbol{\Sigma}}_{\beta}}({\boldsymbol{\theta}}_1,{\boldsymbol{\theta}}_2)^{-1} 
\boldsymbol{W}_{\boldsymbol{\psi}}({\boldsymbol{\Delta}}_1,{\boldsymbol{\Delta}}_2)
$$ with  
$\boldsymbol{W}_{\boldsymbol{\psi}}({\boldsymbol{\Delta}}_1,{\boldsymbol{\Delta}}_2)
=\left[\sqrt{\omega}{\boldsymbol{\Psi}}_1({\boldsymbol{\theta}}_1,{\boldsymbol{\theta}}_2)^T{\boldsymbol{\Delta}}_1 
+ \sqrt{1-\omega}{\boldsymbol{\Psi}}_2({\boldsymbol{\theta}}_1,{\boldsymbol{\theta}}_2)^T{\boldsymbol{\Delta}}_2\right].$
\label{THM:2sample_gen_asympContg}
\end{theorem}

The above theorem directly helps us to obtain the asymptotic power $\widetilde{\pi}_{\beta}({\boldsymbol{\Delta}}_1, {\boldsymbol{\Delta}}_2)$
of our general Wald-type tests based on $\widetilde{T_{m,n}^{(\beta)}}$ under the contiguous alternatives $H_{1,n,m}$ in 
(\ref{EQ:Contiguous_alternative_gen}) as
$$
\widetilde{\pi}_{\beta}({\boldsymbol{\Delta}}_1, {\boldsymbol{\Delta}}_2) = 1 - F_{\chi^2_r(\widetilde{\delta_{\beta}})}(\chi^2_{r,\alpha}).
$$

\subsection{Robustness Properties}\label{SEC:Gen_IF}

Let us now study the robustness properties of the proposed general two-sample Wald-type tests based on 
$\widetilde{T_{m,n}^{(\beta)}}$. We first consider the influence function of the Wald-type test statistics. 
Define the statistical functional corresponding to $\widetilde{T_{m,n}^{(\beta)}}$ 
ignoring the multiplier $\frac{nm}{n+m}$ as 
$$
\widetilde{T_{\beta}}(G_1, G_2) = {\boldsymbol{\psi}}\left(\boldsymbol{U}_{\beta}(G_1),\boldsymbol{U}_{\beta}(G_2)\right)^T
\widetilde{{\boldsymbol{\Sigma}}_{\beta}}^{-1}({\boldsymbol{\theta}}_1, {\boldsymbol{\theta}}_2)
{\boldsymbol{\psi}}\left(\boldsymbol{U}_{\beta}(G_1),\boldsymbol{U}_{\beta}(G_2)\right),
$$ 
where $\boldsymbol{U}_{\beta}$ is the corresponding MDPDE functional. 
Then, we can derive the first and second order influence functions of the Wald-type test statistics 
following the derivations similar to that of Section \ref{SEC:Simple_IF_test}.
So, here we will skip those derivations for brevity and present only the final results in the following theorem.

\begin{theorem}
Consider the notations of Section \ref{SEC:Simple_IF_test}. Under the null hypothesis in (\ref{EQ:7two_sample_Gen})
with $G_1=F_{{\boldsymbol{\theta}}_{10}}$, $G_2=F_{{\boldsymbol{\theta}}_{20}}$ and ${\boldsymbol{\psi}}({\boldsymbol{\theta}}_{10}, {{\boldsymbol{\theta}}_{20}})=0$, 
the first and second order influence functions of our general two-sample Wald-type test statistics are given as follows:
\begin{enumerate}
\item For contamination only in the $i$-th sample ($i=1,2$) at the point $x_i$ ($x_1=x, x_2=y$)
\begin{eqnarray}
IF^{(i)}(x_i; \widetilde{T_{\beta}},  F_{{\boldsymbol{\theta}}_{10}},F_{{\boldsymbol{\theta}}_{20}}) &=& 0,\nonumber\\
IF^{(i)}_2(x_i; \widetilde{T_{\beta}},  F_{{\boldsymbol{\theta}}_{i0}},F_{{\boldsymbol{\theta}}_{20}}) 
&=& 2\mathcal{IF}(x_i; \boldsymbol{U}_{\beta}, F_{{\boldsymbol{\theta}}_{10}})^T 
{\boldsymbol{\Psi}}_i({\boldsymbol{\theta}}_{10},{\boldsymbol{\theta}}_{20})^T
\widetilde{{\boldsymbol{\Sigma}}_{\beta}}({\boldsymbol{\theta}}_{10},{\boldsymbol{\theta}}_{20})^{-1} 
{\boldsymbol{\Psi}}_i({\boldsymbol{\theta}}_{10},{\boldsymbol{\theta}}_{20})
\mathcal{IF}(x_i; \boldsymbol{U}_{\beta}, {\boldsymbol{\theta}}_{i0}).\nonumber
\end{eqnarray}

\item For contamination in both the samples
\begin{eqnarray}
IF(x, y; \widetilde{T_{\beta}},  F_{{\boldsymbol{\theta}}_{10}},F_{{\boldsymbol{\theta}}_{20}}) &=& 0\nonumber\\
IF_2(x, y; \widetilde{T_{\beta}},  F_{{\boldsymbol{\theta}}_{10}},F_{{\boldsymbol{\theta}}_{20}}) &=& 
2 \boldsymbol{Q}_{\beta}(x,y)^T \widetilde{{\boldsymbol{\Sigma}}_{\beta}}({\boldsymbol{\theta}}_{10},{\boldsymbol{\theta}}_{20})^{-1}
\boldsymbol{Q}_{\beta}(x,y).
\nonumber
\end{eqnarray}
with $\boldsymbol{Q}_{\beta}(x,y) 
= {\boldsymbol{\Psi}}_1({\boldsymbol{\theta}}_{10},{\boldsymbol{\theta}}_{20})^T 
\mathcal{IF}(x; \boldsymbol{U}_{\beta}, F_{{\boldsymbol{\theta}}_{10}})
+ {\boldsymbol{\Psi}}_2({\boldsymbol{\theta}}_{10},{\boldsymbol{\theta}}_{20})^T 
\mathcal{IF}(y; \boldsymbol{U}_{\beta}, F_{{\boldsymbol{\theta}}_{20}}).$
\end{enumerate}
\label{THM:Gen_IF_test}
\end{theorem}

Clearly, as in the previous case of simple two sample problem in Section \ref{SEC:Simple_IF_test},
here also the first order IF of the test statistics are always zero and hence non-informative about their robustness.
However, their second order IFs are clearly bounded 
whenever the IF of the corresponding MDPDE is bounded which holds for all $\beta>0$. 
Thus, the proposed general two sample Wald-type tests with any $\beta>0$ 
 yield robust solution under contamination in either of the samples or in both. 
Further, in case of contamination in both the samples, if the IF of the MDPDE is not bounded (at $\beta=0$), 
then also the corresponding second order IF can be bounded generating robust inference 
 provided the term $\boldsymbol{Q}_{\beta}(x,y)$ is bounded. 
 One example of such situation arises in case of the simpler problem of Section \ref{SEC:Simple_problem} 
 under the choice $x=y$, because in that case ${\boldsymbol{\Psi}}_1({\boldsymbol{\theta}}_{10},{\boldsymbol{\theta}}_{20})= -{\boldsymbol{\Psi}}_2({\boldsymbol{\theta}}_{10},{\boldsymbol{\theta}}_{20})=\boldsymbol{I}_p$, 
 the identity matrix of oder $p$,
  and hence $\boldsymbol{Q}_{\beta}(x,y)$ becomes identically zero.

Next, we consider the effect of contamination on the asymptotic power and level of the proposed general Wald-type tests
based on $\widetilde{T_{m,n}^{(\beta)}}$. For this general case, 
we consider the contiguous alternatives $H_{1,m,n}$ as defined in (\ref{EQ:Contiguous_alternative_gen})
but now with the null baseline parameter values as $\boldsymbol{\theta}_{10}$ and $\boldsymbol{\theta}_{20}$
for the two samples respectively instead of the common $\boldsymbol{\theta}_{0}$ 
and define the level and power influence functions using the corresponding contaminated distributions 
as in Section \ref{SEC:Simple_IF_power}.
Following theorem presents the asymptotic distribution of the test statistics under 
the contiguous and contaminated distributions,
where $\widetilde{\boldsymbol{\Delta}}_i$s ($i=1,2$) are as defined in Section \ref{SEC:Simple_IF_power}.

\begin{theorem}
Suppose the model density satisfies the Lehmann and Basu et al.~conditions and Assumptions (A)-(B) hold.
Then, the asymptotic distribution of the general Wald-type test statistics  $\widetilde{T_{m,n}^{(\beta)}}$ 
under any contaminated contiguous alternative distributions $(D_1, D_2)$ is non-central chi-square 
with $r$ degrees of freedom and non-centrality parameter
$\widetilde{\boldsymbol{W}_{\varepsilon}^*}^T
\widetilde{{\boldsymbol{\Sigma}}_{\beta}}({\boldsymbol{\theta}}_1,{\boldsymbol{\theta}}_2)^{-1}
\widetilde{\boldsymbol{W}_{\varepsilon}^*}$, 
where
 
$$
 \begin{array}{r l l}
\widetilde{\boldsymbol{W}_{\varepsilon}^*}
= & \boldsymbol{W}_{\boldsymbol{\psi}}(\widetilde{{\boldsymbol{\Delta}}}_1,{{\boldsymbol{\Delta}}}_2),
 & \mbox{if } ~(D_1, D_2)=(F_{1,n,\varepsilon,x}^P,F_{{\boldsymbol{\theta}}_{2,m}}),\\
 = & \boldsymbol{W}_{\boldsymbol{\psi}}({{\boldsymbol{\Delta}}}_1,\widetilde{{\boldsymbol{\Delta}}}_2), 
 & \mbox{if } ~(D_1, D_2)=(F_{{\boldsymbol{\theta}}_{1,n}},F_{2,m,\varepsilon,y}^P),\\
 = & \boldsymbol{W}_{\boldsymbol{\psi}}(\widetilde{{\boldsymbol{\Delta}}}_2,\widetilde{{\boldsymbol{\Delta}}}_2), 
 & \mbox{if } ~(D_1, D_2)=(F_{1,n,\varepsilon,x}^P,F_{2,m,\varepsilon,y}^P).
 \end{array}
$$
\label{THM:2sample_gen_asympCont}
\end{theorem}

The above theorem can be used to get the asymptotic power of 
the proposed general two-sample Wald-type tests under the contiguous contaminated alternatives 
in terms of an infinite series following Section \ref{SEC:Simple_IF_power} 
(arguments after Theorem \ref{THM:2sample_simple_asympCont}).
This can be also simplified by substituting $\varepsilon = 0 $ or 
$\boldsymbol{\Delta}_1 = \boldsymbol{\Delta}_2 = \boldsymbol{0}_p$
to get asymptotic power under contiguous alternatives or 
the asymptotic level under contiguous contamination respectively.
Further, the resulting infinite series expressions can now be used 
to obtain the power and level influence functions for this general case.
Since the derivations are the same as that of Theorem \ref{THM:2sample_simple_PIF}, 
for brevity, we will only present the resulting expressions skipping  the details 
in the following Theorem.

\begin{theorem}
Suppose the model density satisfies the Lehmann and Basu et al.~conditions, and Assumptions (A)--(B) hold.
Then we have the following results for the proposed Wald-type test functional $\widetilde{T}_{\beta}$
for testing the general two-sample hypothesis in (\ref{EQ:7two_sample_Gen}).
\begin{enumerate}
	\item The power influence functions are given by 
\begin{eqnarray}
PIF^{(1)}(x; \widetilde{{T}_{\beta}}, F_{{\boldsymbol{\theta}}_{10}},F_{{\boldsymbol{\theta}}_{20}})
&=& \sqrt{\omega}K_r^*\left(\widetilde{\delta_{\beta}}\right)
\boldsymbol{W}_{\boldsymbol{\psi}}({\boldsymbol{\Delta}}_1,\boldsymbol{\Delta}_2)^T
\widetilde{\boldsymbol{\Sigma}_{\beta}}({\boldsymbol{\theta}}_0)^{-1}
{\boldsymbol{\Psi}}_1({\boldsymbol{\theta}}_{10},{\boldsymbol{\theta}}_{20})^T
\mathcal{IF}(x; \boldsymbol{U}_{\beta}, F_{{\boldsymbol{\theta}}_{10}}),\nonumber\\
PIF^{(2)}(y; \widetilde{T_{\beta}}, F_{{\boldsymbol{\theta}}_{10}},F_{{\boldsymbol{\theta}}_{20}})
&=& \sqrt{1-\omega}K_r^*\left(\widetilde{\delta_{\beta}}\right)
\boldsymbol{W}_{\boldsymbol{\psi}}({\boldsymbol{\Delta}}_1,\boldsymbol{\Delta}_2)^T
\widetilde{\boldsymbol{\Sigma}_{\beta}}({\boldsymbol{\theta}}_0)^{-1}
{\boldsymbol{\Psi}}_2({\boldsymbol{\theta}}_{10},{\boldsymbol{\theta}}_{20})^T
\mathcal{IF}(y; \boldsymbol{U}_{\beta}, F_{{\boldsymbol{\theta}}_{20}}),\nonumber\\
PIF(x,y; \widetilde{T_{\beta}}, F_{{\boldsymbol{\theta}}_{10}},F_{{\boldsymbol{\theta}}_{20}})
&=& K_r^*\left(\widetilde{\delta_{\beta}}\right)
\boldsymbol{W}_{\boldsymbol{\psi}}({\boldsymbol{\Delta}}_1,\boldsymbol{\Delta}_2)^T
\widetilde{\boldsymbol{\Sigma}_{\beta}}({\boldsymbol{\theta}}_0)^{-1}
\boldsymbol{W}_{\boldsymbol{\psi}}\left(\mathcal{IF}(x; \boldsymbol{U}_{\beta}, F_{{\boldsymbol{\theta}}_{10}}), 
\mathcal{IF}(y; \boldsymbol{U}_{\beta}, F_{{\boldsymbol{\theta}}_{20}})\right),\nonumber
	\end{eqnarray}
where $\widetilde{\delta_{\beta}}$ and $\boldsymbol{W}_{\boldsymbol{\psi}}({\boldsymbol{\Delta}}_1,\boldsymbol{\Delta}_2)$ 
are as defined in Theorem \ref{THM:2sample_gen_asympContg} and $K_r^*(s)$ is as defined in Theorem \ref{THM:2sample_simple_PIF}.

\item Provided the IF of the MDPDE $\boldsymbol{U}_{\beta}$ is bounded, the level influence functions are given by 
\begin{equation}
LIF^{(1)}(x; \widetilde{T_{\beta}}, F_{{\boldsymbol{\theta}}_{10}},F_{{\boldsymbol{\theta}}_{20}})=0, ~~~
LIF^{(2)}(y; \widetilde{T_{\beta}}, F_{{\boldsymbol{\theta}}_{10}},F_{{\boldsymbol{\theta}}_{20}})=0, ~~~
LIF(x,y; \widetilde{T_{\beta}}, F_{{\boldsymbol{\theta}}_{10}},F_{{\boldsymbol{\theta}}_{20}})=0.\nonumber
\end{equation}
\end{enumerate}
	\label{THM:2sample_gen_PIF}
\end{theorem}

Note that for the general two-sample hypothesis (\ref{EQ:7two_sample_Gen}) also, 
the LIFs and the PIFs of our proposed test are bounded whenever the influence function 
of the MDPDE used is bounded which holds for all $\beta >0$.
Thus, our proposal with $\beta >0$ is robust also for testing any general two-sample problem.

\subsection{Special Case: Testing Partial Homogeneity with Nuisance Parameters}

Let us consider a simplified and possibly the most common special case of the general hypothesis in (\ref{EQ:7two_sample_Gen}),
where we test for partial homogeneity of the two samples assuming some parameters to be nuisance. 
Mathematically, let us consider the partition of the parameters
$\boldsymbol{{\boldsymbol{\theta}} }_{1}=
\left( ^{\ast }\boldsymbol{{\boldsymbol{\theta}} }_{1}^{T},^{0}\boldsymbol{{\boldsymbol{\theta}} }_{1}^{T}\right) ^{T}$ 
and $\boldsymbol{{\boldsymbol{\theta}} }_{2}=
\left( ^{\ast }\boldsymbol{{\boldsymbol{\theta}} }_{2}^{T},^{0}\boldsymbol{{\boldsymbol{\theta}} }_{2}^{T}\right) ^{T}$
as in the beginning of Section \ref{SEC:Simple_problem}, but now we assume both,  
$^{0}\boldsymbol{{\boldsymbol{\theta}} }_{1}$ and $^{0}\boldsymbol{{\boldsymbol{\theta}} }_{2}$, to be unknown and nuisance parameters.
Under these notations, we consider the hypothesis of partial homogeneity as given by 
\begin{eqnarray}
H_0 : {^\ast}\boldsymbol{{\boldsymbol{\theta}} }_{1} = {^\ast}\boldsymbol{{\boldsymbol{\theta}} }_{2}
~~~~~\mbox{ against }~~~~
H_1 : {^\ast}\boldsymbol{{\boldsymbol{\theta}} }_{1} \neq {^\ast}\boldsymbol{{\boldsymbol{\theta}} }_{2}, 
\label{EQ:7two_sample_GenP}
\end{eqnarray}
with $^{0}\boldsymbol{{\boldsymbol{\theta}} }_{1}$ and $^{0}\boldsymbol{{\boldsymbol{\theta}} }_{2}$
being unknown under both hypotheses. 
Note that, this special case contains the problem of testing normal mean with unknown variances with 
$^{\ast}\boldsymbol{{\boldsymbol{\theta}} }_{i}$ being the mean and 
$^{0}\boldsymbol{{\boldsymbol{\theta}} }_{i}$ being the variance parameter for each $i=1,2$.
In practice we can either assume $^{0}\boldsymbol{{\boldsymbol{\theta}} }_{1}=^{0}\boldsymbol{{\boldsymbol{\theta}} }_{2}$
(e.g., equal variances) or $^{0}\boldsymbol{{\boldsymbol{\theta}} }_{1}\neq ^{0}\boldsymbol{{\boldsymbol{\theta}} }_{2}$
(e.g., unequal variances). Here, we will consider the general case assuming 
$^{0}\boldsymbol{{\boldsymbol{\theta}} }_{1}\neq ^{0}\boldsymbol{{\boldsymbol{\theta}} }_{2}$; 
other case can also be dealt similarly. 

Note that the hypothesis (\ref{EQ:7two_sample_GenP}) is indeed a special case of the 
general hypothesis in (\ref{EQ:7two_sample_Gen}) with 
$
\boldsymbol{\psi}(\boldsymbol{{\boldsymbol{\theta}} }_{1}, \boldsymbol{{\boldsymbol{\theta}} }_{2})
= {^\ast }\boldsymbol{{\boldsymbol{\theta}} }_{1} - {^\ast }\boldsymbol{{\boldsymbol{\theta}} }_{2}.
$
Hence, the proposed MDPDE based Wald-type test statistics for testing (\ref{EQ:7two_sample_GenP}) is given by 
\begin{equation}
\label{EQ:7_2SDT_genP}
\widetilde{T_{m,n}^{(\beta)}}
= \frac{nm}{n+m} ~ \left( ^{(1)\ast}\widehat{{\boldsymbol{\theta}}}_{\beta} - ^{(2)\ast}\widehat{{\boldsymbol{\theta}}}_{\beta}\right)^T
\left[\omega \boldsymbol{\Sigma}_{\beta}^{11}( ^{(1)}\widehat{{\boldsymbol{\theta}}}_{\beta}) + (1-\omega)\boldsymbol{\Sigma}_{\beta}^{11}(^{(2)}\widehat{{\boldsymbol{\theta}}}_{\beta})\right]^{-1}
\left( ^{(1)\ast}\widehat{{\boldsymbol{\theta}}}_{\beta} - ^{(2)\ast}\widehat{{\boldsymbol{\theta}}}_{\beta}\right),
\end{equation}
where $^{(1)\ast}\widehat{{\boldsymbol{\theta}}}_{\beta}$ and $^{(2)\ast}\widehat{{\boldsymbol{\theta}}}_{\beta}$
are the first $r$-components of the MDPDEs $^{(1)}\widehat{{\boldsymbol{\theta}}}_{\beta}
= (^{(1)\ast}\widehat{{\boldsymbol{\theta}}}_{\beta}^T, ^{(1)0}\widehat{{\boldsymbol{\theta}}}_{\beta}^T)^T$
and $^{(2)}\widehat{{\boldsymbol{\theta}}}_{\beta}
= (^{(2)\ast}\widehat{{\boldsymbol{\theta}}}_{\beta}^T, ^{(2)0}\widehat{{\boldsymbol{\theta}}}_{\beta}^T)^T$ 
of $\boldsymbol{\theta}_1$ and $\boldsymbol{\theta}_2$ respectively
and $\boldsymbol{\Sigma}_{\beta}^{11}(\boldsymbol{\theta})$ denotes 
the $r \times r$ principle minor of the asymptotic variance-covariance matrix 
$\boldsymbol{\Sigma}_{\beta}(\boldsymbol{\theta})=\begin{pmatrix}
\begin{array}{cc}
\boldsymbol{\Sigma}_{\beta}^{11}(\boldsymbol{\theta})   & \boldsymbol{\Sigma}_{\beta}^{12}(\boldsymbol{\theta})\\
\boldsymbol{\Sigma}_{\beta}^{12}(\boldsymbol{\theta})^T & \boldsymbol{\Sigma}_{\beta}^{22}(\boldsymbol{\theta})
\end{array}
\end{pmatrix}$.
Also note that Assumption (B) always holds for the hypothesis (\ref{EQ:7two_sample_GenP}).
Following Theorem \ref{THM:2sample_gen_asympNull}, the asymptotic distribution of $\widetilde{T_{m,n}^{(\beta)}}$
in (\ref{EQ:7_2SDT_genP}) under the null hypothesis in (\ref{EQ:7two_sample_GenP}) is $\chi_r^2$ 
and the test is consistent against any fixed alternatives by Corollary \ref{Corr:gen_const}.
To study the asymptotic contiguous power in this case, we consider the contiguous alternatives
\begin{equation}
H_{1,n,m}': {^\ast}{\boldsymbol{\theta}}_1= {^\ast}{\boldsymbol{\theta}}_{0} + n^{-\frac{1}{2}}{\boldsymbol{\Delta}}_1,~~
{^\ast}{\boldsymbol{\theta}}_2={^\ast}{\boldsymbol{\theta}}_{0} + m^{-\frac{1}{2}}{\boldsymbol{\Delta}}_2, ~~~~~~
({\boldsymbol{\Delta}}_1, {\boldsymbol{\Delta}}_2)\in \mathbb{R}^r\times\mathbb{R}^r-\{(\boldsymbol{0}_r,\boldsymbol{0}_r)\},
\label{EQ:Contiguous_alternative_genP}
\end{equation} 
for some fixed ${^\ast}{\boldsymbol{\theta}}_{0} \in \Theta$. 
Then, by Theorem \ref{THM:2sample_gen_asympContg}, 
the asymptotic distribution of the Wald-type test statistics $\widetilde{T_{m,n}^{(\beta)}}$
in (\ref{EQ:7_2SDT_genP}) under $H_{1,n,m}'$ in (\ref{EQ:Contiguous_alternative_genP}) 
is a non-central chi-square distribution with $r$ degrees of freedom and non-centrality parameter 
${^\ast}\widetilde{\delta_{\beta}} = \boldsymbol{W}(\boldsymbol{\Delta}_1, \boldsymbol{\Delta}_2)^T
\left[\omega \boldsymbol{\Sigma}_{\beta}^{11}( ^{(1)}\widehat{{\boldsymbol{\theta}}}_{\beta}) + (1-\omega)\boldsymbol{\Sigma}_{\beta}^{11}(^{(2)}\widehat{{\boldsymbol{\theta}}}_{\beta})\right]^{-1}
\boldsymbol{W}(\boldsymbol{\Delta}_1, \boldsymbol{\Delta}_2)$
from which the power can be calculated easily.

Next, for examining robustness properties, we define the corresponding test functional 
following Section \ref{SEC:Gen_IF} as given by 
$$
\widetilde{T_{\beta}}(G_1, G_2) = \left(^{\ast}\boldsymbol{U}_{\beta}(G_1) -^{\ast}\boldsymbol{U}_{\beta}(G_2)\right)^T
\left[\omega \boldsymbol{\Sigma}_{\beta}^{11}( ^{(1)}\widehat{{\boldsymbol{\theta}}}_{\beta}) + (1-\omega)\boldsymbol{\Sigma}_{\beta}^{11}(^{(2)}\widehat{{\boldsymbol{\theta}}}_{\beta})\right]^{-1}
\left(^{\ast}\boldsymbol{U}_{\beta}(G_1) - ^{\ast}\boldsymbol{U}_{\beta}(G_2)\right),
$$ 
where $^{\ast}\boldsymbol{U}_{\beta}$ denotes first  $r$-components of the minimum DPD functional $\boldsymbol{U}_{\beta}$.
Then, we can get the IF for this test statistics from Theorem \ref{THM:Gen_IF_test}.
In particular, the first order influence function is identically zero for any kind of contamination
and hence non-informative. And its
second order influence function for contamination in $i$-th sample at the point $x_i$ ($i=1,2$) is given by 
$$
IF^{(i)}_2(x_i; \widetilde{T_{\beta}},  F_{{\boldsymbol{\theta}}_{i0}},F_{{\boldsymbol{\theta}}_{20}}) 
= 2\mathcal{IF}(x_i; ^{\ast}\boldsymbol{U}_{\beta}, F_{{\boldsymbol{\theta}}_{10}})^T 
\left[\omega \boldsymbol{\Sigma}_{\beta}^{11}( ^{(1)}\widehat{{\boldsymbol{\theta}}}_{\beta}) + (1-\omega)\boldsymbol{\Sigma}_{\beta}^{11}(^{(2)}\widehat{{\boldsymbol{\theta}}}_{\beta})\right]^{-1}
\mathcal{IF}(x_i; ^{\ast}\boldsymbol{U}_{\beta}, {\boldsymbol{\theta}}_{i0}).\nonumber
$$
and the same for contamination in both samples is given by 
$$
IF_2(x, y; \widetilde{T_{\beta}},  F_{{\boldsymbol{\theta}}_{10}},F_{{\boldsymbol{\theta}}_{20}}) = 
2 ~{^\ast}\boldsymbol{Q}_{\beta}(x,y)^T 
\left[\omega \boldsymbol{\Sigma}_{\beta}^{11}( ^{(1)}\widehat{{\boldsymbol{\theta}}}_{\beta}) + (1-\omega)\boldsymbol{\Sigma}_{\beta}^{11}(^{(2)}\widehat{{\boldsymbol{\theta}}}_{\beta})\right]^{-1}
{^\ast}\boldsymbol{Q}_{\beta}(x,y),
$$
with $^\ast\boldsymbol{Q}_{\beta}(x,y) 
= \mathcal{IF}(x; {^\ast}\boldsymbol{U}_{\beta}, F_{{\boldsymbol{\theta}}_{10}})
- \mathcal{IF}(y; {^\ast}\boldsymbol{U}_{\beta}, F_{{\boldsymbol{\theta}}_{20}}).$
Similarly, following Theorem \ref{THM:2sample_gen_PIF}, the level influence functions are always zero and 
the power influence functions under contiguous contamination in each sample separately or 
in both the samples are respectively given by 
\begin{eqnarray}
	PIF^{(1)}(x; \widetilde{{T}_{\beta}}, F_{{\boldsymbol{\theta}}_{10}},F_{{\boldsymbol{\theta}}_{20}})
	&=& \sqrt{\omega}K_r^*\left( {^\ast}\widetilde{\delta_{\beta}}\right)
	\boldsymbol{W}({\boldsymbol{\Delta}}_1,\boldsymbol{\Delta}_2)^T
\left[\omega \boldsymbol{\Sigma}_{\beta}^{11}( ^{(1)}\widehat{{\boldsymbol{\theta}}}_{\beta}) + (1-\omega)\boldsymbol{\Sigma}_{\beta}^{11}(^{(2)}\widehat{{\boldsymbol{\theta}}}_{\beta})\right]^{-1}	
\mathcal{IF}(x; {^\ast}\boldsymbol{U}_{\beta}, F_{{\boldsymbol{\theta}}_0}),\nonumber\\
	PIF^{(2)}(y; \widetilde{T_{\beta}}, F_{{\boldsymbol{\theta}}_{10}},F_{{\boldsymbol{\theta}}_{20}})
	&=& \sqrt{1-\omega}K_r^*\left( {^\ast}\widetilde{\delta_{\beta}}\right)
	\boldsymbol{W}({\boldsymbol{\Delta}}_1,\boldsymbol{\Delta}_2)^T
\left[\omega \boldsymbol{\Sigma}_{\beta}^{11}( ^{(1)}\widehat{{\boldsymbol{\theta}}}_{\beta}) + (1-\omega)\boldsymbol{\Sigma}_{\beta}^{11}(^{(2)}\widehat{{\boldsymbol{\theta}}}_{\beta})\right]^{-1}	
\mathcal{IF}(y; {^\ast}\boldsymbol{U}_{\beta}, F_{{\boldsymbol{\theta}}_0}),\nonumber\\
	PIF(x,y; \widetilde{T_{\beta}}, F_{{\boldsymbol{\theta}}_{10}},F_{{\boldsymbol{\theta}}_{20}})
	&=& K_r^*\left( {^\ast}\widetilde{\delta_{\beta}}\right)
	\boldsymbol{W}({\boldsymbol{\Delta}}_1,\boldsymbol{\Delta}_2)^T
\left[\omega \boldsymbol{\Sigma}_{\beta}^{11}( ^{(1)}\widehat{{\boldsymbol{\theta}}}_{\beta}) + (1-\omega)\boldsymbol{\Sigma}_{\beta}^{11}(^{(2)}\widehat{{\boldsymbol{\theta}}}_{\beta})\right]^{-1}	\nonumber\\
&& ~~~~~~ \times \boldsymbol{W}\left(\mathcal{IF}(x; {^\ast}\boldsymbol{U}_{\beta}, F_{{\boldsymbol{\theta}}_0}), 
	\mathcal{IF}(x; {^\ast}\boldsymbol{U}_{\beta}, F_{{\boldsymbol{\theta}}_0})\right),\nonumber
\end{eqnarray}
where ${^\ast}\widetilde{\delta_{\beta}}$ and ${^\ast}\boldsymbol{U}_{\beta}$ are as defined previously in this subsection.
The nature of these PIFs are exactly the same as in the previous cases 
and indicates robustness of our proposals with $\beta>0$.

\medskip
\begin{example}[Testing equality of two Normal means with unknown and unequal variances]\label{EXM:3.1}
	\normalfont{~~~~~\\
We again consider the example of comparing two normal means (say $\mu_1$ and $\mu_2$),
but now with unknown and unequal variances (say $\sigma_1^2$ and $\sigma_2^2$) for the two populations.
Hence the model family is $\mathcal{F}=\{N(\mu,\sigma^2):\boldsymbol{\theta}=(\mu, \sigma)^T \in\mathbb{R}\times[0, \infty)\}$ 
and we want to test for the hypothesis 
\begin{eqnarray}
H_0 : {\mu}_{1} = {\mu}_{2} ~~~~~\mbox{ against }~~~~ H_1 : {\mu}_{1} \neq {\mu}_{2}, 
\label{EQ:7two_sample_GenPeg}
\end{eqnarray}
with $\sigma_1^2$ and $\sigma_2^2$ being unknown under both hypotheses.
Let us denote the MDPDEs based on the $i$-th sample ($i=1,2$) as 
$^{(i)}\widehat{{\boldsymbol{\theta}}}_{\beta} 
= (^{(i)}\widehat{\mu}_{\beta}, ^{(i)}\widehat{\sigma}_{\beta})^T$
and its asymptotic variance matrix ${\boldsymbol{\Sigma}}_{\beta}({\boldsymbol{\theta}})$ is given by 
$$
{\boldsymbol{\Sigma}}_{\beta}(\mu, \sigma) = \begin{pmatrix}
\begin{array}{cc}
\left(1+\frac{\beta^2}{1+2\beta}\right)^{3/2}\sigma^2 & 0\\
0 & \frac{(1+\beta)^2}{(2+\beta^2)^2} \left\{\frac{2 \zeta_{\beta}}{(1+2\beta)^{5/2}} - \beta^2\right\}
\end{array}\end{pmatrix},
$$
with  $\zeta_{\beta} = 1 + 3 \beta + 5\beta^2 + 7\beta^3 + 6\beta^4 + 2 \beta^5$.
Then, noting that the hypothesis (\ref{EQ:7two_sample_GenPeg}) is of the form (\ref{EQ:7two_sample_GenP}),
our proposed generalized Wald-type test statistics (\ref{EQ:7_2SDT_genP}) simplifies to
\begin{eqnarray}
\widetilde{T_{m,n}^{(\beta)}} = \frac{mn}{m+n}\left(1+\frac{\beta^2}{1+2\beta}\right)^{-3/2}
\frac{\left(^{(1)}\widehat{{{\mu}}}_{\beta} - ^{(2)}\widehat{{{\mu}}}_{\beta}\right)^2}{
\left(\omega^{(1)}\widehat{{{\sigma}}}_{\beta}^2 + (1 -\omega) ^{(2)}\widehat{{{\sigma}}}_{\beta}^2\right)},
\label{EQ:7_2SDT_genPeg}
\end{eqnarray}
whose null asymptotic distribution is $\chi_1^2$ from Theorem \ref{THM:2sample_gen_asympNull}.
In the particular case of $\beta=0$, we have 
$$
\widetilde{T_{m,n}^{(0)}} = \frac{mn}{m+n}\frac{\left(^{(1)}\widehat{{{\mu}}}_{0} - ^{(2)}\widehat{{{\mu}}}_{0}\right)^2}{
	\left(\omega {^{(1)}}\widehat{{{\sigma}}}_{0}^2 + (1 -\omega) {^{(2)}}\widehat{{{\sigma}}}_{0}^2\right)}
= \frac{mn}{m+n}\frac{\left(\bar{X} - \bar{Y}\right)^2}{\left(\omega s_X^2 + (1 -\omega) s_Y^2\right)},
$$ 
where $\bar{X}$ and $\bar{Y}$ are the sample means and $s_X^2$ and $s_Y^2$ are the sample variances 
of $X_1, \ldots, X_n$ and $Y_1, \ldots, Y_m$ respectively, 
and this is nothing but the classical MLE based Wald test statistic.

We can now study the asymptotic and robustness properties of these proposed Wald-type tests 
following the theoretical results derived in this section.
However, due to the asymptotic independence of the MDPDEs of $\mu$ and $\sigma$ under normal model, 
all the properties of the Wald-type test statistics in (\ref{EQ:7_2SDT_genPeg}) 
turn out to be similar in nature to those of the proposed Wald-type test with known $\sigma$ 
as discussed in Examples \ref{EXM:1}, \ref{EXM:3.2} and \ref{EXM:3.3}
with the common variance $\sigma^2$ there replaced by 
$\left[\omega\sigma_1^2 + (1-\omega)\sigma_2^2\right]$ in the present case.
This fact can also be observed intuitively by noting that the Wald-type test statistics in (\ref{EQ:7_2SDT_genPeg})
have a similar form as the corresponding Wald-type test statistics for known common $\sigma^2$ case (in Example \ref{EXM:1})
with the known value there being replaced by 		
$\left[\omega {^{(1)}}\widehat{{{\sigma}}}_{\beta}^2 + (1 -\omega) {^{(2)}}\widehat{{{\sigma}}}_{\beta}^2\right]$.
So, we will skip these details for the present general case for brevity. 
However, examining them, one can easily verify that, in this case of unknown and unequal variances also, 
the asymptotic contiguous power of the proposed Wald-type test decreases only slightly as $\beta$ increases
(exactly in the same rate as in Table \ref{TAB:ContPower}) but the robustness increases significantly 
having bounded (second order) influence functions of the Wald-type test statistics and 
bounded power and level influence functions for all $\beta>0$.
	}
\end{example}

\section{The Cases of One-Sided Alternatives}
\label{SEC:One-sided_problem}

As we have mentioned in the introduction (Section \ref{SEC:intro}),
majority of common practical applications of the two-sample problems are 
in comparing the treatment and control groups in any experimental or clinical trials 
or any observational studies among two such groups of population. 
However, in most of such cases, researchers want to test 
weather there is any improvement in the treatment group over the control groups due to the treatment effects.
For example, one might be interested to test if the success rate of cure (modeled by binomial probability model) is reduced, 
or if the number of attacks of a disease (modeled by Poisson model) decreases in the treat group,
or some continuous biomarkers like blood pressure etc.~(modeled by normal model) changes in the targeted direction 
from control to treatment group. 
All of them lead to the one-sided alternatives in contrast to the omnibus two-sided alternatives considered so far in this paper. 
Although the case of general one-sided alternatives with vector parameters are much difficult to define 
and dealt with and hence need more targeted future research, our proposal of robust Wald-type tests in this paper
can be easily extended for comparing any scalar parameters with one-sided alternatives. 
Noting that all the above motivating practical scenarios indeed deal with scalar parameter comparison, 
in this section we extend our proposal to these particular one sample problems.

In general, we consider the class of one-sided version of (\ref{EQ:7two_sample_Gen}) with $r=1$. 
So, $\psi(\boldsymbol{\theta}_1, \boldsymbol{\theta}_2)$ is a real function of the parameters
and we develop the robust test for the one-sided hypothesis given by 
\begin{equation}
H_0 : {{\psi}}({\boldsymbol{\theta}}_1, {\boldsymbol{\theta}}_2)=0 ~~~ \mbox{against} ~~~~ 
H_1 : {{\psi}}({\boldsymbol{\theta}}_1, {\boldsymbol{\theta}}_2) > 0.
\label{EQ:7two_sample_GenO}
\end{equation}
Note that the one sided version of the simple two-sided hypothesis in (\ref{EQ:7two_sample}) 
with scalar parameters ($p=1$), that contains the motivating examples for Poisson and binomial models 
and normal model with known variances, belong to this general class (\ref{EQ:7two_sample_GenO}).
Also, this general class of hypotheses contains many more useful cases like 
testing for increase (or decrease) in normal means with unknown variances.

For testing the one sided hypothesis (\ref{EQ:7two_sample_GenO}), 
we define the corresponding robust Wald-type test statistics by taking a signed square-root of 
our two-sided Wald-type test statistics $\widetilde{T_{m,n}^{(\beta)}}$ in (\ref{EQ:7_2SDT_gen})
\begin{equation}
\label{EQ:7_2SDT_genO}
\widetilde{T_{m,n}^{(\beta)P}}
= sgn\left({{\psi}}\left( ^{(1)}\widehat{{\boldsymbol{\theta}}}_{\beta} , ^{(2)}\widehat{{\boldsymbol{\theta}}}_{\beta}\right)\right) \sqrt{\widetilde{T_{m,n}^{(\beta)}}}
= \sqrt{\frac{nm}{n+m}} \frac{{\psi}\left( ^{(1)}\widehat{{\boldsymbol{\theta}}}_{\beta} , ^{(2)}\widehat{{\boldsymbol{\theta}}}_{\beta}\right)}{
\sqrt{\widetilde{{\Sigma}_{\beta}}(^{(1)}\widehat{{\boldsymbol{\theta}}}_{\beta}, ^{(2)}\widehat{{\boldsymbol{\theta}}}_{\beta})}},
\end{equation}
where $sgn(\cdot)$ denotes the sign function and note that
$\widetilde{{\Sigma}_{\beta}}({\boldsymbol{\theta}}_{\beta},{\boldsymbol{\theta}}_{\beta})$ is a scalar for $r=1$.
Then, we have the following null asymptotic distribution.

\begin{theorem}
Under the assumptions of Theorem \ref{THM:2sample_gen_asympNull}, 
the asymptotic null distribution of the one-sided test statistics $\widetilde{T_{m,n}^{(\beta)P}}$
for testing (\ref{EQ:7two_sample_GenO}) is standard normal.
\label{THM:2sample_genO_asympNull}
\end{theorem}

Following the above theorem, the level-$\alpha$ critical region for testing the one-sided hypothesis in (\ref{EQ:7two_sample_GenO})
is given by $\left\{\widetilde{T_{m,n}^{(\beta)P}} > z_{1-\alpha}\right\}$, 
where $z_{1-\alpha}$ denotes the $(1-\alpha)$-th quantile of the standard normal distribution.

Further, as in the case of two-side alternatives, we can also derive an power approximation of these proposed Wald-type tests
at any fixed alternative $(\boldsymbol{\theta}_1, \boldsymbol{\theta}_2)$ satisfying 
$\psi(\boldsymbol{\theta}_1, \boldsymbol{\theta}_2)>0$ as follows:
\begin{eqnarray}
\widetilde{\pi_{m,n,\alpha}}^{(\beta)P} ({\boldsymbol{\theta}}_1, {\boldsymbol{\theta}}_2) 
&=& P\left(\widetilde{T_{m,n}^{(\beta)P}} > z_{1-\alpha}\right) \nonumber\\
&=& P\left(\sqrt{\frac{nm}{n+m}} 
\frac{\left[{\psi}\left( ^{(1)}\widehat{{\boldsymbol{\theta}}}_{\beta} , ^{(2)}\widehat{{\boldsymbol{\theta}}}_{\beta}\right)
- \psi\left(\boldsymbol{\theta}_1, \boldsymbol{\theta}_2\right)\right]}{
\sqrt{\widetilde{{\Sigma}_{\beta}}(^{(1)}\widehat{{\boldsymbol{\theta}}}_{\beta}, ^{(2)}\widehat{{\boldsymbol{\theta}}}_{\beta})}}  > z_{1-\alpha} - \sqrt{\frac{nm}{n+m}} \frac{{\psi}\left(\boldsymbol{\theta}_1 , \boldsymbol{\theta}_2\right)}{
\sqrt{\widetilde{{\Sigma}_{\beta}}(\boldsymbol{\theta}_1, \boldsymbol{\theta}_2)}}\right) \nonumber\\
&=& 1 - \Phi_n \left( z_{1-\alpha} - \sqrt{\frac{nm}{n+m}} \frac{{\psi}\left(\boldsymbol{\theta}_1 , \boldsymbol{\theta}_2\right)}{
\sqrt{\widetilde{{\Sigma}_{\beta}}(\boldsymbol{\theta}_1, \boldsymbol{\theta}_2)}} \right), \nonumber
\end{eqnarray}
for a sequence of distributions $\Phi_n(\cdot)$ tending uniformly to the standard normal distribution $\Phi(\cdot)$,
since under the alternative parameter values $(\boldsymbol{\theta}_1, \boldsymbol{\theta}_2)$
$$
\sqrt{\frac{nm}{n+m}} 
\frac{\left[{\psi}\left( ^{(1)}\widehat{{\boldsymbol{\theta}}}_{\beta} , ^{(2)}\widehat{{\boldsymbol{\theta}}}_{\beta}\right)
	- \psi\left(\boldsymbol{\theta}_1, \boldsymbol{\theta}_2\right)\right]}{
	\sqrt{\widetilde{{\Sigma}_{\beta}}(^{(1)}\widehat{{\boldsymbol{\theta}}}_{\beta}, ^{(2)}\widehat{{\boldsymbol{\theta}}}_{\beta})}} 
\displaystyle\underset{m, n\rightarrow \infty }{\overset{\mathcal{L}}{\longrightarrow }}  N(0, 1).
$$
Now, since ${\psi}\left(\boldsymbol{\theta}_1 , \boldsymbol{\theta}_2\right)>0$ under the alternatives in (\ref{EQ:7two_sample_GenO}),
we have 
$\widetilde{\pi_{m,n,\alpha}}^{(\beta)P} ({\boldsymbol{\theta}}_1, {\boldsymbol{\theta}}_2) \rightarrow 1$
as $m,n \rightarrow\infty$ and hence the proposed Wald-type tests are consistent for the one-sided alternatives also.

Next to study the contiguous power of the proposed Wald-type tests, we can consider the class of contiguous alternatives 
in (\ref{EQ:Contiguous_alternative_gen}) but now with $(\boldsymbol{\Delta}_1, \boldsymbol{\Delta}_2)$ 
being such that ${\psi}\left(\boldsymbol{\theta}_{1,n} , \boldsymbol{\theta}_{2,m}\right)>0$ for all $m, n$.
This can be equivalently (asymptotic) expressed in terms of the sequence of alternatives
\begin{eqnarray}
H_{1,m,n}^P~:~  {\psi}\left(\boldsymbol{\theta}_{1,n} , \boldsymbol{\theta}_{2,m}\right) = \sqrt{\frac{m+n}{mn}}d,
\label{EQ:Contiguous_alternative_genO}
\end{eqnarray}
with $d = W_{\psi}\left(\boldsymbol{\Delta}_1, \boldsymbol{\Delta}_2\right)>0$. 
The following theorem then gives the asymptotic distribution of our Wald-type test statistics 
under the contiguous alternatives in (\ref{EQ:Contiguous_alternative_genO}) and the corresponding asymptotic power.

\begin{theorem}
Under the assumptions of Theorem \ref{THM:2sample_gen_asympContg}, the asymptotic distribution of 
$\widetilde{T_{m,n}^{(\beta)P}}$  in (\ref{EQ:7_2SDT_genO}) under the sequence of contiguous alternatives 
in (\ref{EQ:Contiguous_alternative_genO}) is normal with mean 
$d/\sqrt{\widetilde{{\Sigma}_{\beta}}(\boldsymbol{\theta}_1, \boldsymbol{\theta}_2)}$ and  variance 1.
Hence, the corresponding asymptotic contiguous power of the proposed Wald-type tests is given by 
\begin{eqnarray}
\widetilde{\pi}_{\beta}^P({\boldsymbol{\Delta}}_1, {\boldsymbol{\Delta}}_2) 
=\widetilde{\pi}_{\beta}^P(d) 
= 1 - \Phi\left(z_{1-\alpha} - d\big/\sqrt{\widetilde{{\Sigma}_{\beta}}(\boldsymbol{\theta}_1, \boldsymbol{\theta}_2)}\right).
\nonumber
\end{eqnarray}
\label{THM:2sample_gen_asympContgO}
\end{theorem}

Now we can also derive the robustness properties of the proposed Wald-type tests against one-sided alternatives 
by defining the corresponding statistical function as
$$
\widetilde{T_{\beta}}^P(G_1, G_2) = {\psi}\left(\boldsymbol{U}_{\beta}(G_1),\boldsymbol{U}_{\beta}(G_2)\right)
\bigg/\sqrt{\widetilde{\Sigma_{\beta}}({\boldsymbol{\theta}}_{10},{\boldsymbol{\theta}}_{20})}.
$$ 
Then, under the assumptions of Theorem \ref{THM:Gen_IF_test} 
with contamination in only $i$-th sample at the point $x_i$ ($i=1,2$), 
the first order influence function of the proposed Wald-type test statistics at the null hypothesis in (\ref{EQ:7two_sample_GenO})
is given by 
\begin{eqnarray}
IF^{(i)}(x_i; \widetilde{T_{\beta}}^P,  F_{{\boldsymbol{\theta}}_{10}},F_{{\boldsymbol{\theta}}_{20}}) &=& 
{\boldsymbol{\Psi}}_i({\boldsymbol{\theta}}_{10},{\boldsymbol{\theta}}_{20})^T
\mathcal{IF}(x_i; \boldsymbol{U}_{\beta}, {\boldsymbol{\theta}}_{i0})
\bigg/\sqrt{\widetilde{\Sigma_{\beta}}({\boldsymbol{\theta}}_{10},{\boldsymbol{\theta}}_{20})},\nonumber
\end{eqnarray}
and the same for contamination in both the samples is given by
\begin{eqnarray}
IF(x_1, x_2; \widetilde{T_{\beta}}^P,  F_{{\boldsymbol{\theta}}_{10}},F_{{\boldsymbol{\theta}}_{20}}) &=& 
{Q}_{\beta}(x_1,x_2)\bigg/\sqrt{\widetilde{\Sigma_{\beta}}({\boldsymbol{\theta}}_{10},{\boldsymbol{\theta}}_{20})},
\nonumber
\end{eqnarray}
with ${Q}_{\beta}(\cdot,\cdot)$ being as defined in Theorem \ref{THM:Gen_IF_test} (but is a scalar now).
Note that, unlike the two-sided hypotheses, here the first order influence function 
of the proposed Wald-type test statistics is non-zero.
Further, it is bounded whenever te IF of the corresponding MDPDE is bounded, i.e., only for $\beta>0$
and unbounded at $\beta=0$ implying the robustness of our proposal with $\beta>0$.

In order to derive the corresponding level and power influence functions, 
we consider the same set of hypothesis as in Section \ref{SEC:Gen_IF} 
but now with the restriction ${\psi}\left(\boldsymbol{\theta}_{1,n} , \boldsymbol{\theta}_{2,m}\right) >0$ 
for all $m, n$ under the alternative sequence, which is ensured by assuming
$W_{\psi}\left(\boldsymbol{\Delta}_1, \boldsymbol{\Delta}_2\right)>0$.
Then, the following theorem gives the asymptotic distribution of the one-sided test statistics
$\widetilde{T_{m,n}^{(\beta)P}}$ under the contiguous contaminated distributions.

\begin{theorem}
Under the assumptions of Theorem \ref{THM:2sample_gen_asympCont}, 
the asymptotic distribution of $\widetilde{T_{m,n}^{(\beta)P}}$ 
under any contaminated contiguous alternative distributions $(D_1, D_2)$ is normal with mean 
$\widetilde{\boldsymbol{W}_{\varepsilon}^*}\big/\sqrt{\widetilde{\Sigma_{\beta}}({\boldsymbol{\theta}}_1,{\boldsymbol{\theta}}_2)}$
and variance 1, where $\widetilde{\boldsymbol{W}_{\varepsilon}^*}$ is as defined in Theorem \ref{THM:2sample_gen_asympCont}
for different $(D_1, D_2)$.
\label{THM:2sample_genO_asympCont}
\end{theorem}

Using above theorem and following the arguments similar to those for the two-sided alternatives in Section \ref{SEC:Gen_IF},
we can get the power influence functions for this case of one-sided alternatives also, 
which is presented in the next theorem.

\begin{theorem}
Under the assumptions of Theorem \ref{THM:2sample_gen_PIF}, 
the power influence functions of our proposed Wald-type test functional $\widetilde{T}_{\beta}^P$
for testing the one-sided hypothesis in (\ref{EQ:7two_sample_GenO}) are given by
\begin{eqnarray}
PIF^{(1)}(x; \widetilde{{T}_{\beta}}, F_{{\boldsymbol{\theta}}_{10}},F_{{\boldsymbol{\theta}}_{20}})
&=& \frac{\sqrt{\omega}}{\sqrt{\widetilde{\Sigma_{\beta}}({\boldsymbol{\theta}}_1,{\boldsymbol{\theta}}_2)}}
\phi\left(z_{1-\alpha} - \frac{W_{\psi}\left(\boldsymbol{\Delta}_1, \boldsymbol{\Delta}_2\right)}{
	\sqrt{\widetilde{\Sigma_{\beta}}({\boldsymbol{\theta}}_1,{\boldsymbol{\theta}}_2)}}\right)
{\boldsymbol{\Psi}}_1({\boldsymbol{\theta}}_{10},{\boldsymbol{\theta}}_{20})^T
\mathcal{IF}(x; \boldsymbol{U}_{\beta}, F_{{\boldsymbol{\theta}}_{10}}),\nonumber\\
PIF^{(2)}(y; \widetilde{T_{\beta}}, F_{{\boldsymbol{\theta}}_{10}},F_{{\boldsymbol{\theta}}_{20}})
&=& \frac{\sqrt{1-\omega}}{\sqrt{\widetilde{\Sigma_{\beta}}({\boldsymbol{\theta}}_1,{\boldsymbol{\theta}}_2)}}
\phi\left(z_{1-\alpha} - \frac{W_{\psi}\left(\boldsymbol{\Delta}_1, \boldsymbol{\Delta}_2\right)}{
	\sqrt{\widetilde{\Sigma_{\beta}}({\boldsymbol{\theta}}_1,{\boldsymbol{\theta}}_2)}}\right)
{\boldsymbol{\Psi}}_2({\boldsymbol{\theta}}_{10},{\boldsymbol{\theta}}_{20})^T
\mathcal{IF}(y; \boldsymbol{U}_{\beta}, F_{{\boldsymbol{\theta}}_{20}}),\nonumber\\
PIF(x,y; \widetilde{T_{\beta}}, F_{{\boldsymbol{\theta}}_{10}},F_{{\boldsymbol{\theta}}_{20}})
&=& \frac{\sqrt{1}}{\sqrt{\widetilde{\Sigma_{\beta}}({\boldsymbol{\theta}}_1,{\boldsymbol{\theta}}_2)}}
\phi\left(z_{1-\alpha} - \frac{W_{\psi}\left(\boldsymbol{\Delta}_1, \boldsymbol{\Delta}_2\right)}{
	\sqrt{\widetilde{\Sigma_{\beta}}({\boldsymbol{\theta}}_1,{\boldsymbol{\theta}}_2)}}\right)
{W}_{\psi}\left(\mathcal{IF}(x; \boldsymbol{U}_{\beta}, F_{{\boldsymbol{\theta}}_{10}}), 
\mathcal{IF}(y; \boldsymbol{U}_{\beta}, F_{{\boldsymbol{\theta}}_{20}})\right).\nonumber
\end{eqnarray}
\label{THM:2sample_genO_PIF}
\end{theorem}

Note that, the nature of these PIFs with respect to the contamination points $x$ and $y$ are 
exactly same as those in the case of two-sided alternatives
except for a multiplicative constant.
In particular, they are bounded whenever the influence function of the MDPDE used is bounded, 
i.e., at $\beta>0$, implying robustness of our proposal.

Finally, we can get the level influence functions from the above theorem by substituting 
$\boldsymbol{\Delta}_1= \boldsymbol{\Delta}_2 = \boldsymbol{0}$ in the expressions of PIFs. 
Note that, in this case of one-sided hypothesis testing, the LIFs are not identically zero,
but they are bounded only for $\beta>0$ implying again the level stability of our proposed Wald-type tests.

For illustration, we will again present the case of normal model with one-sided alternatives in the following example. 
Other motivating models  with relevant data examples will be provided in the next section.

\begin{example}[Comparing two Normal means against one-sided alternatives]\label{EXM:4.1}
\normalfont{~~~~~\\
Let us again consider the two-sample problem under normal model with unknown and unequal variances as in Example \ref{EXM:3.1},
but now with the one-sided alternatives so that our target hypothesis is 
\begin{eqnarray}
H_0 : {\mu}_{1} = {\mu}_{2} ~~~~~\mbox{ against }~~~~ H_1 : {\mu}_{1} > {\mu}_{2}, 
\label{EQ:7two_sample_GenOeg}
\end{eqnarray}
with the variance parameters $\sigma_1$ and $\sigma_2$ being unknown for both hypotheses.
Considering the notations of Example \ref{EXM:3.1}, 
our proposed test statistics $\widetilde{T_{m,n}^{(\beta)P}}$ is then given by 
\begin{eqnarray}
\widetilde{T_{m,n}^{(\beta)P}} = \sqrt{\frac{mn}{m+n}}\left(1+\frac{\beta^2}{1+2\beta}\right)^{-3/4}
\frac{\left(^{(1)}\widehat{{{\mu}}}_{\beta} - ^{(2)}\widehat{{{\mu}}}_{\beta}\right)}{
\sqrt{\omega {^{(1)}}\widehat{{{\sigma}}}_{\beta}^2 + (1 -\omega) {^{(2)}}\widehat{{{\sigma}}}_{\beta}^2}},
\label{EQ:7_2SDT_genOeg}
\end{eqnarray}
which has standard normal asymptotic distribution under the null.
Clearly this statistic also coincides with the corresponding classical Wald test statistic at $\beta=0$.
Since the test is consistent at any fixed alternatives, we consider the contiguous alternatives
$H_{1,m,n}^P : \psi(\boldsymbol{\theta}_1, \boldsymbol{\theta}_2) = \mu_1 - \mu_2 = \sqrt{\frac{m+n}{mn}}d$
with $d>0$, under which the test statistics has asymptotic distribution as normal with mean 
$\left(1+\frac{\beta^2}{1+2\beta}\right)^{-3/4}d
\left[\omega \sigma_1^2 + (1 -\omega) \sigma_2^2\right]^{-\frac{1}{2}}$
and variance 1. Corresponding asymptotic contiguous power at different values of $d$ and $\beta$ with $\sigma_1^2=\sigma_2^2 = 1$
and $\omega=0.5$ ($n=m$) is presented in Table \ref{TAB:ContPowerO}.
Note that, as expected this power decreases only slightly as $\beta$ increases 
(note the similarity with Table \ref{TAB:ContPower}). 

\begin{table}[!th] 
\caption{Asymptotic contiguous power of the proposed Wald-type tests at 95\% level 
	for testing equality of two normal means against one-sided alternatives as in Example \ref{EXM:4.1}}
\centering
\begin{tabular}{|l|ccccccc|} \hline 
& \multicolumn{7}{c|}{$\beta$} \\ 	
$d$ &	0	&	0.1	&	0.3	&	0.5	&	0.7	&	0.9	&	1	\\\hline\hline
0	&	0.050	&	0.050	&	0.050	&	0.050	&	0.050	&	0.050	&	0.050	\\
1	&	0.260	&	0.258	&	0.247	&	0.233	&	0.219	&	0.207	&	0.201	\\
2	&	0.639	&	0.634	&	0.608	&	0.574	&	0.538	&	0.503	&	0.487	\\
3	&	0.912	&	0.909	&	0.891	&	0.865	&	0.833	&	0.798	&	0.780	\\
5	&	1.000	&	1.000	&	0.999	&	0.998	&	0.997	&	0.994	&	0.991	\\\hline
\end{tabular}
\label{TAB:ContPowerO}
\end{table}

Further, the influence function of the proposed Wald-type test statistics in this case of one-sided alternatives 
simplifies to
$$
IF_2^{(i)}(x_i; \widetilde{T_{\beta}}^P, F_{{\boldsymbol{\theta}}_{10}}, F_{{\boldsymbol{\theta}}_{20}})
=\left[\omega \sigma_{10}^2 + (1 -\omega) \sigma_{20}^2\right]^{-\frac{1}{2}}
\left(1+2\beta\right)^{3/4}(x_i-\mu_{i0}) e^{-\frac{\beta(x_i-\mu_{i0})^2}{2\sigma_{i0}^2}},
$$ 
and
$$
IF_2(x_1,x_2; \widetilde{T_{\beta}}^P, F_{{\boldsymbol{\theta}}_{10}}, F_{{\boldsymbol{\theta}}_{20}})
=\frac{\left(1+2\beta\right)^{3/4}}{\sqrt{\omega \sigma_{10}^2 + (1 -\omega) \sigma_{20}^2}}
\left[(x_1-\mu_{10}) e^{-\frac{\beta(x_1-\mu_{10})^2}{2\sigma_{10}^2}}
- (x_2-\mu_{20}) e^{-\frac{\beta(x_2-\mu_{20})^2}{2\sigma_{20}^2}}\right].
$$
Note that these influence functions are square roots of 
the corresponding influence functions under two-sided alternatives in Example \ref{EXM:3.2}
except for a multiplicative constant. Further, by the general theory developed above, 
the corresponding PIFs and LIFs in this case can be shown to be also 
a constant multiplication of the corresponding PIFs in the two-sided case presented in Example \ref{EXM:3.3}.
Therefore, the boundedness nature of all these influence functions for the one-sided alternative 
will  be similar to those presented in Figures \ref{FIG:IF2_testStat} and \ref{FIG:PIF_test2},
i.e., bounded at $\beta>0$ and unbounded at $\beta=0$.
These again imply the robustness of our proposal with $\beta>0$ over the classical Wald test at $\beta=0$.  
}
\end{example}

\section{Real Life Applications}
\label{SEC:real_data}

\subsection{Poisson Model for Clinical Trial: Adverse Events Data}

In our first example we will consider the application of the proposed Wald-type tests 
with Poisson model to the adverse event data in an Asthma clinical trial conducted by  \citet[][Table 3]{Kerstjens/etc:2012}.
In this two phase randomized controlled trials, 912 patients having asthma and receiving inhaled glucocorticoids and LABAs
had been divided into treatment and control groups of the two trials and 
were randomly assigned a total dose of 5 μg tiotropium (treatment group) or suitable placebo (control group) once daily for 48 weeks. 
Then, \cite{Kerstjens/etc:2012} investigated the effect of this combined treatment on patient's lung function and exacerbations.

\begin{table}[!th] 
	\caption{No of Different adverse events reported in Trial 2 of the \cite{Kerstjens/etc:2012} clinical trail study}
	\centering
	\begin{tabular}{|l|ccccccccccccccccccc|} \hline 
		Treatment	&	91	& 49	& 19	& 12	& 12	& 3	 & 13	& 10	& 6	& 3	& 3	& 7	& 6	& 5	& 4	& 4	& 3	& 2	& 0\\\hline
		Control		&	109	& 58	& 20	& 13	& 10	& 10 & 6	& 4 	& 5	& 7	& 5	& 1	& 2	& 4	& 4	& 5	& 2	& 2	& 1	\\\hline
	\end{tabular}
	\label{TAB:Adverse_event_data}
\end{table}

Here we will consider the data on 19 reported adverse effect on the patients in trail 2 of this study,
presented in Table \ref{TAB:Adverse_event_data}, that can be modeled by a Poisson distribution with mean $\theta$.
Note that the first two entry for both the groups (corresponding to the events of Asthma and Decreased rate of peak expiratory flow)
clearly stands out as outliers from the remaining observations. Hence, in presence of these two observations
the MLE of the Poisson parameters $\theta_1$ and $\theta_2$ in treatment and control groups ($15$ and $18.47$ respectively) 
turns out to be drastically different from the MLEs without them ($8.82$ and $9.65$ respectively). 
However the robust MDPDEs with larger $\beta$ remains stable (see Table \ref{TAB:Adverse_event_MDPDE}). 
Clearly, the number of average adverse effect decreases from control to treatment group; 
but to check how significant this change is, one might be interested in testing the one-side hypothesis 
\begin{equation}
H_0 : \theta_2 = \theta_1 ~~~ \mbox{against} ~~~~ 
H_1 : \theta_2 > \theta_1.
\label{EQ:Hyp_CT}
\end{equation} 
We have applied our proposed Wald-type tests for this problem, as developed in Section \ref{SEC:One-sided_problem},
to both the full dataset and after deleting the first two outliers from both the groups; 
the resulting p-values are presented in Figure \ref{FIG::Adverse_event_PV}.
Clearly, the classical Wald test results in completely different inference due to the inclusion of these outlying observations
-- it's p-value becomes significant from non-significant inference without them (at 95\% level).
On the other hand,  proposed  MDPDE based robust Wald-type tests with $\beta> 0$ gives stable results 
(accept the null hypothesis) even in presence of outlying observations. 

\begin{table}[!th] 
	\caption{MDPDEs of Poisson parameter $\theta$ for the Adverse Events Data in Table \ref{TAB:Adverse_event_data}}
	\centering
\begin{tabular}{|l|l|ccccccc|} \hline 
& & \multicolumn{7}{c|}{$\beta$} \\ 	
& 	Group & 0	&	0.1	&	0.3	&	0.5	&	0.7	&	0.9	&	1\\\hline
With &  Treatment	& 15.00	&	7.25	&	6.94	&	6.35	&	5.86	&	6.05	&	5.70\\
Outlier & Control	& 18.47	&	8.25	&	7.75	&	7.56	&	7.53	&	7.41	&	7.81\\\hline
Without & Treatment	& 8.82	&	7.47	&	6.44	&	6.20	&	6.14	&	5.58	&	6.58 \\
Outlier& Control	& 9.65	&	7.97	&	7.63	&	7.61	&	7.56	&	7.68	&	7.75 \\\hline
	\end{tabular}
	\label{TAB:Adverse_event_MDPDE}
\end{table}

\subsection{Poisson Model for Experimental Trial: Drosophila Data}
\label{SEC:7_2SDT_Poisson_mean}

We next consider another application to the Poisson model with data 
from an controlled experimental trial with Drosophila flies producing occasional spurious counts.
The dataset contains two independent samples on the numbers of recessive lethal mutations 
observed among the daughters of male flies who are exposed either 
to a certain degree of chemical to be screened (treatment group) or to control conditions. 
This dataset has been previously analyzed by many statisticians 
including \cite{Woodruff/etc:1984, Simpson:1989, Basu/etc:2013}
who have shown that the response data can be modeled by Poisson distribution,
but there are two outlying observations in one sample 
that affects the likelihood based inference and so the classical Wald test. 
See \citet[][Table 7]{Basu/etc:2013} for the dataset and 
the MDPDEs of the Poisson parameters.

\begin{figure}[!th]
	\centering
	\subfloat[Adverse Events Data]{
		\includegraphics[width=0.35\textwidth]{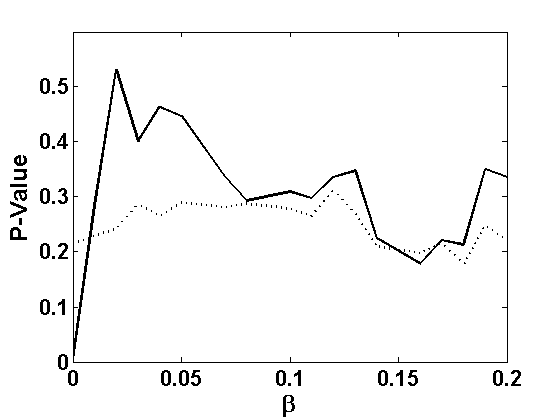}
		\label{FIG::Adverse_event_PV}} ~
	\subfloat[Droshophila Data]{
		\includegraphics[width=0.35\textwidth]{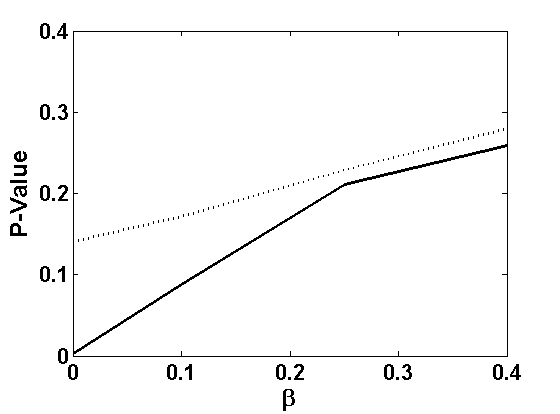}
		\label{FIG:Drosophila Data_PV}} \\
	\subfloat[Infant Platelet Count Data]{
		\includegraphics[width=0.35\textwidth]{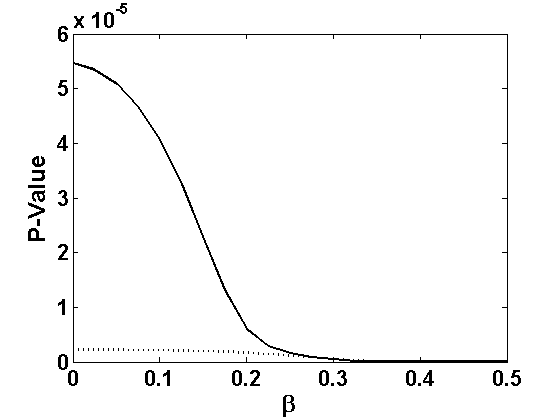}
		\label{FIG:Infant_Platelet_Data}} ~
	\subfloat[Hair Zn Content Data]{
		\includegraphics[width=0.35\textwidth]{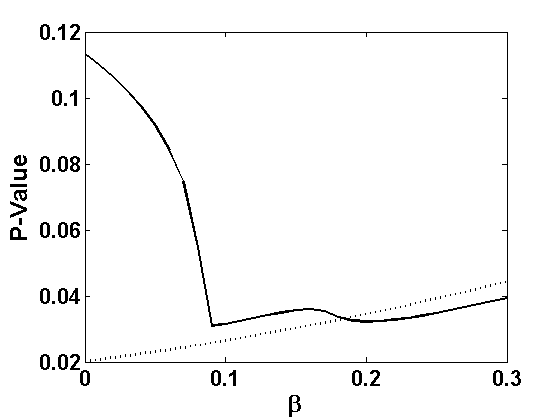}
		\label{FIG:Zn_content_PV}} \\
	\subfloat[Cloth Manufacturing data]{
		\includegraphics[width=0.35\textwidth]{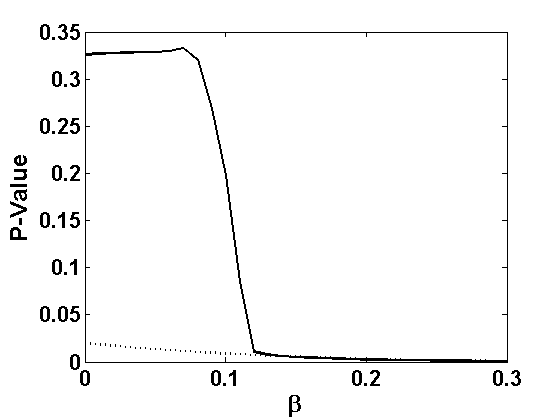}
		\label{FIG:Cloth_manuf_PV}} ~
	\subfloat[Components Life-time Data]{
		\includegraphics[width=0.35\textwidth]{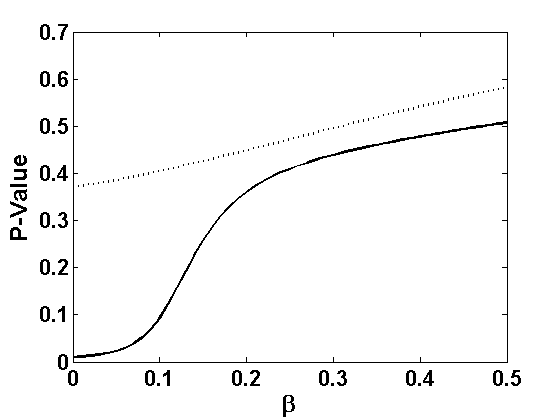}
		\label{FIG:Life-time_Data}}
	\caption{P-values of the proposed Wald-type tests under the real data examples with outliers (solid line) 
		and without outliers (doted line)}
	\label{FIG:real_data}%
\end{figure}

Here, we will apply the proposed Wald-type tests for comparing the Poisson parameters for the two samples,
say $\theta_1$ and $\theta_2$, through testing the one-sided hypothesis in \ref{EQ:Hyp_CT}.
The resulting p-values are presented in Figure \ref{FIG:Drosophila Data_PV}.
Clearly, in presence of outliers, the classical rejects the null hypothesis indicating that 
the average number of mutation is significantly more for the second sample, 
which is the opposite of the true inference obtained after removing these outliers from the second sample. 
But, the proposed MDPDE based Wald-type tests with $\beta\geq 0.1$ produce robust results even in presence of outliers 
accepting the null hypothesis.

\subsection{Normal Model for Clinical Trial: Infant Platelet Count Data}

We will now present another clinical trial example from \cite{Karpatkin/etc:1981} 
to illustrate the applications under the normal model.
This clinical trial was conducted to study if the infant platelet count can be increased 
by giving steroids to the mothers with autoimmune thrombocytopenia during pregnancy. 
The study consists of 19 mothers with 12 being given steroid (treatment group)
and 7 not given steroid (control group) and 
the corresponding infant platelet counts (in thousands, per mm$^3$) after delivery are given in Table \ref{TAB:Infant_Platelet_Data}.
These can be modeled by a normal model with means $\theta_1$, $\theta_2$ and the variances $\sigma_1^2$, $\sigma_2^2$
for the treatment and control groups respectively. Then, the primary research problem can be solved  
by testing the one-sided hypothesis in (\ref{EQ:Hyp_CT}) with $\sigma_1^2$ and $\sigma_2^2$ being unknown.

\begin{table}[!th] 
	\caption{Infant Platelet count after delivery (in thousands, per mm$^3$) in the  \cite{Karpatkin/etc:1981} clinical trail study}
	\centering
	\begin{tabular}{|l|cccccccccccc|} \hline 
		Treatment	&	120	& 124	& 215	& 90	& 67	& 126	 & 95	& 190	& 180	& 135	& 399	& 65 \\\hline
		Control		&	12	& 20	& 112	& 32	& 60	& 40 & 18	&  	& 	& 	& 	& 	\\\hline
	\end{tabular}
	\label{TAB:Infant_Platelet_Data}
\end{table}

The p-values for this testing problem obtained by applying the proposed Wald-type tests,
as described in Example \ref{EXM:4.1}, are presented in Figure \ref{FIG:Infant_Platelet_Data} for different $\beta\geq 0$. 
One can easily observe that there is a large outlier value of 399 (thousands) in the treatment group 
that affects the classical Wald test (at $\beta=0$). However, our MDPDE based proposal with $\beta>0$ 
produces stable p-value ignoring the effect of the outlying observation.

\subsection{Normal Model for Health Study: Hair Zn Content data}

Two-sample test under the normal model has many possible applications
from which we now present a health study to examine the impact of polluted urban environment 
over individual health in Sri Lanka. The dataset consist of the zinc (Zn) content of the hair of 
two independent samples taken from urban (polluted) and rural (unpolluted) Sri Lanka
and our target is to check if the Zn content is more for polluted urban residents impacting their health conditions. 
The dataset was presented in \citet[][Table 6]{Basu/etc:2015b} and it has been shown their that 
each sample can be modeled by normal distributions with means $\theta_i$ and variance $\sigma_i^2$ 
($i=1,2$ for rural and urban groups respectively) except for two possible outliers. 
There is one outlier in each of the samples that affects the MLE based inference
while testing for the targeted hypothesis (\ref{EQ:Hyp_CT}) of comparing 
$\theta_1$ and $\theta_2$ with unknown $\sigma_1^2$ and $\sigma_2^2$.

We have applied the proposed MDPDE based Wald-type test for this problem following Example \ref{EXM:4.1}
and the resulting p-values are presented in Figure \ref{FIG:Zn_content_PV}.
Clearly, the significance increase of the zinc contents in urban residents cannot be identified 
by the classical Wald-test in presence of outliers, but our proposal with $\beta\geq 0.1$ 
gives  stable and correct inference ignoring the effect of the outliers.

\subsection{Normal Model for Quality Control: Cloth Manufacturing data}

Our third and final example with normal model will be in the context of quality control 
based on the data from the Levi-Strauss clothing manufacturing plant. 
The dataset consists of 22 measurements on run-up (a percentage measure of wastage in cloth)
for each of two particular mills supplying cloths to the plant \cite[][Table 1]{Basu/etc:2015b}.
To control the quality of the cloths, the plant want to test for the consistency of the run-up measures 
from the two mills. Since the sample from each mill can be modeled by normal distribution with mean $\theta_i$ 
and variance $\sigma_i^2$ ($i=1,2$), the objective is then to test for the both sided hypothesis 
\begin{equation}
H_0 : \theta_1 = \theta_2 ~~~ \mbox{against} ~~~~ 
H_1 : \theta_1 \neq \theta_2,
\label{EQ:Hyp_CT2}
\end{equation} 
with $\sigma_1^2$ and $\sigma_2^2$ being unknown under both cases. 
However, as illustrated in \cite{Basu/etc:2015b}, the dataset contains 3 potential outliers 
that make the MLE based inference highly non-robust.
Hence the classical Wald test rejects the null hypothesis in presence of outliers 
whereas it accept the null after removing the outliers.
When we apply the proposed MDPDE based Wald-type problem, following the description as in Example \ref{EXM:3.1},
the corresponding p-values (reported in Figure \ref{FIG:Cloth_manuf_PV}) becomes highly stable for $\beta\geq 1.5$
rejecting the null hypothesis even in presence of the outliers.

\subsection{Exponential Model for Reliability Testing: Components Life-time Data}

We will end this section with an example of exponential model used in reliability testing 
between two sets of products' lifetimes. We will use the (simulated) data from \cite{Perng:1978} 
which consist of the lifetimes (in thousand of hours) of a particular electronic components  
produced by two different processes (see Table \ref{TAB:lifetime_data}).  
Each sample can be then modeled by exponential distributions with mean $\theta_i$ ($i=1,2$).
Our objective in reliability testing of the manufacturing process is to test whether the lifetimes for 
both the process have the same distributions, i.e., if $\theta_1=\theta_2$ against the both-sided alternatives
as in the hypothesis (\ref{EQ:Hyp_CT2}). It has been observed that there is no significant difference in 
the distributions of both the processes and so the null hypothesis should be accepted by any standard test.

\begin{table}[!th] 
	\caption{Lifetimes (in thousand of hours) of a particular electronic components produced by two different processes \citep{Perng:1978}}
	\centering
	\begin{tabular}{|l|cccccccccccc|} \hline 
		Process 1	&	.044 & .134	& .142	& .158	& .216	& .625	 & .649	& .658	& 1.062	& 1.140	& 1.159	& 1.238\\\hline
		Process 2	&	.060 & .174	& .237	& .272	& .335	& .391   & .670	& .902 	& 1.543	& 1.615	& 2.013	& 2.309	\\\hline
	\end{tabular}
	\label{TAB:lifetime_data}
\end{table}
  
Since there is no outliers in this dataset, in order to study the robustness aspect of our proposal
we add one outlying value of 20 (assuming a decimal is misplaced by one digit from 2.0) in the second sample. 
The resulting p-values obtained by the proposed Wald-type tests for both 
the pure data and with this artificial outlier are presented in Figure \ref{FIG:Life-time_Data} for different $\beta$.
Clearly, the classical Wald test changes drastically by rejecting null due to insertion of only one outlying observations,
but our proposed Wald-type tests with $\beta\geq0.1$ remains stable and still accept the null hypothesis robustly in presence of the outlier.

\section{Simulation Study and the Choice of Tunning Parameter $\beta$}
\label{SEC:numerical}

Finally to examine the finite sample performances of our Wald-type tests, 
we have performed several simulation studies with all the models considered in the previous section for real datasets.
However, noting the similarity of the results for different models, for brevity,
here we will report the results from only one simulation study under normal model with two-sided alternatives.

\begin{figure}[h]
	\centering
	\subfloat[Sizes, known variances]{
		\includegraphics[width=0.35\textwidth]{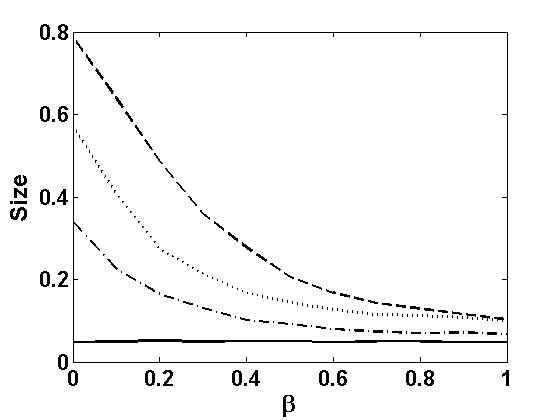}
		\label{FIG:SizeE_knownS}} ~
	\subfloat[Sizes, unknown variances]{
		\includegraphics[width=0.35\textwidth]{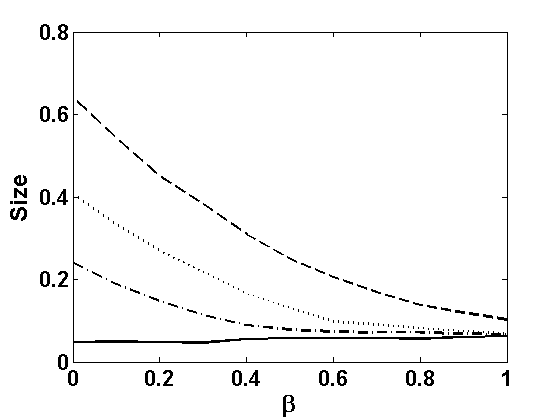}
		\label{FIG:SizeE_unknownS}} \\
	\subfloat[Powers, known variances]{
		\includegraphics[width=0.35\textwidth]{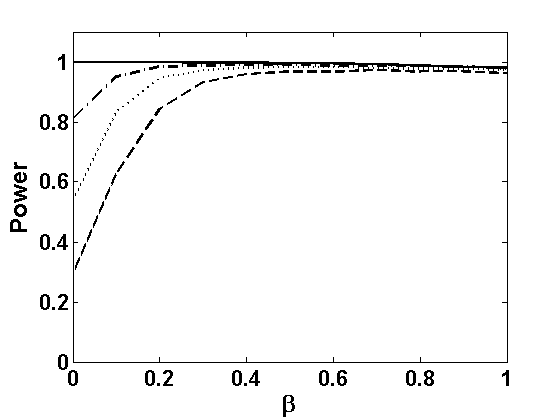}
		\label{FIG:PowerE_knownS}} ~
	\subfloat[Powers, unknown variances]{
		\includegraphics[width=0.35\textwidth]{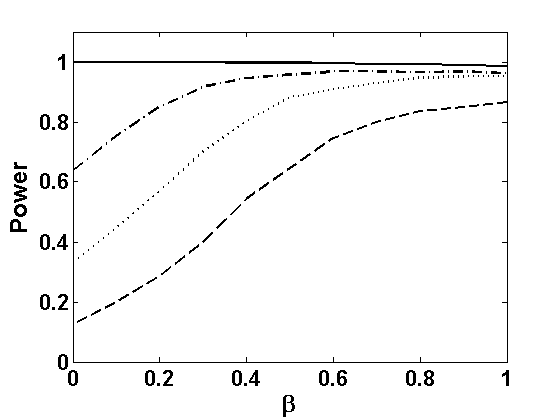}
		\label{FIG:PowerE_unknownS}}
	\caption{Empirical sizes and powers of the proposed Wald-type tests for testing equality of two normal means
		with both the known and unknown variance case at sample size $n=50$ under pure data (solid line) 
		and with contamination of 10\% (dash-doted line), 15\% (doted line)
		and 20\% (dashed line)}
	\label{FIG:Simlation_norm}%
\end{figure}

We simulate 1000 pair of samples, each of size $n=50$, independently drawn from $N(\theta_i,1)$ distributions ($i=1,2$) 
and perform the proposed Wald-type tests for testing $H_0: \theta_1=\theta_2$ against the two-sided alternative 
$H_1: \theta_1 \neq \theta_2$, once assuming both variances to be known (equal 1) and 
then assuming variances to be unknown and unequal following Examples \ref{EXM:1} and \ref{EXM:3.1} respectively.
Then, we compute the empirical sizes and powers of the proposed test under these pure data over $1000$ iterations,
where for size calculation we have taken $\theta_1=\theta_2=0$ and for power calculation $\theta_1=0$, $\theta_2=1$.
Next, to study the robustness performances, we contaminate $100\varepsilon\%$ of second sample in each iteration 
(for $\varepsilon=0.1, 0.15, 0.2$) by observations from $N(\theta_c,1)$ distributions 
and repeat the above simulation to compute empirical sizes and power under contamination.
We have taken $\theta_c=3$ and $-3$ for studying the robustness of size and power respectively.
Note that these contamination distributions are not very far from the corresponding true distributions 
and hence generate reasonably common practical situations. 
Resulting empirical sizes and powers are reported in Figure \ref{FIG:Simlation_norm}.

It can be easily observed from Figure \ref{FIG:Simlation_norm} that
the size and power of the proposed Wald-type tests under pure data change (increases and decreases respectively) 
only very slightly with increasing $\beta$, but their stability increases significantly.
In particular, under contamination, both size and power of the tests near $\beta=0$, 
the classical Wald test, changes drastically. But they become stable at larger positive values of $\beta$ 
for both the cases of known and unknown  variances. 
However, for the cases of known (and correctly specified) variances we get highly stable results near $\beta \approx 0.3, 0.4$,
whereas we need $\beta \approx 0.5, 0.6$ for the case of unknown variances.
This is intuitively expected since under the present contamination schemes the variance estimates also changes
and so we need more robustness power to get overall stable inference with larger values of $\beta$.

Throughout all our example and simulations above, we have notices that the tuning parameter $\beta$ 
controls between robustness of the proposed Wald-type tests and its asymptotic contiguous power under pure data.
So, we need to chose $\beta$ properly for any practical applications.
In particular we note that, in most of the example models, the loss in power is not significant enough at small positive $\beta$,
whereas we get highly robust inferences for $\beta \geq 0.3$ 
(except for few cases with very high contaminations where we may need $\beta \approx 0.4, 0.5$).
Therefore, an empirical suggestion for the choice of $\beta$ in any application suspecting some contamination 
could be within the range $\beta\in[0.3, 0.5]$ for generating robust inference without significant loss in power.

Although this ad hoc empirical choice of $\beta$ works well enough in most practical datasets suspectable to outliers, 
many practitioners will prefer a data-driven choice of $\beta$ in case of no idea on the level of contamination in dataset
that might produce a better trade-off. In this respect, we note that the performance of the proposed Wald-type tests directly
depends on that of the MDPDE (with tuning parameter $\beta$) used in constructing the test statistics.
In particular the asymptotic contiguous power of the proposed test has the same nature as the asymptotic efficiency
of the corresponding MDPDE whereas all the robustness measures of our tests directly depend on the robustness of the MDPDE 
through its influence function. So, a suitable data-driven choice of $\beta$ for our Wald-type test statistics also can be 
equivalently formed by adjusting the trade-off between efficiency and robustness of the MDPDE used. 
For this second problem, \cite{Warwick/Jones:2005} proposed to minimize an estimator of MSE of the MDPDE to chose optimum $\beta$.
Based on the first sample $X_1, \ldots, X_n$, they proposed to minimize the estimated MSE
\begin{eqnarray}
\widehat{MSE}_n(\beta) = \left( {^{(1)}}\widehat{{\boldsymbol{\theta}}}_{\beta} - {\boldsymbol{\theta}}_{\beta}^P\right)^T
\left(^{(1)}\widehat{{\boldsymbol{\theta}}}_{\beta} - {\boldsymbol{\theta}}_{\beta}^P\right)
+ \frac{1}{n} Trace\left(\widehat{\boldsymbol{J}}_{\beta,n}^{-1}\widehat{\boldsymbol{K}}_{\beta,n}
\widehat{\boldsymbol{J}}_{\beta,n}^{-1}\right)
\label{EQ:Est_MSE}
\end{eqnarray}
over $\beta$, where ${\boldsymbol{\theta}}_{\beta}^P$ is a pilot estimator of the target parameter 
and $\widehat{\boldsymbol{J}}_{\beta,n}$ and $\widehat{\boldsymbol{K}}_{\beta,n}$
are estimators of the matrices $\boldsymbol{J}_{\beta }$ and $\boldsymbol{K}_{\beta }$ respectively,
which can be easily obtained from their expressions by substituting $\boldsymbol{\theta}$ by the MDPDE and 
integrations by sample means.
Although there is no direct choice for ${\boldsymbol{\theta}}_{\beta}^P$, \cite{Warwick/Jones:2005} suggested,
based on an extensive simulation studies, that the MDPDE with $\beta=1$ can serve the purpose well for the i.i.d.~set-up
and we will stick to that suggestion for the present case also
(the non-i.i.d.~cases have been studied in \cite{Ghosh/Basu:2013,Ghosh/Basu:2015}).
However, the problem in the present two-sample case is that, the optimum $\beta$ obtained by minimizing  $\widehat{MSE}_n(\beta)$
based on the first sample may not be the same as that obtained for the second sample due to possible different level of contaminations.  
As a standard solution, we propose the minimization of the total estimated MSE, 
the sum of the MSE estimates based on two samples separately, 
over $\beta\in[0,1]$ to obtain the optimum choice of the tuning parameter for the present two-sample testing problem.

\begin{figure}[h]
	\centering
	\subfloat[Known var., No contamination]{
		\includegraphics[width=0.3\textwidth]{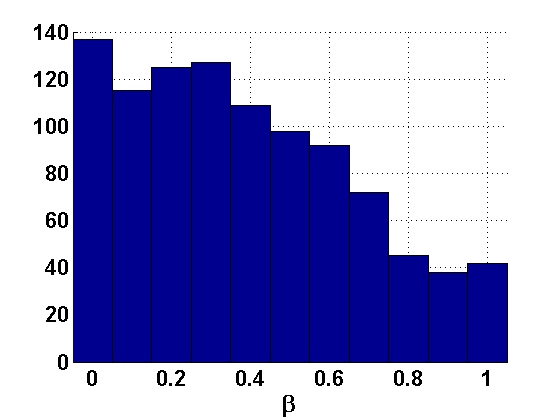}
		\label{FIG:Simlation_norm_OptA_00_knownS}} ~
	\subfloat[Known var., 10\% contamination]{
		\includegraphics[width=0.3\textwidth]{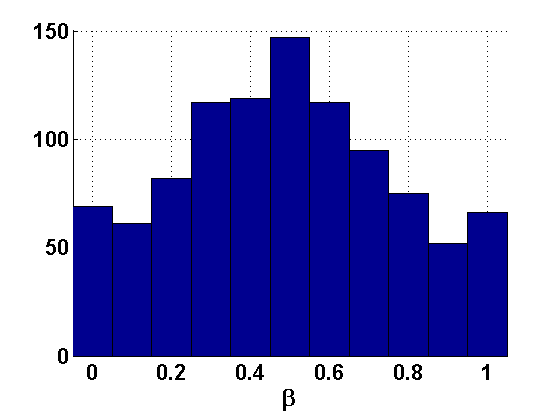}
		\label{FIG:Simlation_norm_OptA_10_knownS}} ~
	\subfloat[Known var., 20\% contamination]{
		\includegraphics[width=0.3\textwidth]{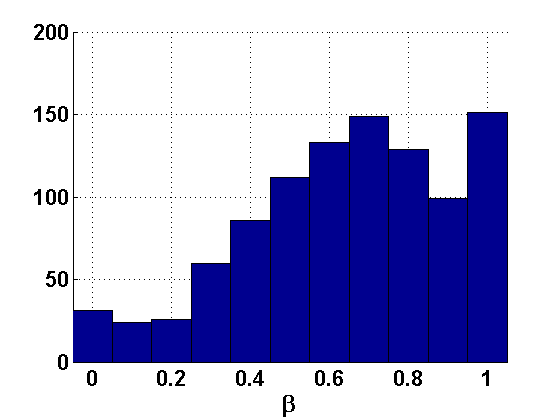}
		\label{FIG:Simlation_norm_OptA_20_knownS}}
	\caption{Histograms for optimally chosen tuning parameter $\beta$ under normal models with different contamination levels}
	\label{FIG:Simlation_norm_OptA}%
\end{figure}

We have implemented this proposal for the above simulation study with normal model to check its effectiveness. 
Figure \ref{FIG:Simlation_norm_OptA} presents the histograms of the 1000 selected optimum $\beta$ following this proposal 
for the normal model with known and equal variances under the simulation scheme 
used for studying size stability above (in Figure \ref{FIG:Simlation_norm}).
Clearly, the mode of these optimum $\beta$s shift from 0 to 1 as the contamination proportion increases
yielding the expected trade-off between the power and robustness based on the level of contaminations.


\section{Concluding remarks}
\label{SEC:conclusion}

In this paper, we have considered the problem of testing with two independent samples of i.i.d.~observations
and proposed a class of robust Wald-type tests for both simple and composite hypothesis testing. 
These Wald-type tests are constructed using the robust minimum density power divergence estimators 
of the underlying parameters in each sample.
The asymptotic and robustness properties of the proposed Wald-type tests have been discussed
along with their applications to several important real-life problems like clinical trial,
medical experiment, reliability testing and many more.

Although we have discussed all possible types of general two-sample hypotheses, 
in this paper, we have restricted our attention to the cases 
where each of the two independent samples is identically distributed. 
The natural extension  of this work will be to develop robust tests for hypotheses involving 
two independent samples from non-homogeneous populations; 
this also has many practical applications including comparing the regression lines between 
two groups of patients in a fixed design clinical trial. 
Also, one could further explore the possibility of robust hypothesis testing 
using the minimum density power divergence estimators for two paired samples 
or for more than two sample cases. we hope to pursue some of this possible extensions in our future research.

\appendix
\section{Proof of Results}
\label{APP:Proof}

\subsection{Proof of Theorem \ref{THM:2sample_simple_asympNull}}
 Using the asymptotic distribution of 
  $\sqrt{n}(^{(1)}\widehat{{\boldsymbol{\theta}}}_{\beta} - {\boldsymbol{\theta}}_1)$ and $\sqrt{n}(^{(2)}\widehat{{\boldsymbol{\theta}}}_{\beta} - {\boldsymbol{\theta}}_2)$, we have
  $$\sqrt{\frac{mn}{m+n}} \left( ^{(1)}\widehat{{\boldsymbol{\theta}}}_{\beta} - {\boldsymbol{\theta}}_1 \right) 
	  \displaystyle\underset{m, n\rightarrow \infty }{\overset{\mathcal{L}}{\longrightarrow }}  
	  N(\boldsymbol{0}_p, \omega {\boldsymbol{\Sigma}}_{\beta}({\boldsymbol{\theta}}_1))$$
  and
  $$\sqrt{\frac{mn}{m+n}} \left( ^{(2)}\widehat{{\boldsymbol{\theta}}}_{\beta} - {\boldsymbol{\theta}}_2 \right) 
  \displaystyle\underset{m, n\rightarrow \infty }{\overset{\mathcal{L}}{\longrightarrow }}  
  N(\boldsymbol{0}_p, (1-\omega) {\boldsymbol{\Sigma}}_{\beta}({\boldsymbol{\theta}}_2)).$$
  Hence under $H_0 : {\boldsymbol{\theta}}_1 = {\boldsymbol{\theta}}_2={\boldsymbol{\theta}}_0$, we get 
  $$
  \sqrt{\frac{mn}{m+n}} \left( ^{(1)}\widehat{{\boldsymbol{\theta}}}_{\beta} - ^{(1)}\widehat{{\boldsymbol{\theta}}}_{\beta} \right) 
  \underset{m, n\rightarrow \infty }{\overset{\mathcal{L}}{\longrightarrow }}  
  N(\boldsymbol{0}_p,  {\boldsymbol{\Sigma}}_{\beta}({\boldsymbol{\theta}}_0)).
  $$
  Further, under $H_0$, $^{(0)}\widehat{{\boldsymbol{\theta}}}_{\beta} ~\displaystyle{\overset{\mathcal{P}}{\rightarrow }} ~ {\boldsymbol{\theta}}_0$ as $m+n \rightarrow\infty$.
  Then the theorem follows using the continuity of the matrix ${\boldsymbol{\Sigma}}_{\beta}({\boldsymbol{\theta}})$.
  \hfill{$\square$}

\subsection{Proof of Theorem \ref{THM:7two_sample_power1}}
Note that, $^{(0)}\widehat{{\boldsymbol{\theta}} }_{\beta }\underset{n,m\rightarrow \infty }%
{\overset{\mathcal{P}}{\longrightarrow }}{\boldsymbol{\theta}} _{3}$ and hence the
asymptotic distribution of $l_{^{(0)}\widehat{{\boldsymbol{\theta}} }_{\beta },\beta
}^{\ast }(^{(1)}\widehat{{\boldsymbol{\theta}} }_{\beta },^{(2)}\widehat{{\boldsymbol{\theta}} }_{\beta })$
is the same as that of $l_{{\boldsymbol{\theta}} _{3},\beta }^{\ast }(^{(1)}\widehat{{\boldsymbol{\theta}} 
}_{\beta },^{(2)}\widehat{{\boldsymbol{\theta}} }_{\beta })$. Now, a suitable Taylor series
expansion leads to 
\begin{eqnarray}
 l_{{\boldsymbol{\theta}} _{3},\beta }^{\ast }(^{(1)}\widehat{{\boldsymbol{\theta}} }_{\beta },^{(2)}%
\widehat{{\boldsymbol{\theta}} }_{\beta })-l_{{\boldsymbol{\theta}} _{3},\beta }^{\ast }({\boldsymbol{\theta}}
_{1},{\boldsymbol{\theta}} _{2})  &=&\left( ^{(1)}\widehat{{\boldsymbol{\theta}} }_{\beta }-{\boldsymbol{\theta}}
_{1}\right) ^{T}\frac{\partial }{\partial {\boldsymbol{\theta}} _{1}}l_{{\boldsymbol{\theta}} _{3},\beta
}^{\ast }({\boldsymbol{\theta}} _{1},{\boldsymbol{\theta}} _{2})+\left( ^{(2)}\widehat{{\boldsymbol{\theta}} }_{\beta
}-{\boldsymbol{\theta}} _{2}\right) ^{T}\frac{\partial }{\partial {\boldsymbol{\theta}} _{2}}l_{{\boldsymbol{\theta}}
_{3},\beta }^{\ast }({\boldsymbol{\theta}} _{1},{\boldsymbol{\theta}} _{2})  \nonumber \\
&&+o_{P}\left( ||^{(1)}\widehat{{\boldsymbol{\theta}} }_{\beta }-{\boldsymbol{\theta}} _{1}||^{2}\right)
+o_{P}\left( ||^{(2)}\widehat{{\boldsymbol{\theta}} }_{\beta }-{\boldsymbol{\theta}} _{2}||^{2}\right)  
\nonumber \\
&=&2\left( ^{(1)}\widehat{{\boldsymbol{\theta}} }_{\beta }-{\boldsymbol{\theta}} _{1}\right) ^{T}{\boldsymbol{\Sigma}}
_{\beta }({\boldsymbol{\theta}} _{3})^{-1}({\boldsymbol{\theta}} _{1}-{\boldsymbol{\theta}} _{2})-2\left( ^{(2)}\widehat{%
	{\boldsymbol{\theta}} }_{\beta }-{\boldsymbol{\theta}} _{2}\right) ^{T}{\boldsymbol{\Sigma}} _{\beta }({\boldsymbol{\theta}}
_{3})^{-1}({\boldsymbol{\theta}} _{1}-{\boldsymbol{\theta}} _{2})  \nonumber \\
&&+o_{P}\left( ||^{(1)}\widehat{{\boldsymbol{\theta}} }_{\beta }-{\boldsymbol{\theta}} _{1}||^{2}\right)
+o_{P}\left( ||^{(2)}\widehat{{\boldsymbol{\theta}} }_{\beta }-{\boldsymbol{\theta}} _{2}||^{2}\right)  
\nonumber \\
&=&2\left[ \left( ^{(1)}\widehat{{\boldsymbol{\theta}} }_{\beta }-^{(2)}\widehat{{\boldsymbol{\theta}} }%
_{\beta }\right) -\left( {\boldsymbol{\theta}} _{1}-{\boldsymbol{\theta}} _{2}\right) \right] ^{T}{\boldsymbol{\Sigma}}
_{\beta }({\boldsymbol{\theta}} _{3})^{-1}({\boldsymbol{\theta}} _{1}-{\boldsymbol{\theta}} _{2})  \nonumber \\
&&+o_{P}\left( ||^{(1)}\widehat{{\boldsymbol{\theta}} }_{\beta }-{\boldsymbol{\theta}} _{1}||^{2}\right)
+o_{P}\left( ||^{(2)}\widehat{{\boldsymbol{\theta}} }_{\beta }-{\boldsymbol{\theta}} _{2}||^{2}\right) . 
\nonumber
\end{eqnarray}%
Then, the theorem follows from the above expression by noting that 
\[
\sqrt{\frac{mn}{m+n}}\left[ \left( ^{(1)}\widehat{{\boldsymbol{\theta}} }_{\beta }-^{(2)}%
\widehat{{\boldsymbol{\theta}} }_{\beta }\right) -\left( {\boldsymbol{\theta}} _{1}-{\boldsymbol{\theta}} _{2}\right) %
\right] \underset{n,m\rightarrow \infty }{\overset{\mathcal{L}}{%
		\longrightarrow }}\mathcal{N}\left( 0,\left[ \omega {\boldsymbol{\Sigma}} _{\beta }({\boldsymbol{\theta}}
_{1})+(1-\omega ){\boldsymbol{\Sigma}} _{\beta }({\boldsymbol{\theta}} _{2})\right] \right) ,
\]%
as $m,n\rightarrow \infty $ at any ${\boldsymbol{\theta}} _{1}\neq {\boldsymbol{\theta}} _{2}$. Here, the
last convergence follows from the asymptotic distributions of the MDPDEs $%
^{(1)}\widehat{{\boldsymbol{\theta}} }_{\beta }$ and $^{(2)}\widehat{{\boldsymbol{\theta}} }_{\beta }$.
\hfill{$\square$}

\subsection{Proof of Theorem \ref{THM:2sample_simple_asympCont1}}
  Using the asymptotic distribution of 
  $\sqrt{n}(^{(1)}\widehat{{\boldsymbol{\theta}}}_{\beta} - {\boldsymbol{\theta}}_{1,n})$ and 
  $\sqrt{n}(^{(2)}\widehat{{\boldsymbol{\theta}}}_{\beta} - {\boldsymbol{\theta}}_{2,m})$ under $H_{1,n,m}$
  and continuity of ${\boldsymbol{\Sigma}}_{\beta}({\boldsymbol{\theta}}_0)$, we have 
  $^{(2)}\widehat{{\boldsymbol{\theta}}}_{\beta} \underset{m\rightarrow \infty }{\overset{\mathcal{P}}{\longrightarrow }} {\boldsymbol{\theta}}_0$,
  $$
  \sqrt{\frac{mn}{m+n}} \left( ^{(1)}\widehat{{\boldsymbol{\theta}}}_{\beta} - {\boldsymbol{\theta}}_0 \right) 
  \underset{m, n\rightarrow \infty }{\overset{\mathcal{L}}{\longrightarrow }}
   N(\sqrt{\omega}{\boldsymbol{\Delta}}_1, \omega {\boldsymbol{\Sigma}}_{\beta}({\boldsymbol{\theta}}_0))
  $$
  and
  $$
  \sqrt{\frac{mn}{m+n}} \left( ^{(2)}\widehat{{\boldsymbol{\theta}}}_{\beta} - {\boldsymbol{\theta}}_0 \right) 
  \underset{m, n\rightarrow \infty }{\overset{\mathcal{L}}{\longrightarrow }} 
  N(\sqrt{1-\omega}{\boldsymbol{\Delta}}_2, (1-\omega) {\boldsymbol{\Sigma}}_{\beta}({\boldsymbol{\theta}}_0)).
  $$
  Hence,  under $H_{1,n,m}$, we get 
  $$
  \sqrt{\frac{mn}{m+n}} \left( ^{(1)}\widehat{{\boldsymbol{\theta}}}_{\beta} - ^{(1)}\widehat{{\boldsymbol{\theta}}}_{\beta} \right) 
  \underset{m, n\rightarrow \infty }{\overset{\mathcal{L}}{\longrightarrow }} 
  N(\sqrt{\omega}{\boldsymbol{\Delta}}_1 - \sqrt{1-\omega}{\boldsymbol{\Delta}}_2,  {\boldsymbol{\Sigma}}_{\beta}({\boldsymbol{\theta}}_0)),
  $$
from which the theorem follows immediately.
  \hfill{$\square$}

\subsection{Proof of Theorem \ref{THM:2sample_simple_asympCont}}
We will only prove the case $(D_1, D_2)=(F_{1,m,\varepsilon,x}^P,F_{2,n,\varepsilon,y}^P)$. Other two cases will follow similarly. 

Let us denote ${\boldsymbol{\theta}}_{1,n}^\ast = \boldsymbol{U}_{\beta}(F_{1,m,\varepsilon,x}^P)$ and 
 ${\boldsymbol{\theta}}_{2,m}^\ast = \boldsymbol{U}_{\beta}(F_{2,n,\varepsilon,y}^P)$.
 Then using the  continuity of ${\boldsymbol{\Sigma}}_{\beta}({\boldsymbol{\theta}}_0)$, we get under 
$(D_1, D_2)=(F_{1,m,\varepsilon,x}^P,F_{2,n,\varepsilon,y}^P)$,
the asymptotic distribution of 
  $\sqrt{n}(^{(1)}\widehat{{\boldsymbol{\theta}}}_{\beta} - {\boldsymbol{\theta}}_{1,n}^\ast)$ and 
  $\sqrt{n}(^{(2)}\widehat{{\boldsymbol{\theta}}}_{\beta} - {\boldsymbol{\theta}}_{2,m}^\ast)$  
 are both $p$-variate normal with mean zero and variance ${\boldsymbol{\Sigma}}_{\beta}({\boldsymbol{\theta}}_0)$. 
Further, suitable Taylor series expansion yields
 \begin{eqnarray}
 {\boldsymbol{\theta}}_{1,n}^\ast &=& {\boldsymbol{\theta}}_{1,n}+ \frac{\varepsilon}{\sqrt{n}}
 \mathcal{IF}(x; \boldsymbol{U}_{\beta}, F_{{\boldsymbol{\theta}}_{1,n}}) + o(n^{-1/2})\nonumber\\
 &=& {\boldsymbol{\theta}}_0 + \frac{{\boldsymbol{\Delta}}_1}{\sqrt{n}}+ \frac{\varepsilon}{\sqrt{n}}
 \mathcal{IF}(x; \boldsymbol{U}_{\beta}, F_{{\boldsymbol{\theta}}_{1,n}}) + o(n^{-1/2})\nonumber\\
 &=& {\boldsymbol{\theta}}_0 + \frac{\widetilde{{\boldsymbol{\Delta}}_1}}{\sqrt{n}}+ o(n^{-1/2}).\nonumber
 \end{eqnarray}
 Similarly, we have 
 $$
 {\boldsymbol{\theta}}_{2,m}^\ast = {\boldsymbol{\theta}}_0 + \frac{\widetilde{{\boldsymbol{\Delta}}_2}}{\sqrt{n}}+ o(n^{-1/2}).
 $$
 Combining all these, we get
 $$
  \sqrt{\frac{mn}{m+n}} \left( ^{(1)}\widehat{{\boldsymbol{\theta}}}_{\beta} - {\boldsymbol{\theta}}_0 \right) 
  \displaystyle\underset{m, n\rightarrow \infty }{\overset{\mathcal{L}}{\longrightarrow }}  N(\sqrt{\omega}\widetilde{{\boldsymbol{\Delta}}}_1, \omega {\boldsymbol{\Sigma}}_{\beta}({\boldsymbol{\theta}}_0))
  $$
  and
  $$
  \sqrt{\frac{mn}{m+n}} \left( ^{(2)}\widehat{{\boldsymbol{\theta}}}_{\beta} - {\boldsymbol{\theta}}_0 \right) 
  \displaystyle\underset{m, n\rightarrow \infty }{\overset{\mathcal{L}}{\longrightarrow }}  N(\sqrt{1-\omega}\widetilde{{\boldsymbol{\Delta}}}_2, (1-\omega) {\boldsymbol{\Sigma}}_{\beta}({\boldsymbol{\theta}}_0)).
  $$
Hence,  under $(D_1, D_2)=(F_{1,m,\varepsilon,x}^P,F_{2,n,\varepsilon,y}^P)$, we get 
$$
\sqrt{\frac{mn}{m+n}} \left( ^{(1)}\widehat{{\boldsymbol{\theta}}}_{\beta} - ^{(1)}\widehat{{\boldsymbol{\theta}}}_{\beta} \right) 
  \underset{m, n\rightarrow \infty }{\overset{\mathcal{L}}{\longrightarrow }}  N(\sqrt{\omega}\widetilde{{\boldsymbol{\Delta}}}_1 - \sqrt{1-\omega}\widetilde{{\boldsymbol{\Delta}}}_2,  {\boldsymbol{\Sigma}}_{\beta}({\boldsymbol{\theta}}_0))
 $$
 and hence the theorem follows immediately.
  \hfill{$\square$}

\subsection{Proof of Theorem \ref{THM:2sample_gen_asympNull}}
Using suitable Taylor series expansion, we get 
\begin{eqnarray}
 \boldsymbol{\psi}(^{(1)}\widehat{{\boldsymbol{\theta}}}_{\beta}, ^{(2)}\widehat{{\boldsymbol{\theta}}}_{\beta})
 &=& \boldsymbol{\psi}({\boldsymbol{\theta}}_1,{\boldsymbol{\theta}}_2) 
+\boldsymbol{\Psi}_1({\boldsymbol{\theta}}_1,{\boldsymbol{\theta}}_2)^T(^{(1)}\widehat{{\boldsymbol{\theta}}}_{\beta}-{\boldsymbol{\theta}}_1)
+\boldsymbol{\Psi}_2({\boldsymbol{\theta}}_1,{\boldsymbol{\theta}}_2)^T(^{(2)}\widehat{{\boldsymbol{\theta}}}_{\beta}-{\boldsymbol{\theta}}_2)
\nonumber\\
&& ~~~~~~~~  
+  o_P\left(||^{(1)}\widehat{{\boldsymbol{\theta}}}_{\beta}-{\boldsymbol{\theta}}_1||\right)
   +  o_P\left(||^{(2)}\widehat{{\boldsymbol{\theta}}}_{\beta}-{\boldsymbol{\theta}}_2||\right).
 \end{eqnarray}
Now, from the asymptotic distribution of 
  $\sqrt{n}(^{(1)}\widehat{{\boldsymbol{\theta}}}_{\beta} - {\boldsymbol{\theta}}_1)$ and $\sqrt{n}(^{(2)}\widehat{{\boldsymbol{\theta}}}_{\beta} - {\boldsymbol{\theta}}_2)$ it follows that 
  $$
  \sqrt{\frac{mn}{m+n}} \boldsymbol{\Psi}_1({\boldsymbol{\theta}}_1,{\boldsymbol{\theta}}_2)^T 
  \left( ^{(1)}\widehat{{\boldsymbol{\theta}}}_{\beta} - {\boldsymbol{\theta}}_1 \right) 
  \displaystyle\underset{m, n\rightarrow \infty }{\overset{\mathcal{L}}{\longrightarrow }}  N(0, \omega \boldsymbol{\Psi}_1({\boldsymbol{\theta}}_1,{\boldsymbol{\theta}}_2)^T{\boldsymbol{\Sigma}}_{\beta}({\boldsymbol{\theta}}_1)
  \boldsymbol{\Psi}_1({\boldsymbol{\theta}}_1,{\boldsymbol{\theta}}_2))
  $$
  and
  $$
  \sqrt{\frac{mn}{m+n}} \boldsymbol{\Psi}_2({\boldsymbol{\theta}}_1,{\boldsymbol{\theta}}_2)^T\left( ^{(2)}\widehat{{\boldsymbol{\theta}}}_{\beta} - {\boldsymbol{\theta}}_2 \right) 
  \displaystyle\underset{m, n\rightarrow \infty }{\overset{\mathcal{L}}{\longrightarrow }}  N(0, (1-\omega) \boldsymbol{\Psi}_2({\boldsymbol{\theta}}_1,{\boldsymbol{\theta}}_2)^T{\boldsymbol{\Sigma}}_{\beta}({\boldsymbol{\theta}}_2)
  \boldsymbol{\Psi}_2({\boldsymbol{\theta}}_1,{\boldsymbol{\theta}}_2)).
  $$
  Hence under $H_0 : \boldsymbol{\psi}({\boldsymbol{\theta}}_1, {\boldsymbol{\theta}}_2)=\boldsymbol{0}_r$, we get 
  $$
  \sqrt{\frac{mn}{m+n}} \boldsymbol{\psi}\left( ^{(1)}\widehat{{\boldsymbol{\theta}}}_{\beta}, 
  ^{(2)}\widehat{{\boldsymbol{\theta}}}_{\beta} \right) 
  \underset{m, n\rightarrow \infty }{\overset{\mathcal{L}}{\longrightarrow }}  
  N(\boldsymbol{0}_r,  \widetilde{{\boldsymbol{\Sigma}}_{\beta}}({\boldsymbol{\theta}}_1, {\boldsymbol{\theta}}_2)).
  $$
  Finally, by the consistency of the MDPDEs and the continuity of the matrices $\boldsymbol{\Psi}_1$, $\boldsymbol{\Psi}_2$ 
  and ${\boldsymbol{\Sigma}}_{\beta}$, it follows that $\widetilde{{\boldsymbol{\Sigma}}_{\beta}}(^{(1)}\widehat{{\boldsymbol{\theta}}}_{\beta} ^{(2)}\widehat{{\boldsymbol{\theta}}}_{\beta})
  ~\displaystyle{\overset{\mathcal{P}}{\rightarrow }} ~ 
  \widetilde{{\boldsymbol{\Sigma}}_{\beta}}({\boldsymbol{\theta}}_1, {\boldsymbol{\theta}}_2)$ as $m+n \rightarrow\infty$, from which the theorem follows immediately.
  \hfill{$\square$}

\subsection{Proof of Theorem \ref{THM:2sample_Gen_powerApprox}}
Using an appropriate Taylor series expansion, we get
\begin{eqnarray}
  && \widetilde{l^*}( ^{(1)}\widehat{{\boldsymbol{\theta}}}_{\beta},^{(2)}\widehat{{\boldsymbol{\theta}}}_{\beta})
  - \widetilde{l^*}({\boldsymbol{\theta}}_1,{\boldsymbol{\theta}}_2 =
  \left(^{(1)}\widehat{{\boldsymbol{\theta}}}_{\beta}-{\boldsymbol{\theta}}_1\right)^T\frac{\partial}{\partial{\boldsymbol{\theta}}_1}\widetilde{l^*}({\boldsymbol{\theta}}_1,{\boldsymbol{\theta}}_2)
  +\left(^{(2)}\widehat{{\boldsymbol{\theta}}}_{\beta}-{\boldsymbol{\theta}}_2\right)^T\frac{\partial}{\partial{\boldsymbol{\theta}}_2}\widetilde{l^*}({\boldsymbol{\theta}}_1,{\boldsymbol{\theta}}_2) \nonumber\\
&& ~~~~~~~~~~~~~~~~~~~~~~~~~~~~ 
+o_P\left(||^{(1)}\widehat{{\boldsymbol{\theta}}}_{\beta}-{\boldsymbol{\theta}}_1||^2\right)
+o_P\left(||^{(2)}\widehat{{\boldsymbol{\theta}}}_{\beta}-{\boldsymbol{\theta}}_2||^2\right)\nonumber\\
&=& 
  2\left(^{(1)}\widehat{{\boldsymbol{\theta}}}_{\beta}-{\boldsymbol{\theta}}_1\right)^T
\boldsymbol{\Psi}_1({\boldsymbol{\theta}}_1,{\boldsymbol{\theta}}_2)\widetilde{{\boldsymbol{\Sigma}}_{\beta}}({\boldsymbol{\theta}}_1, {\boldsymbol{\theta}}_2)^{-1}\boldsymbol{\psi}({\boldsymbol{\theta}}_1, {\boldsymbol{\theta}}_2)
+2\left(^{(2)}\widehat{{\boldsymbol{\theta}}}_{\beta}-{\boldsymbol{\theta}}_2\right)^T
\boldsymbol{\Psi}_2({\boldsymbol{\theta}}_1,{\boldsymbol{\theta}}_2)\widetilde{{\boldsymbol{\Sigma}}_{\beta}}({\boldsymbol{\theta}}_1, {\boldsymbol{\theta}}_2)^{-1}\boldsymbol{\psi}({\boldsymbol{\theta}}_1, {\boldsymbol{\theta}}_2)
\nonumber\\
&&~~~~~~~~~~~~~~~~~~~~~~~~~~
 +o_P\left(||^{(1)}\widehat{{\boldsymbol{\theta}}}_{\beta}-{\boldsymbol{\theta}}_1||^2\right)
+o_P\left(||^{(2)}\widehat{{\boldsymbol{\theta}}}_{\beta}-{\boldsymbol{\theta}}_2||^2\right)\nonumber\\
&=& 2\left[\boldsymbol{\Psi}_1({\boldsymbol{\theta}}_1,{\boldsymbol{\theta}}_2)^T
\left(^{(1)}\widehat{{\boldsymbol{\theta}}}_{\beta}-{\boldsymbol{\theta}}_1\right) +\boldsymbol{\Psi}_2({\boldsymbol{\theta}}_1,{\boldsymbol{\theta}}_2)^T
\left(^{(2)}\widehat{{\boldsymbol{\theta}}}_{\beta}-{\boldsymbol{\theta}}_2\right)\right]^T\widetilde{{\boldsymbol{\Sigma}}_{\beta}}({\boldsymbol{\theta}}_1, {\boldsymbol{\theta}}_2)^{-1}\boldsymbol{\psi}({\boldsymbol{\theta}}_1, {\boldsymbol{\theta}}_2)
\nonumber\\
&& ~~~~~~~~~~~~~~~~~~~~~~~~~~~~
+o_P\left(||^{(1)}\widehat{{\boldsymbol{\theta}}}_{\beta}-{\boldsymbol{\theta}}_1||^2\right)
+o_P\left(||^{(2)}\widehat{{\boldsymbol{\theta}}}_{\beta}-{\boldsymbol{\theta}}_2||^2\right).\nonumber
\end{eqnarray}  
Then, the theorem follows from the asymptotic distributions of the MDPDEs
$^{(1)}\widehat{{\boldsymbol{\theta}}}_{\beta}$ and $^{(2)}\widehat{{\boldsymbol{\theta}}}_{\beta}$.   
\hfill{$\square$}

\subsection{Proof of Theorem \ref{THM:2sample_gen_asympContg}}
  Using the asymptotic distribution of 
  $\sqrt{n}(^{(1)}\widehat{{\boldsymbol{\theta}}}_{\beta} - {\boldsymbol{\theta}}_{1,n})$ and 
  $\sqrt{n}(^{(2)}\widehat{{\boldsymbol{\theta}}}_{\beta} - {\boldsymbol{\theta}}_{2,m})$ under $H_{1,n,m}$
  and continuity of ${\boldsymbol{\Sigma}}_{\beta}({\boldsymbol{\theta}}_0)$, we have, as $m,n \rightarrow\infty$, 
  $^{(2)}\widehat{{\boldsymbol{\theta}}}_{\beta} \displaystyle\mathop{\rightarrow}^\mathcal{P}{\boldsymbol{\theta}}_0$,
  $$
  \sqrt{\frac{mn}{m+n}} \left( ^{(1)}\widehat{{\boldsymbol{\theta}}}_{\beta} - {\boldsymbol{\theta}}_{10} \right) 
  \displaystyle\underset{m, n\rightarrow \infty }{\overset{\mathcal{L}}{\longrightarrow }}  N(\sqrt{\omega}{\boldsymbol{\Delta}}_1, \omega {\boldsymbol{\Sigma}}_{\beta}({\boldsymbol{\theta}}_1))
  $$
  and
  $$
  \sqrt{\frac{mn}{m+n}} \left( ^{(2)}\widehat{{\boldsymbol{\theta}}}_{\beta} - {\boldsymbol{\theta}}_{20} \right) 
  \displaystyle\underset{m, n\rightarrow \infty }{\overset{\mathcal{L}}{\longrightarrow }}  N(\sqrt{1-\omega}{\boldsymbol{\Delta}}_2, (1-\omega) {\boldsymbol{\Sigma}}_{\beta}({\boldsymbol{\theta}}_2)).
  $$
  Hence,  following the proof of Theorem \ref{THM:2sample_simple_asympCont1}, we get under $H_{1,n,m}$ 
  $$
  \sqrt{\frac{mn}{m+n}}\boldsymbol{\psi}\left( ^{(1)}\widehat{{\boldsymbol{\theta}}}_{\beta},^{(2)}\widehat{{\boldsymbol{\theta}}}_{\beta} \right) 
  \underset{m, n\rightarrow \infty }{\overset{\mathcal{L}}{\longrightarrow }}  
  N\left(\left[\sqrt{\omega}\boldsymbol{\Psi}_1({\boldsymbol{\theta}}_1,{\boldsymbol{\theta}}_2)^T{\boldsymbol{\Delta}}_1 
  + \sqrt{1-\omega}\boldsymbol{\Psi}_2({\boldsymbol{\theta}}_1,{\boldsymbol{\theta}}_2)^T{\boldsymbol{\Delta}}_2\right],  
  \widetilde{{\boldsymbol{\Sigma}}_{\beta}}({\boldsymbol{\theta}}_1,{\boldsymbol{\theta}}_2)\right),
  $$
from which the theorem follows immediately.
  \hfill{$\square$}

\subsection{Proof of Theorems \ref{THM:2sample_gen_asympCont}  and \ref{THM:2sample_gen_PIF}}
These proofs are similar to that of Theorems \ref{THM:2sample_simple_asympCont}  and 
\ref{THM:2sample_simple_PIF} and hence omitted.
 \hfill{$\square$}


\end{document}